\documentclass[3p,preprint,10pt,authoryear]{elsarticle}
\usepackage[dvipsnames]{xcolor}

\newcommand{\rev}[1]{\textcolor{black}{#1}}

\usepackage{amssymb}
\usepackage{amsmath}
\usepackage{color}
\usepackage{lineno}
\usepackage{subcaption}
\usepackage{bbm}
\usepackage[sans]{dsfont}
\usepackage[mathscr]{eucal}
\renewcommand{\d}{\textrm{d}}
\newcommand{\tripledot}{\mathrel{\substack{\cdot \\[-3.5pt] \cdot \\[-3.5pt] \cdot}}}
\allowdisplaybreaks

\journal{Journal of the Mechanics and Physics of Solids}

\begin{document}

\begin{frontmatter}



\title{The mechanics of anisotropic active plates with applications to cell alignment on curved substrates} 


\author[label1]{Gabriele Fioretto}
\author[label2]{Giulio Lucci}
\author[label1]{Chiara Giverso} 
\author[label1]{Luigi Preziosi} 

\affiliation[label1]{organization={Dept. of Mathematical Sciences “G.L. Lagrange”, Politecnico di Torino},
            addressline={Corso Duca degli Abruzzi 24}, 
            city={Turin},
            postcode={10129}, 
            country={Italy}}

\affiliation[label2]{organization={Dept. of Structural and Geotechnical Engineering, Sapienza University of Rome},
            addressline={Via Eudossiana 18}, 
            city={Rome},
            postcode={00184}, 
            country={Italy}}

\begin{abstract}
We develop a continuum mechanics framework for active anisotropic plates within the Föppl–von Kármán limit, incorporating a preferential direction and inelastic active contractions in geometrically nonlinear plate theory. Through asymptotic expansion, we derive coupled equilibrium equations for plates with transversely isotropic and possibly inhomogeneous reinforcement undergoing spatially varying active contractions through their thickness. The framework highlights the coupling between material anisotropy and active deformations, with target curvatures that compete with geometric constraints. To demonstrate its capabilities, we apply the model to curvature-induced cell alignment, where substrate geometry, cytoskeletal anisotropy, and contractility interact to determine orientation. For cylindrical substrates, the model predicts a supercritical bifurcation in preferred orientation, from perpendicular to parallel through an oblique orientation, governed by the ratio of active contractility to substrate curvature. For ellipsoidal geometries, we capture stable parallel, perpendicular, and oblique configurations set by principal curvatures, whereas spherical substrates show no preferred alignment. These predictions qualitatively reproduce experimental observations across cell types, providing a mechanistic interpretation of the distinct behaviors of contractile epithelial cells and stiffer fibroblasts. As a further illustration, we analyze the buckling of active anisotropic rings, showing how reinforcement and contractility jointly modulate the instability threshold. More broadly, the model applies to thin fiber-reinforced active structures arising in soft robotics, morphogenesis, and tissue engineering.
\end{abstract}







\end{frontmatter}



\section{Introduction}

Thin structures exhibiting both anisotropic material response and active internal strains represent a class of mechanical systems with applications spanning multiple scales and domains. From liquid crystal elastomers and shape-morphing thin films \citep{mihai2020plate, pezzulla2016morphing} to cell layers and biological tissues \citep{wang2024curling}, understanding the interplay between anisotropic elasticity, geometric nonlinearity, and active strain generation is crucial for predicting morphogenetic processes and designing programmable materials. While the mechanics of passive anisotropic plates and active isotropic systems have been studied extensively, a comprehensive theoretical framework unifying these features, particularly within the geometrically nonlinear Föppl--von Kármán (FvK) regime, has remained elusive. 

A compelling application motivating such a framework arises from cellular mechanics, specifically the behavior of adherent cells on curved substrates. Experimental observations reveal that cells actively reorganize their cytoskeleton in response to substrate geometry: preferred orientation angles depend on curvature magnitude and on the trade-off between passive elastic deformation and active contractility along stress fibers. Mechanically, alignment arises from a competition between bending elasticity, which resists substrate-induced deformation, and directionally organized contractile stresses. Because stress fibers and microtubules form aligned bundles, cells behave effectively as anisotropic materials with direction-dependent stiffness. Experiments on cylindrical substrates have systematically demonstrated curvature-dependent orientation across cell types. For example, \citet{bade2017curvature} reported axial alignment of human vascular smooth muscle cells (hVSMCs) and fibroblasts as wire radius decreased, with reorientation thresholds that differ between cell types (mouse embryonic fibroblasts showed stronger alignment than hVSMCs, plausibly related to higher F-actin content). In contrast, \citet{yevick2015architecture} observed that epithelial monolayers tend to align circumferentially on cylinders with small radii, while \citet{yu2018substrate} and \citet{bade2017curvature} highlighted the decisive role of active contractility: Rho activation can induce thick circumferential stress fibers (reversing alignment tendencies), whereas inhibition of Rho kinase reduces myosin-driven contractility and promotes perpendicular orientation. These studies report deflections comparable to or exceeding cell thickness, and radii of curvature down to $R=200\ \mu\text{m}$, justifying a moderately large-deflection framework and emphasizing the interplay between substrate curvature, anisotropic cytoskeletal organization, and activity in determining cellular alignment. This complexity goes beyond the more studied problem of reorientation on planar substrates, which has been investigated both experimentally \citep{livne2014cell, roshanzadeh2020sf} and theoretically \citep{ lucci2021nonlinear, lucci2021visco, ciambella2022reorientation},  requiring a formulation that explicitly handles intrinsic geometry (see also \citep{giverso2023cell} and references therein).

One of the few theoretical frameworks developed to capture these observations was proposed by \cite{biton2009cellular} who postulated that an interplay between passive bending deformation and active cellular contractility gives rise to an optimal orientation. While pioneering, this model incorporates stress fiber contraction in an experimentally motivated way, without fully resolving the coupling between geometry, anisotropy, and activity. Although over a decade has passed since its publication, to the best of our knowledge no refinements of this framework based on 3D continuum mechanics have been proposed, while biological interest in this phenomenon remains high \citep{callens2020substrate, schamberger2023curvature}.

In this work, we develop a comprehensive continuum mechanical framework for transversely isotropic active plates undergoing moderately large deflections. Our approach is grounded in three-dimensional elasticity with distortions, accounting for the decomposition of the total deformation gradient into passive elastic and active components, thereby extending the classical FvK theory to incorporate both material anisotropy and active strains through a systematic asymptotic reduction. This construction generalizes growth-based models \citep{dervaux2008prl, dervaux2009morphogenesis, wang2024morphomechanics, ackermann2022bilayer, iakunin2018biofilms, liang2009leaf},  formulations for nematic elastomers \citep{mihai2020plate} and tissues as active liquid crystal shells \citep{khoromskaia2023epithelial}, as well as surfaces with prescribed incompatible metrics \citep{efrati2009noneuclidean, klein2007metrics, zurlo2017prl} or pre-stress \citep{ciarletta2022prestress}, by providing a framework that rigorously couples activity to anisotropic elasticity. 

The resulting model consists of coupled nonlinear partial differential equations governing in-plane membrane stresses and out-of-plane deflections, with constitutive parameters reflecting both passive elastic anisotropy and active strain magnitude. The coupling between these two aspects naturally emerges in the final equations, which we write both for the case of spatially varying and constant anisotropic director fields. Furthermore, we demonstrate that introducing differential active strains along the plate thickness, essential for describing the contraction of cell stress fibers, produces a significant source of incompatibility and results in active bending moments. 

In the context of cellular mechanics, this framework allows us to  analyze how bending rigidity, anisotropic cytoskeletal stiffness, and active contractility jointly determine the orientation of cells adhering to curved substrates. The theory predicts curvature- and contractility-dependent alignment transitions on different prototypical surfaces, revealing that the preferred orientation arises from a minimization of the energy involving bending resistance and active contraction. Predictions are qualitatively compared against experimental observations, suggesting how differences in cytoskeletal organization -- for example, the higher F-actin content of fibroblasts compared to epithelial cells – may give rise to distinct mechanical responses under identical geometric conditions. 

Although this study focuses on adherent cell mechanics, the framework also applies more generally to thin active anisotropic structures encountered in solid mechanics, including liquid crystal elastomer sheets and layered active composites. In such systems, much like in cells, the interplay between direction-dependent stiffness and internal stress generation governs shape adaptation and mechanical response.

The remainder of the paper is organized as follows. Section~\ref{sec:active_formalism} introduces the three-dimensional formulation, notation, and constitutive assumptions for transversely isotropic active bodies. Section~\ref{sec:plate_model} performs the asymptotic expansion to derive the equilibrium equations for active anisotropic FvK plates. \rev{Section~\ref{sec:example} illustrates an application of the framework by analyzing the buckling of an active anisotropic ring.} Then, in Section~\ref{sec:applications} we apply the model to substrates of cylindrical and ellipsoidal shape to determine cell preferential alignment and compare our results with experimental literature. Finally, Section~\ref{sec:conclusions} summarizes the results and outlines directions for future research.

\section{Theoretical framework}
\label{sec:active_formalism}

In this section, we introduce the general theoretical framework for a three-dimensional hyperelastic active body with anisotropic reinforcement. This framework forms the basis for the dimensional reduction performed in the next section, which yields a Föppl--von Kármán plate formulation. The nonlinearity of the model necessitates careful treatment of the two contributing deformation mechanisms: the passive bending, imposed by the substrate geometry, and the active cellular response. 
Active deformations are typically modeled using one of two approaches \citep{giantesio2019comparison, nardinocchi2007active}: (i) a multiplicative decomposition of the deformation gradient, or (ii) an additive decomposition of the stress tensor. Both approaches are mathematically consistent and widely used in modeling active materials, although the active stress approach may present certain constitutive issues \citep{ambrosi2012active}. In this work, we favor the multiplicative decomposition due to its conceptual clarity. This approach has also been employed in previous studies on Föppl--von Kármán inelastic plates \citep{dervaux2009morphogenesis, xu2020leaves, wang2024morphomechanics, ackermann2022bilayer} to model growth processes. 
Yet the coupling between activity and anisotropy in this plate limit remains unexplored. 

\subsection{Kinematics}

Let $\mathcal{B}_r\subset\mathcal{E}$ be the \textit{reference configuration} of the body, identified with a regular subregion of the Euclidean space $\mathcal{E}$,  and let $\chi: \mathcal{B}_r\to \mathcal{E}$ be the vector field referred to as \textit{deformation map}, assigning to each point $\mathbf{X} \in \mathcal{B}_r$ a position $\mathbf{x} = \chi(\mathbf{X})$. The image of the body through this map, i.e., the set $\mathcal{B} = \chi(\mathcal{B}_r)$, is referred to as \textit{deformed configuration}. As usual, the \textit{displacement field} is denoted by $\mathbf{\mathbf{u}}: \mathcal{B}_r\to \mathcal{E}$, and defined by $\mathbf{u}(\mathbf{X}) = \mathbf{x}(\mathbf{X}) - \mathbf{X}$. 

The deformation of the body from the reference configuration to the deformed one is described by the \textit{deformation gradient} $\mathbb{F} = \nabla \chi = \mathbb{I} + \nabla \mathbf{u}$. This tensor characterizes the local variation of length and orientation of material line elements and is purely kinematic in nature. 
For a purely elastic material, the deformation gradient $\mathbb{F}$ defined above is sufficient to characterize the mechanical response of the body. However, the framework must be extended when there are changes in the microstructure of the material as occurs, for instance, in plasticity, growth processes, and active behavior. In these cases, the material undergoes distortions that are not directly associated with energy storage but rather with a variation in the reference configuration itself, requiring an additional level of complexity \citep{nardinocchi2007active}.  Mathematically, this is achieved by introducing an additional kinematic variable $\mathbb{F}_{\rm a}$, and the deformation gradient $\mathbb{F}$ follows the multiplicative decomposition  \citep{rodriguez1994stress,sadik2017mult}
\begin{equation}
    \mathbb{F} = \mathbb{F}_{\rm e}\mathbb{F}_{\rm a} \, .
    \label{eq: gradient_decomposition}
\end{equation}
In particular, the elastic part of the deformation $\mathbb{F}_{\rm e}$ accounts for \textit{elastic}  distortions and is the only contribution that stores energy in the body, while $\mathbb{F}_{\rm a}$ describes the \textit{active} behavior and is prescribed constitutively, possibly following its own evolution law. This multiplicative decomposition can be visualized through the introduction of an intermediate \textit{natural state} $\mathcal{B}_\nu$, as shown in Fig.~\ref{fig:schematic_plate}, and has been widely used in contexts such as plasticity \citep{lubarda2004review}, viscoelasticity \citep{sadik2024visco,ciambella2024visco}, biological growth \citep{grillo2023growth,fraldi2018growth}, and active materials mechanics \citep{giantesio2019comparison,nardinocchi2007active}, just to mention a few examples. The determinants will be denoted by $J = \det\mathbb{F} > 0$, $J_{\rm a} = \det\mathbb{F}_{\rm a} > 0$, and $J_{\rm e} = \det\mathbb{F}_{\rm e} >0$, such that $J = J_{\rm e} J_{\rm a}$.

Note that $\mathbb{F}$ is a \textit{gradient} by definition, whereas $\mathbb{F}_{\rm a}$ and $\mathbb{F}_{\rm e}$ are generally not, that is, they may not be integrable. In particular, assuming the body to be simply connected, compatibility of $\mathbb{F}_{\rm a}$ is granted if and only if $\operatorname{curl}\mathbb{F}_{\rm a} = \mathbb{O}$. Therefore, an \textit{incompatible} tensor $\mathbb{F}_{\rm a}$ implies that, locally, an elastic deformation $\mathbb{F}_{\rm e}$ is needed to restore overall compatibility and to ensure that their product is a gradient. Indeed, if the active deformations could be described by a gradient of a displacement field, there would only be an evolution of the reference configuration. In general, however, the active deformation tensor appears as incompatible, leading to the development of residual stresses \citep{dervaux2009morphogenesis, efrati2009noneuclidean}. 

The multiplicative decomposition introduces the standard strain measures $\mathbb{C} := \mathbb{F}^{\rm T}\mathbb{F}$ and $\mathbb{B} := \mathbb{F}\mathbb{F}^{\rm T}$, where $\mathbb{C}$ and $\mathbb{B}$ are respectively the right and left Cauchy--Green strain tensors, and the corresponding elastic strain measures, computed as
\begin{equation}
    \mathbb{C}_{\rm e} = \mathbb{F}_{\rm e}^{\rm T}\mathbb{F}_{\rm e} =  \mathbb{F}_{\rm a}^{\rm -T}\mathbb{C} \mathbb{F}_{\rm a}^{-1} \, , \qquad \mathbb{B}_{\rm e} = \mathbb{F}_{\rm e}\mathbb{F}_{\rm e}^{\rm T} =  \mathbb{F}\mathbb{C}_{\rm a}^{-1}\mathbb{F}^{\rm T} \, ,
    \label{eq:cauchy_elastic}
\end{equation}
where $\mathbb{C}_{\rm a} := \mathbb{F}_{\rm a}^{\rm T}\mathbb{F}_{\rm a}$ describes the strains induced by the active deformation field.

The elastic Green--Lagrange strain tensor $\mathbb{E}_{\rm e}$ is then
\begin{equation}
    \mathbb{E}_{\rm e}  := \frac{1}{2}(\mathbb{C}_{\rm e} - \mathbb{I}) =  \frac{1}{2} \bigg( \mathbb{F}_{\rm a}^{\rm -T}\mathbb{F}_{\rm a}^{-1} + \mathbb{F}_{\rm a}^{\rm -T} \nabla\mathbf{u}^{\rm T}\mathbb{F}_{\rm a}^{-1} + \mathbb{F}_{\rm a}^{\rm -T} \nabla\mathbf{u}\mathbb{F}_{\rm a}^{-1} + \mathbb{F}_{\rm a}^{\rm -T} \nabla\mathbf{u}^{\rm T}\nabla\mathbf{u}\mathbb{F}_{\rm a}^{-1} - \mathbb{I} \bigg) \, .
    \label{eq:GreenLagrange_def}
\end{equation}
As specified before, the active deformation tensor $\mathbb{F}_{\rm a}$ in general may follow its own dynamics, and its expression is necessary to compute the macroscopic deformation field. Typically, material symmetries or physical observations are used to impose restrictions on the structure of $\mathbb{F}_{\rm a}$, which is assumed \emph{a priori}. The general structure of this tensor can be written as 
\begin{equation*}
    \mathbb{F}_{\rm a} = \alpha_{\rm a} \mathbf{v}_1 \otimes \mathbf{v}_1 +\beta_{\rm a} \mathbf{v}_2\otimes \mathbf{v}_2 + \gamma_{\rm a} \mathbf{v}_3 \otimes\mathbf{v}_3 \, ,
\end{equation*}
where $\mathbf{v}_i$, $i = 1, \dots, 3$ are the principal directions of activation, often related to the alignment of active fibers, while $\alpha_{\rm a}$, $\beta_{\rm a}$, $\gamma_{\rm a}$ are scalar fields that may depend on external stimuli. For instance, a simple structure where there is only one preferential direction for activation has been used in \citep{riccobelli2019activation} to model muscle contraction, whereas an orthotropic fiber alignment has been employed in heart mechanics \citep{nardinocchi2013cardio}.

\subsection{Constitutive assumptions}
We postulate the existence of a strain energy function per unit volume of the natural state \(\mathscr{W}\) dependent only on the \textit{elastic} distortion tensor \(\mathbb{F}_{\rm e}\) and such that there exists a minimal energy state \citep{nardinocchi2007active,giantesio2019comparison}. This is summarized as
\begin{equation}
    \mathscr{W}: \mathbb{F}_{\rm e} \to \mathscr{W}(\mathbb{F}_{\rm e}), \quad \mathscr{W}(\mathbb{I}) = 0 \, .
\end{equation}
It is important to note that the minimal energy state can be reached for a non-vanishing visible deformation:
\begin{equation}
    \mathbb{I} = \mathbb{F}_{\rm e} = \mathbb{F}\mathbb{F}_{\rm a}^{-1} \quad \iff \quad \mathbb{F} = \mathbb{F}_{\rm a} \, .
\end{equation}
This identity carries an important physical meaning: if the active distortion \(\mathbb{F}_{\rm a}\), which changes the reference configuration through the material microstructure, is the gradient of a given vector field, then the observed macroscopic deformation is identified by that gradient. Hence, the deformed state is the minimum of the energy, i.e., it is a stress-free configuration. 

In the following, we will consider a transversely isotropic material, with a preferential direction locally identified by the unit vector \(\mathbf{m}(\mathbf{X})\) in the reference configuration. The corresponding structural tensor is denoted by \(\mathbb{M} = \mathbf{m}\otimes\mathbf{m}\). Therefore, the strain energy density function, as a result of frame indifference and material symmetry, may be written as:
\begin{equation}
\label{eq:energy_W_invariants}
    \mathscr{W}(\mathbb{F}_{\rm e}, \mathbb{M}) = \widehat{\mathscr{W}}(I_1, I_2, I_3, I_4, I_5) \, ,
\end{equation}
where \(I_1, \dots, I_5\)  are the isotropic and anisotropic elastic invariants: 
\begin{equation*}
    I_1 := \mathbb{C}_{\rm e} : \mathbb{I} \, , \quad I_2 := \mathbb{C}_{\rm e}^*:\mathbb{I} \, , \quad I_3 := \det\mathbb{C}_{\rm e} \, , \quad I_4 := \mathbb{C}_{\rm e} : \mathbb{M} \, , \quad I_5 := \mathbb{C}_{\rm e}^2 : \mathbb{M} \, , 
\end{equation*}
with \(\mathbb{C}_{\rm e}\) defined in Eq.~\eqref{eq:cauchy_elastic}, $\mathbb{C}_{\rm e}^* := (\det\mathbb{C}_{\rm e})\mathbb{C}_{\rm e}^{\rm -1}$, and the operator \(:\) denoting double contraction, i.e. the scalar product between second-order tensors. 
We remark that, in writing Eq.~\eqref{eq:energy_W_invariants}, we have implicitly assumed that the material structure remains unaltered by active distortions.

The material will be assumed as \textit{elastically incompressible}, a condition described by the constraint
\begin{equation}
    \mathcal{C}(\mathbb{F}_{\rm e}) := \det\mathbb{F}_{\rm e} - 1 = 0 \quad \Rightarrow \quad \det\mathbb{F} = \det\mathbb{F}_{\rm a} = J \, ,
    \label{eq:incompr_constrain}
\end{equation}
which introduces a Lagrange multiplier \(p\) associated with this constraint. Then, we define an incompressible strain energy density function \(\widetilde{\mathscr{W}}\) per unit volume of the reference state as
\begin{equation}
    \widetilde{\mathscr{W}}(\mathbb{F}, \mathbb{F}_{\rm a}, \mathbb{M}) = J_{\rm a}\widehat{\mathscr{W}}(\mathbb{F}\mathbb{F}_{\rm a}^{-1}, \mathbb{M}) - J_{\rm a} \, p\left(\det(\mathbb{F}\mathbb{F}_{\rm a}^{-1}) - 1\right).
    \label{eq:tildeW}
\end{equation}
In what follows, we assume that the strain energy per unit volume of the natural state is given by the sum of a Neo-Hookean model and a quadratic reinforcement:  
\begin{equation}
    \widehat{\mathscr{W}}(I_1, I_4) = \frac{\mu}{2} (I_1 - 3)  + \frac{k_{44}}{2}(I_4 - 1)^2 \, .
    \label{eq:hatW}
\end{equation}
Note that we have omitted the dependence on the second invariant $I_2$ and on the fifth invariant $I_5$. This choice is motivated by the fact that, as we will show later, due to displacements scalings one can prove that the contribution of $I_2$ can be absorbed into $I_1$ \citep{dervaux2009morphogenesis}, while the contribution of $I_5$ is equivalent to that of $I_4$.

Therefore, the \textit{first Piola-Kirchhoff stress tensor} \(\mathbb{P}\) reads 
\begin{equation}
    \mathbb{P} = \frac{\partial \widetilde{\mathscr{W}}}{\partial \mathbb{F}} = J_{\rm a}\frac{\partial \widehat{\mathscr{W}}}{\partial \mathbb{F}} - J_{\rm a}  p \frac{\partial \mathcal{C}(\mathbb{F}_{\rm e})}{\partial \mathbb{F}} = J\frac{\partial \widehat{\mathscr{W}}}{\partial \mathbb{F}_{\rm e}}\mathbb{F}_{\rm a}^{\rm -T} - Jp\mathbb{F}^{\rm -T} \, .
\end{equation}
The \textit{second Piola-Kirchhoff stress tensor} \(\mathbb{S}\) is defined by \(\mathbb{S} = \mathbb{F}^{-1}\mathbb{P}\). Finally, the \textit{Cauchy stress tensor} \(\mathbb{T}\) is related to the first Piola-Kirchhoff stress tensor by:
\begin{equation}
    \mathbb{T} = J^{-1}\mathbb{P}\mathbb{F}^{\rm T} =  \frac{\partial \widehat{\mathscr{W}}}{\partial \mathbb{F}_{\rm e}}\mathbb{F}_{\rm e}^{\rm T} - p\mathbb{I} \, .
\end{equation}
The choice of the strain energy density in Eq.~\eqref{eq:hatW} implies a Cauchy stress tensor of the form:
\begin{equation}
    \mathbb{T} = \mu \mathbb{B}_{\rm e} + 2 k_{44}(I_4 - 1)\mathbb{F}_{\rm e} \mathbb{M}\mathbb{F}_{\rm e}^{\rm T} - p\mathbb{I} \,,
    \label{eq:CauchyStress}
\end{equation}
where we used the relations 
\begin{equation}
    \frac{\partial I_1}{\partial \mathbb{F}_{\rm e}} = 2\mathbb{F}_{\rm e} \, , \quad \quad \frac{\partial I_4}{\partial\mathbb{F}_{\rm e}} = 2\mathbb{F}_{\rm e} \mathbb{M} \, .
\end{equation}

\section{Active anisotropic plate model}
\label{sec:plate_model}

Building on the framework of Sect.~\ref{sec:active_formalism}, the goal of this section is to exploit the geometrical properties of the body to perform a dimensional reduction. Many natural systems exhibit a characteristic length scale that is much smaller than the other two, so that the body effectively behaves as a plate. 
Starting from the three-dimensional continuum formulation summarized above, several strategies can be used to perform this dimensional reduction, including \(\Gamma\)--convergence \citep{lewicka2011foppl}, asymptotic expansions of the equilibrium equations \citep{kaplunov1998dynamics}, or the identification of leading-order contributions in the three-dimensional stored energy \citep{audoly2005elasticity}. For the present purposes, we adopt the last approach, following related mathematical models in the recent literature \citep{dervaux2008prl,dervaux2009morphogenesis,mihai2020plate,wang2024morphomechanics, xu2020leaves, ackermann2022bilayer, ciarletta2022prestress}. In particular, we assume that the vertical displacement of the plate is comparable to its thickness, corresponding to a moderate-deflection regime. This setting is commonly referred to as the Föppl--von Kármán  plate limit, which describes the situation in which the  stretching and bending energies of the plate are of the same order of magnitude, often resulting in systems of equations with the following characteristic structure: 
\begin{equation}
\begin{cases}
D \Delta^2 \xi - H[\xi, \phi] = f_{\xi,\text{inc}} \\[0.3ex]
\Delta^2 \phi + E[\xi, \xi] = f_{\phi,\text{inc}}
\end{cases}
\label{eq:Foppl_von_Karman_syst}
\end{equation}
where $\xi$ represents the deflection, $\phi$ is the Airy stress function, while $D, E$ depend on the material properties and the terms $f_{\xi, \rm inc}$, $f_{\phi, \rm inc}$ encode the effect of incompatible deformations, such as growth or active remodeling. 
The nonlinear coupling between equations arises through the bilinear operator $[\cdot,\cdot]$, defined as 
\begin{equation}
     [a,b] := \frac{1}{2}\frac{\partial^2 a}{\partial X^2}\,\frac{\partial^2 b}{\partial Y^2} + \frac{1}{2}\frac{\partial^2 a}{\partial Y^2}\,\frac{\partial^2 b}{\partial X^2} - \frac{\partial^2 a}{\partial X\partial Y}\,\frac{\partial^2 b}{\partial X\partial Y} \, .
     \label{eq:operator_sqbracket}
\end{equation}
Under these conditions, we derive the equilibrium equations for an anisotropic active plate of FvK type. To our knowledge, current geometrically nonlinear plate models with incompatible strains do not incorporate constitutive anisotropy. However, the presence of reinforcing fibers must be accounted for in several contexts and is essential for capturing the phenomenon of cell alignment on curved surfaces. Our model extends existing frameworks by incorporating constitutive anisotropy that can vary spatially throughout the plate and by capturing its interplay with active deformation, which naturally emerges in the derivation.


\subsection{Setup and assumptions}
\label{sec:setup}

\begin{figure}
    \centering
    \includegraphics[width=\textwidth]{plate_config_2.png}
    \caption{Schematic of the reference configuration $\mathcal{B}_r$ of the plate and of the decomposition $\mathbb{F} = \mathbb{F}_{\rm e}\mathbb{F}_{\rm a}$. The middle surface of the plate is denoted by $\mathcal{S}$.} 
    \label{fig:schematic_plate}
\end{figure}

Let the reference configuration be a flat plate with thickness \(H\), that is, \(\mathcal{B}_r = \mathcal{S} \times \left[-\frac{H}{2}, \frac{H}{2}\right]\), where \(\mathcal{S}\) is the middle surface that describes the shape of the plate, as shown in Fig.~\ref{fig:schematic_plate}. In particular, we assume that the lateral dimensions of the plate are of order $L$, with $H \ll L$. Introducing a system of Cartesian coordinates, a material point will be denoted by
\begin{equation*}
    \mathbf{X} = X\mathbf{e}_1 + Y \mathbf{e}_2 + Z\mathbf{e}_3 \, ,
\end{equation*}
where \(\mathbf{e}_\alpha\) (\(\alpha = 1, 2\)) are the in-plane directions, while \(\mathbf{e}_3\) is the thickness direction, hereafter referred to as the \textit{transverse direction}. Thus, the displacement field writes as
\begin{equation}
    \mathbf{u}(X,Y,Z) = u_\alpha(X,Y,Z) \mathbf{e}_\alpha + u_3(X,Y,Z)\mathbf{e}_3 \, .
\end{equation}
In the equation above, and throughout the remainder of this work, we adopt the Einstein summation convention. Specifically, Greek indices refer to the $1$ and $2$ coordinates, whereas Latin indices are used when considering both in-plane and transverse components. 
 
Let us denote the transverse deflection of the central surface \(Z=0\) of the plate by the field \(\xi\), such that \(u_3(X,Y,Z=0) = \xi(X,Y)\). Hence, the transverse component of the displacement can be  split as $u_3(X,Y,Z) = W(X,Y,Z) + \xi(X,Y)$, with $W(X,Y,Z=0) = 0$. Moreover, we assume that the anisotropy of the material does not have a transverse component, so that \(m_3 = 0\). This reflects the hypothesis that the material structure is confined \textit{in-plane} and notably simplifies the problem, as will be discussed later on (see Remark 3). 


Since the parameter \(\varepsilon = H/L\) is small, it is convenient to introduce a new transverse coordinate $\bar{Z} := Z/\varepsilon$, which varies in $[-L/2,L/2]$. An arbitrary field \(f:\mathcal{B}_r\to\mathcal{E}\) defined on the three-dimensional body can then be expanded as follows:
\begin{equation*}
    f(X,Y,Z) = \bar{f}(X,Y,\bar{Z}) = \bar{f}^{[0]}(X,Y,\bar{Z}) + \bar{f}^{[1]}(X,Y,\bar{Z}) \varepsilon + \bar{f}^{[2]}(X,Y,\bar{Z}) \varepsilon^2 + \mathcal{O}(\varepsilon^3) \, ,
\end{equation*}
where the symbol \(\mathcal{O}(\cdot)\) indicates that the quantity is, at most, of the order of its argument. This can be interpreted as a Taylor expansion of \(f\) around \(Z=0\), where \(Z\) is small under the plate assumptions.

Therefore, by considering the expansion of the terms involved in the strain energy density of Eq.~\eqref{eq:tildeW}, i.e., $I_1$, $I_4$, and $p$, the total energy of the body breaks down into two components \citep{dervaux2009morphogenesis}:
\begin{equation}
    \mathscr{W}_{\rm tot} = \int_S \int_{-H/2}^{H/2} J\left[ \widehat{\mathscr{W}}(I_1,I_4) - p(\det\mathbb{F}_{\rm e} - 1)  \right] \d Z \, \d A \approx H\int_S \psi^{[0]}\, \d A + \frac{H^3}{12} \int_S\psi^{[2]}\, \d A \, ,
    \label{eq:totenergy_intoSt_and_Be}
\end{equation}
where \(\psi^{[0]}\) and \(\psi^{[2]}\) contain the \(0\)$^{\rm th}$-order and \(2\)$^{\rm nd}$-order terms in \(\varepsilon\), respectively. This separation highlights the two contributions to the energy. The first term, proportional to $H$, accounts for the in-plane deformation of the middle surface \(\mathcal{S}\) and is referred to as the \textit{stretching energy}. On the other hand, the second term, which is proportional to \(H^3\), is referred to as the \textit{bending energy} and contains information about the deformation of the plate in the third dimension. The explicit form of these two contributions will be derived in Sect.~\ref{subsection:energy_equilibrium}. 

 In order to set the problem in a specific plate limit, it is necessary to make some assumptions on the scaling of the quantities involved. This reflects the fact that, depending on the scaling of the external forces applied to the body, we observe different final behaviors \citep{audoly2005elasticity}. Regarding the displacement field, we assume that
\begin{equation}
    u_{\alpha}(X,Y,Z) = \mathcal{O}\left(\frac{\xi^2}{L}\right) \, , \qquad \xi(X,Y) = \mathcal{O}(H)\, , \qquad  W(X,Y,Z) = \mathcal{O}\left(\frac{\xi^2 H}{L^2}\right) \, .
    \label{eq:scalings}
\end{equation}
The second hypothesis is of geometric nature: it describes the order of magnitude of the deflection \(\xi\), which in the classical Föppl–von Kármán scaling satisfies $\xi \sim H$. This implies that we are considering finite but moderate rotations, which will be the reason for geometric nonlinearity in the equilibrium equations. Instead, the first scaling assumption is necessary to ensure that the two energy components, bending and stretching, have a comparable order of magnitude. Note that the nonlinearities that arise in the equilibrium equations are not a consequence of large deformations, since the strains remain small (at least of order \(\varepsilon\), which is small by assumption). Rather, they have a geometric origin, since they arise from the compatibility between the strain and displacement fields in the model. 

Regarding the active deformation tensor, we take it in the form 
\begin{equation}
    \mathbb{F}_{\rm a} = \mathbb{I} + \mathbb{A} = \mathbb{I} + \mathbbm{a} + Z \mathbbm{k}_{\rm a} \, .
    \label{eq:active_deformation_1}
\end{equation}
In particular, we make the following assumptions on the structure of $\mathbb{F}_{\rm a}$: 
\begin{enumerate}
    \item $\mathbbm{a} = \mathbbm{a}(X,Y)$ and $\mathbbm{k}_{\rm a}=\mathbbm{k}_{\rm a}(X,Y)$, so that the dependence on the transverse coordinate $Z$ is linear and explicit; 
    \item $a_{33} = \mathbf{e}_3 \cdot \mathbbm{a}\mathbf{e}_3$ is constant in space; 
    \item $\mathbbm{k}_{\rm a}$ is assumed to have no transverse components, that is, $k_{\alpha 3} = k_{33} = 0$. 
\end{enumerate}
\rev{We further note that, as can be easily shown, only $\operatorname{sym}(\mathbbm{k}_{\rm a})$ contributes to the FvK reduction performed here. This stems from the fact that only $\operatorname{sym}(\mathbbm{k}_{\rm a})$ enters the target metric $\mathbb{C}_{\rm a}$ at first order in $Z$ (see Sect.~\ref{sec:compatibility}). Without loss of generality, we therefore take $\mathbbm{k}_{\rm a}$ symmetric in the following, keeping in mind that every occurrence of $k_{12}$ could in principle be replaced by $\frac{1}{2}(k_{12}+k_{21})$.} Following these choices, to highlight the spatial dependencies and the structure of $\mathbb{A}$, its expression can be rewritten as: 
\begin{equation*}
    \mathbb{A} = \left[\mathbbm{a}_{\rm 2D}(X,Y) + Z\mathbbm{k}_a(X,Y)\right] + \mathbf{a}_1(X,Y) \otimes \mathbf{e}_3 + \mathbf{e}_3 \otimes \mathbf{a}_3(X,Y) + a_{33} \mathbf{e}_3 \otimes \mathbf{e}_3 \, ,
\end{equation*}
where we explicitly highlighted the planar component $\mathbbm{a}_{\rm 2D}(X,Y)$ of $\mathbbm{a}$ and introduced $\mathbf{a}_1 := (a_{13}, a_{23}, 0)^{\rm T}$, $\mathbf{a}_3 := (a_{31}, a_{32}, 0)^{\rm T}$. \\
In addition, we impose the following scaling assumptions on the in-plane and transverse components of $\mathbb{A}$:
\begin{equation}
    a_{\alpha\beta} = \mathcal{O}(\xi^2/L^2) \, ,  \quad a_{\alpha 3} \hspace{2pt}, a_{3 \beta} =  \mathcal{O}(\xi/L) \, , \quad a_{33} = \mathcal{O}(\xi^2/L^2) \, , \quad k_{\alpha\beta} = \mathcal{O}(\xi/L^2) \, .
\label{eq:active_scaling}    
\end{equation}
These hypotheses on the active part ensure that the contribution of $\mathbb{F}_{\rm a}$ remains compatible with the assumptions of the Föppl--von Kármán limit.
In the particular case where $\mathbb{A} = \mathbbm{a}$, so that the active deformation field $\mathbb{F}_{\rm a}$ (or its approximation) is independent of the transverse coordinate $Z$,
the active part of the model is equivalent to the growth setting considered in  \citep{dervaux2009morphogenesis}.

\paragraph{Remark 1 (on the compatibility of $\mathbb{F}_{\rm a}$)} Following the definition of Eq.~\eqref{eq:active_deformation_1}, we have 
\begin{equation} \label{eq:curl_Fa}
\operatorname{curl}\mathbb{F}_{\rm a} = \operatorname{curl}\mathbbm{a} + Z\operatorname{curl}\mathbbm{k}_{\rm a} + [(\mathbf{e}_3)_{\times}]\rev{\mathbbm{k}_{\rm a}} \, ,
\end{equation}
where $[(\mathbf{e}_3)_{\times}]$ is the skew tensor whose axial vector is $\mathbf{e}_3$. Clearly, if $\mathbbm{k}_{\rm a} = \mathbb{O}$, the compatibility of $\mathbb{F}_{\rm a}$ is equivalent to the compatibility of $\mathbbm{a}$. Otherwise, it follows that: 
\begin{enumerate}
    \item if $\mathbbm{k}_{\rm a}$ is compatible, then it is possible to find $\mathbbm{a}$ (possibly incompatible) such that $\mathbb{F}_{\rm a}$ is compatible; 
    \item if $\mathbbm{a}$ is compatible, then the only way to obtain a compatible $\mathbb{F}_{\rm a}$ is $\mathbbm{k}_{\rm a} = \mathbb{O}$; 
    \item compatibility of both $\mathbbm{a}$ and $\mathbbm{k}_{   \rm a}$ does not guarantee a compatible $\mathbb{F}_{\rm a}$, in general.  
\end{enumerate}
In particular, the ability of the plate to exhibit $Z$-dependent contractions introduces a significant additional source of incompatibility into the model. 
\paragraph{\rev{Remark 2 (on the compatibility of $\mathbb{C}_{\rm a}$)}} \rev{We recall that, in a simply connected domain, as the one we consider here, a compatible (curl-free) active deformation $\mathbb{F}_{\rm a}$ implies that the associated metric tensor $\mathbb{C}_{\rm a} = \mathbb{F}_{\rm a}^{\rm T}\mathbb{F}_{\rm a}$ is also compatible, that is, its Riemann curvature tensor vanishes: $\mathsf{R}(\mathbb{C}_{\rm a}) = \mathsf{0}$ \citep{ciarlet2005, argento2021shaping, colorado2022morphing}. The induced metric is therefore flat and can be embedded in the Euclidean space without residual stresses. The converse is true in the sense that, if $\mathbb{C}_{\rm a}$ is a metric such that $\mathsf{R}(\mathbb{C}_{\rm a}) = \mathsf{0}$, then there exists an embedding $\chi$ such that $\nabla \chi^{\rm T} \nabla \chi = \mathbb{C}_{\rm a}$, but the choice of $\mathbb{F}_{\rm a}$ is not unique. In fact, if we set $\mathbb{F}_{\rm a} = \nabla \chi$, any superposed local rotation provides the same metric tensor $\mathbb{C}_{\rm a}$. However, the energy is insensitive to such rotation, as it can be absorbed into the elastic part of the deformation without altering any of the elastic invariants. The geometric meaning of metric compatibility in the plate limit is discussed in Sect.~\ref{sec:compatibility}.} \\

In the absence of externally applied forces, the boundary conditions for the Cauchy stress tensor along the transverse direction \(\mathbf{n}\), that is, the direction orthogonal to the central surface in the deformed configuration,  are summarized by the interface conditions
\begin{equation} \label{eq:Tn0}
    \mathbb{T}\,\mathbf{n} = \mathbf{0} \, . 
\end{equation}
\rev{At this point, we adopt the membrane assumption\footnote{\rev{This formulation accommodates elastic thickness changes through the incompressibility constraint, but does not introduce an independent kinematic descriptor for significant through-thickness distension; richer thin-body theories \citep{deseri2008derivation, deseri2013stretching, dicarlo2001shells} would be needed to capture such effects in regimes of highly inhomogeneous active deformation.}}  for moderate deflections \citep{dervaux2009morphogenesis, audoly2005elasticity}: since the plate is thin and the boundary is stress-free, cf. Eq.~\eqref{eq:Tn0}, we assume that $\mathbb{T}\mathbf{e}_3 =~\mathbf{0}$ throughout the plate at leading order.}
\rev{This assumption implies that} \((\mathbb{T}\,\mathbf{e}_3)_1 = (\mathbb{T}\,\mathbf{e}_3)_2 = 0 \) everywhere at leading order. These stress relations can be used to make the dependence on the transverse coordinate explicit and to reduce the dimensionality of the problem. 
Specifically, using the scaling assumptions of Eq.~\eqref{eq:scalings}, the definition of the elastic deformation $\mathbb{F}_{\rm e}$ from Eq.~\eqref{eq: gradient_decomposition}, and the Cauchy stress expression from Eq.~\eqref{eq:CauchyStress}, the vanishing of the out-of-plane shear stresses at leading order yields: 
\begin{align}
& \frac{\partial u_1}{\partial Z} = -\left(\frac{\partial \xi}{\partial X} - a_{13} - a_{31}\right) + \mathcal{O}\left(\xi^3 / L^3\right) \, , \label{eq:u1_Z} \\[0.5ex]
& \frac{\partial u_2}{\partial Z} = -\left(\frac{\partial \xi}{\partial Y} - a_{23} - a_{32}\right) + \mathcal{O}\left(\xi^3 / L^3\right) \label{eq:u2_Z} \, .
\end{align}
Note that the gradient of $W$ has been neglected, since it is of higher order compared to $\xi$. Moreover, the anisotropic term within the Cauchy stress \(\mathbb{T}\) contributes only at higher order due to the presence of the anisotropic invariant \(I_4 - 1\), which are therefore included in the term \(\mathcal{O}(\xi^3/L^3)\). This follows from the planarity assumption for the structure tensor \(\mathbb{M}\). If \(\mathbf{m}\) had an out-of-plane component, additional terms would appear at leading order, significantly complicating the analysis. Consequently, this ensures that the transverse components of the Cauchy stress, \(\mathbb{T}_{\alpha 3}\), vanish up to order \(\xi^3/L^3\) \citep{audoly2005elasticity}. 

The leading-order terms on the right-hand side of Eqs.~\eqref{eq:u1_Z}--\eqref{eq:u2_Z} consist of quantities that do not depend on the transverse coordinate. Indeed, \(\xi\) is defined as the vertical displacement of the middle surface \(\mathcal{S}\), while \(a_{\alpha3}\) are independent of \(Z\) by assumption. Therefore, upon integration, we find that the in-plane displacements exhibit the classical linear dependence on \(Z\):
\begin{equation}
\begin{split}
        u_1(X,Y,Z) & = -Z\left(\frac{\partial \xi}{\partial X} - a_{13} - a_{31}\right) + u_1^{(0)}(X,Y) \, , \\[0.5ex]
        u_2(X,Y,Z) & = -Z\left(\frac{\partial \xi}{\partial Y} - a_{23} - a_{32}\right) + u_2^{(0)}(X,Y) \, .
    \label{eq:displacemnt_Kirch}
    \end{split}
\end{equation}
In general, depending on the scaling of the in-plane displacements \(u_\alpha^{(0)}\), different plate models can be derived. Under the present scaling assumptions from Eq.~\eqref{eq:scalings}, we have \(u_\alpha^{(0)} = \mathcal{O}(\xi^2/L)\), effectively rendering all terms on the right-hand side of Eq.~\eqref{eq:displacemnt_Kirch} of the same order. We remark that the kinematic relations of Eq.~\eqref{eq:displacemnt_Kirch} are equivalent to those derived in \citep{dervaux2009morphogenesis}. However, as will become clear, in the following the presence of anisotropy and $Z$-dependent activity will significantly affect $u_1^{(0)}$, $u_2^{(0)}$, and $\xi$, leading to marked differences in the model compared to others. 

The last condition about transverse  stresses, which is \((\mathbb{T}\,\mathbf{e}_3)_3 = 0\), allows the computation of the unknown Lagrange multiplier \(p\):
\begin{equation*}
    p = \mu \,(\mathbb{B}_{\rm e})_{33} + 2 k_{44}(I_4 - 1) \, (\mathbb{F}_{\rm e} \mathbb{M} \mathbb{F}_{\rm e}^{\rm T})_{33} \, .
\end{equation*}
This expression can be simplified further by making the following observations:
\begin{itemize}
    \item  The displacement fields computed in Eq.~\eqref{eq:displacemnt_Kirch} from the membrane assumption allow us to conclude that, at leading order,
\begin{equation*}
    (\mathbb{B}_{\rm e})_{33} = (\mathbb{F}_{\rm e}\mathbb{F}_{\rm e}^{\rm T})_{33} \approx (\mathbb{F}_{\rm e}^{\rm T}\mathbb{F}_{\rm e})_{33} = (\mathbb{C}_{\rm e})_{33} \, .
\end{equation*}
  
\item  As noted previously, the planar assumption on \(\mathbb{M}\) implies that the anisotropic component of the Cauchy stress is proportional to
  \begin{equation*}
       (\mathbb{F}_{\rm e} \mathbb{M}\mathbb{F}_{\rm e}^{\rm T})_{33} = \mathcal{O}(\xi^2/L^2) \, ,
  \end{equation*} 
  which, after multiplication by \((I_4 - 1)\), results in a term which is at least of order \(\xi^4/L^4\). Therefore, this term is negligible compared to the isotropic one.
\end{itemize}
Thus, exploiting the identity \(\mathbb{C}_{\rm e} = 2\mathbb{E}_{\rm e} + \mathbb{I}\), where \(\mathbb{E}_{\rm e}\) is the elastic Green-Lagrange deformation tensor defined in Eq.~\eqref{eq:GreenLagrange_def}, the pressure may be approximated as:
\begin{equation}
    p \approx \mu (\mathbb{B}_{\rm e})_{33} \approx \mu\, (\mathbb{C}_{\rm e})_{33} = \mu \Bigl(2(\mathbb{E}_{\rm e})_{33} + 1\Bigr) \, .
    \label{eq:pressure_field}
\end{equation}
Note that the pressure can be divided into two components: one contains an isotropic term independent of the displacement field, while the other depends on the deformation through \((\mathbb{E}_{\rm e})_{33}\). Moreover, the anisotropic part does not contribute because of the in-plane structure hypothesis on $\mathbb{M}$. We observe that
\[
\det\mathbb{F}_{\rm e} = 1 \iff \det\mathbb{C}_{\rm e} = 1 \implies 2\operatorname{tr}\mathbb{E}_{\rm e} = 0 \quad \text{(at leading order)},
\]
and, consequently,
\begin{equation}
    p \approx \mu\Bigl(2(\mathbb{E}_{\rm e})_{33} + 1\Bigr) \approx \mu\Bigl(1 - 2(\mathbb{E}_{\rm e})_{\alpha\alpha}\Bigr) = \mu\Bigl(1 - 2(\mathbb{E}_{\rm e})_{11} - 2(\mathbb{E}_{\rm e})_{22} \Bigr) \, .
    \label{eq:pressure_approx}
\end{equation}
As discussed earlier, the strain energy function may, in general, depend on the full set of invariants, in particular on $I_2$ and $I_5$. However, the scalings associated with the Föppl--von Kármán regime~\eqref{eq:scalings} imply that, as shown in \citep{dervaux2009morphogenesis}, the contribution of $I_2$ becomes asymptotically equivalent to that of $I_1$. Moreover, to leading order, we have
\[       I_5 - 1 = \mathbb{C}_{\rm e}^2 : \mathbb{M} - 1 
       \approx 2(\mathbb{C}_{\rm e} : \mathbb{M} - 1) = 2(I_4 - 1) \, ,
\]
so that the $I_5$ contribution effectively reduces to that of $I_4$.  Therefore, our choice of the strain energy density $\widehat{\mathscr{W}}$ in the natural state 
represents a fairly general constitutive form for a transversely isotropic material in the FvK limit. 

The membrane assumption and the expansions of the relevant terms allow us to reduce the kinematics of the problem from 3D to 2D, with unknowns $u_1^{(0)}(X,Y)$, $u_{2}^{(0)}(X,Y)$, and $\xi(X,Y)$. In what follows, we derive the governing equations by variational arguments. 

\subsection{Leading order of the energy}
\label{subsection:energy_equilibrium}

Recalling Eq.~\eqref{eq:totenergy_intoSt_and_Be}, we now compute both the in-plane stretching energy density $\psi^{[0]}$ and the bending energy density $\psi^{[2]}$ by using the energetic conjugation between the elastic Green-Lagrange tensor and the second Piola--Kirchhoff stress tensor pushed forward in the natural state $\mathcal{B}_\nu$: 
\begin{equation}
\mathbb{S}_n = J_{\rm a}^{-1} \mathbb{F}_{\rm a} \mathbb{S} \mathbb{F}_{\rm a}^{\rm T} = J_{\rm e} \mathbb{F}_{\rm e}^{-1} \mathbb{T}\mathbb{F}_{\rm e}^{\rm -T} \, .
\end{equation}
Using the expression of the Cauchy stress and the fact that $J_{\rm e} = 1$, we find 
\begin{equation}
\mathbb{S}_n = \mu\,\mathbb{I} + 2 k_{44}(I_4 - 1)\mathbb{M} - p\,\mathbb{C}_{\rm e}^{-1} \,.
\end{equation}
The hypothesis from Eq.~\eqref{eq:scalings} about the leading order of the displacement,   combined with the leading order of the pressure field in Eq.~\eqref{eq:pressure_approx}, allows us to further simplify this tensor as 
\begin{equation} \label{eq:reduced_constitutive_relations}
\mathbb{S}_n = 2 \mu \Bigl[ \mathbb{E}_{\rm e} + \bigl( (\mathbb{E}_{\rm e})_{11} + (\mathbb{E}_{\rm e})_{22}\bigr)\mathbb{I}\Bigr] + 4k_{44} J_4 \mathbb{M} \, ,  
\end{equation}
where $J_4 := \mathbb{E}_{\rm e} : \mathbb{M} = \frac{1}{2}(I_4 - 1)$. 
Note that the in-plane components of this stress tensor depend solely on in-plane quantities. This result is typical of reduced plate models, and Eq.~\eqref{eq:reduced_constitutive_relations} is referred to as the \emph{effective constitutive equation}. 

Before deriving the energy expression, it is useful to compute the expansion along the transverse direction of the elastic strain tensor $\mathbb{E}_{\rm e}$. We write 
\begin{equation}
    \mathbb{E}_{\rm e} \approx \mathbb{E}_{\rm e}^{[0]} + Z \mathbb{E}_{\rm e}^{[1]} + Z^2 \mathbb{E}_{\rm e}^{[2]} 
    \label{eq:approx_greenLagrange}
\end{equation}
and, by using the scaling assumptions together with the kinematic definitions of Eq.~\eqref{eq:displacemnt_Kirch}, we find the following leading orders: 
\begin{align}
& \left(\mathbb{E}_{\rm e}^{[0]}\right)_{\alpha\beta} = \frac{1}{2} \left( -a_{\alpha\beta} - a_{\beta\alpha} - a_{3\alpha}a_{3\beta} + \frac{\partial u_\alpha^{(0)}}{\partial X_\beta} + \frac{\partial u_\beta^{(0)}}{\partial X_\alpha} + \frac{\partial \xi}{\partial X_\alpha}\frac{\partial \xi}{\partial X_\beta} \right) \, , \label{eq:E0}\\[0.8ex]  
& \left(\mathbb{E}_{\rm e}^{[1]}\right)_{\alpha\beta} = \frac{1}{2} \left( -k_{\alpha\beta} - k_{\beta\alpha} - 2\frac{\partial^2 \xi}{\partial X_\alpha \partial X_\beta} + \frac{\partial}{\partial X_\beta}(a_{\alpha 3} + a_{3\alpha}) + \frac{\partial}{\partial X_\alpha}(a_{\beta 3} + a_{3\beta})\right) \, , \label{eq:E1} \\[0.8ex]
& \left(\mathbb{E}_{\rm e}^{[2]}\right)_{\alpha\beta} = \frac{1}{2}\left( \frac{\partial^2\xi}{\partial X_\alpha \partial X_\gamma} - \frac{\partial}{\partial X_\alpha}(a_{\gamma 3} + a_{3\gamma}) \right) \left( \frac{\partial^2 \xi}{\partial X_\beta \partial X_\gamma} - \frac{\partial}{\partial X_\beta}(a_{\gamma 3} + a_{3\gamma}) \right) \, . \label{eq:E2}
\end{align}
Therefore, the total energy can be written as 
\begin{equation} \label{eq:Wtot_lin}
\mathscr{W}_{\rm tot} = \frac{1}{2} \int_{\mathcal{B}_n} \mathbb{S}_n : \mathbb{E}_{\rm e} \, \d V_n \approx \underbrace{\frac{H}{2}\int_{S} \mathbb{S}_n^{[0]} : \mathbb{E}_{\rm e}^{[0]} \, \d A}_{\mathscr{W}_{\rm str}} + \underbrace{\frac{H^3}{24}\int_S \mathbb{S}_n^{[1]} : \mathbb{E}_{\rm e}^{[1]} \, \d A}_{\mathscr{W}_{\rm bend}} \, .  
\end{equation}
Note that, in writing the bending energy, we have neglected the terms $\mathbb{S}_n^{[0]} : \mathbb{E}_{\rm e}^{[2]}$ and $\mathbb{S}_n^{[2]} : \mathbb{E}_{\rm e}^{[0]}$. Indeed, from Eqs.~\eqref{eq:E0}--\eqref{eq:E2}, we have $(\mathbb{E}_{\rm e}^{[0]})_{\alpha\beta} = \mathcal{O}(\xi^2/L^2)$, \((\mathbb{E}_{\rm e}^{[1]})_{\alpha\beta} = \mathcal{O}(\xi/L^2)\) and \((\mathbb{E}_{\rm e}^{[2]})_{\alpha\beta} = \mathcal{O}(\xi^2/L^4)\). Thus, the reduced constitutive equation \eqref{eq:reduced_constitutive_relations} yields $\mathbb{S}_n^{[0]}:\mathbb{E}_{\rm e}^{[2]} = \mathbb{S}_n^{[2]}:\mathbb{E}_{\rm e}^{[0]} = \mathcal{O}(\xi^4 / L^6)$, and these terms are of higher order. 

\paragraph{Remark 3 (on the dependence of $\mathbb{M}$ on $Z$)}
In the previous section, we made a single assumption on the vector field $\mathbf{m}$, 
namely that its transverse component vanishes, i.e., $m_3 = 0$. 
No additional hypotheses were introduced concerning the scaling of the in-plane components. 
Hence, in general, $\mathbf{m} = \mathbf{m}(\mathbf{X})$ may depend on the transverse coordinate $Z$. 
Accordingly, as done for the other fields, the structural tensor can be expanded as
\begin{equation*}
    \mathbb{M}(X,Y,Z) = \bar{\mathbb{M}}(X,Y,\bar{Z}) 
    = \bar{\mathbb{M}}^{[0]}(X,Y,\bar{Z}) 
    + \bar{\mathbb{M}}^{[1]}(X,Y,\bar{Z})\,\varepsilon 
    + \bar{\mathbb{M}}^{[2]}(X,Y,\bar{Z})\,\varepsilon^2 
    + \mathcal{O}(\varepsilon^3) \,.
\end{equation*}
If one assumes that $\bar{\mathbb{M}}^{[0]} = \mathcal{O}(1)$, 
which must hold since $\mathbb{I}:\mathbb{M} = 1$ by the definition of the unit vector $\mathbf{m}$, one obtains
\begin{equation}
    \frac{1}{2}\left(I_4 - 1\right) 
    = \frac{1}{2}\left(\mathbb{C}_{\rm e}:\mathbb{M} -1\right) 
    = \mathbb{E}_{\rm e}:\mathbb{M} 
    = (\mathbb{E}_{\rm e})_{\alpha\beta}\mathbb{M}_{\alpha\beta} \,,
\end{equation}
where, in the last step, we used $m_3 = 0$. The expansion above, combined with the approximation~\eqref{eq:approx_greenLagrange}, yields
\begin{equation*}
    \mathbb{E}_{\rm e} : \mathbb{M} 
    \approx \mathbb{E}_{\rm e}^{[0]} : \mathbb{M}^{[0]} + \bigl(\mathbb{E}_{\rm e}^{[1]} : \mathbb{M}^{[0]} + \mathbb{E}_{\rm e}^{[0]} : \mathbb{M}^{[1]}\bigr) Z \,.
\end{equation*}
Hence, the leading-order term is of order $\mathcal{O}(\xi^2/L^2)$.  Given the previously established scalings for $\mathbb{E}_{\rm e}$, a non-negligible contribution from $\mathbb{M}^{[1]}$ would require $\mathbb{M}^{[1]} = \mathcal{O}(1/\xi)$, implying a singular behavior across the thickness. In the following, we exclude such  behavior and assume
$\mathbf{m}$ to be independent of $Z$, so that $\mathbb{M}(\mathbf{X}) = \mathbb{M}^{[0]}(X,Y)$. \\

The detailed computation of the bending energy at leading order is shown in \ref{sec:Appendix_A}. Here, we report and discuss the final expression: 
\begin{equation} \label{eq:final_bending_energy}
\mathscr{W}_{\rm bend} = \frac{\mu H^3}{6} \int_S \Bigl[ \left(\Delta \xi - C_m \right)^2 - \nabla \xi \cdot \mathbf{b}_{\rm a} - \mathcal{A} - \mathcal{F} \Bigr] \, \d A + \frac{k_{44}H^3}{6} \int_S \left(\mathbb{M} : \mathbb{W}_{\rm a}\right)^2 \, \d A  \, , 
\end{equation}
where $\Delta$ denotes the in-plane Laplacian, $\mathcal{A}$ collects boundary contributions and $\mathcal{F}$ gathers terms independent of $\xi$. For convenience, we define
\begin{align}
& C_m := \frac{\partial}{\partial X}(a_{13}+a_{31}) + \frac{\partial}{\partial Y}(a_{23}+a_{32}) - k_{11} - k_{22} = \operatorname{div} (\mathbf{a}_1+\mathbf{a}_3) - k_{11} - k_{22} \, , \label{eq:Cm} \\[2ex] 
& \mathbb{W}_{\rm a} := \begin{pmatrix}
    \dfrac{\partial^2 \xi}{\partial X^2} - \dfrac{\partial}{\partial X}(a_{13}+a_{31}) + k_{11} & W_{12}  \\[2.5ex] W_{12} & \dfrac{\partial^2 \xi}{\partial Y^2} - \dfrac{\partial}{\partial Y}(a_{23}+a_{32}) + k_{22} 
    \end{pmatrix} \, , 
     \label{eq:Wa_ba} \\[2ex]
& W_{12} = \dfrac{\partial^2 \xi}{\partial X \partial Y} - \dfrac 12 \dfrac{\partial}{\partial X}(a_{23} + a_{32}) - \dfrac{1}{2}\dfrac{\partial}{\partial Y}(a_{13}+a_{31}) + k_{12} \, , \\[2.5ex]
& \mathbf{b}_{\rm a} := \begin{pmatrix}
    \dfrac{\partial k_{12}}{\partial Y} - \dfrac{\partial k_{22}}{\partial X} \\[2.5ex]
    \dfrac{\partial k_{12}}{\partial X} - \dfrac{\partial k_{11}}{\partial Y}
\end{pmatrix} = \rev{\begin{pmatrix}
    -(\operatorname{curl}\mathbbm{k}_{\rm a})_{32} \\[0.5ex]
    (\operatorname{curl}\mathbbm{k}_{\rm a})_{31}
\end{pmatrix}} \, .
\end{align}
\rev{where $\operatorname{div}$ is the in-plane divergence operator and, with a slight abuse of notation, we denote $\mathbf{a}_1 = (a_{13}, a_{23})^{\rm T}$ and $\mathbf{a}_3 = (a_{31}, a_{32})^{\rm T}$.} 
\rev{It is also useful to introduce the symmetric tensor $\mathbbm{f} := \operatorname{sym}\nabla(\mathbf{a}_1 + \mathbf{a}_3)$ and the \emph{target curvature tensor} $\mathbbm{h} := \mathbbm{f} - \mathbbm{k}_{\rm a}$, in terms of which $C_m = \operatorname{tr}\mathbbm{h}$ and $\mathbb{W}_{\rm a} = \mathbb{H}(\xi)-\mathbbm{h}$, with $\mathbb{H}$ the Hessian matrix.}

\subsection{\rev{Geometric interpretation of $\mathbf{b}_{\rm a}$ and target metric compatibility}} \label{sec:compatibility}

The bending energy of Eq.~\eqref{eq:final_bending_energy} has some interesting features and clearly highlights several novel aspects of our approach. For the first term in the isotropic part we note that, in the FvK limit, the Laplacian $\Delta\xi$ represents twice the mean curvature of the deformed mid-surface. The term $C_m$, on the other hand, encodes the \emph{target (or spontaneous) mean curvature} generated by out-of-plane active distortions. Thus, the quantity $\Delta\xi - C_m$ measures the mismatch between the geometric mean curvature of the plate and the natural curvature imposed by activity. This interpretation is consistent with classical theories of incompatible growth or swelling, in which out-of-plane distortions induce residual bending and drive the sheet toward a preferred shape (e.g., \citep{dervaux2008prl, efrati2009noneuclidean}). In our formulation, the same mechanism arises directly from the incompatible active strain, and the structure of Eq.~\eqref{eq:final_bending_energy} makes explicit how activity determines a curvature field that the plate attempts to accommodate. 
\rev{Regarding the term $\mathbf{b}_{\rm a}$, which represents one of the main novelties in our model, we note that, if $\mathbbm{k}_{\rm a}$ is curl-free (i.e., compatible), then $\mathbf{b}_{\rm a} = \mathbf{0}$. This vector field is therefore a measure of the incompatibility of active contractions along $Z$, and the contribution $\nabla\xi\cdot\mathbf{b}_{\rm a}$ can be thought of as the work done by active bending moments generated by incompatible deformations through the thickness. Since this kind of distortion cannot be accommodated in the plane without introducing stresses, it drives out-of-plane bending, as expected. This term also admits a clear differential-geometric interpretation in terms of the target metric $\mathbb{C}_{\rm a} = \mathbb{F}_{\rm a}^{\rm T}\mathbb{F}_{\rm a}$ \citep{ciarlet2005, argento2021shaping, colorado2022morphing}. To illustrate it, consider first the case $\mathbbm{a} = \mathbb{O}$, in which $\mathbb{F}_{\rm a} = \mathbb{I} + Z\mathbbm{k}_{\rm a}$. The target metric then expands as  $\mathbb{C}_{\rm a} = \mathbb{I} + 2Z\operatorname{sym}\mathbbm{k}_{\rm a} + \mathcal{O}(Z^2)$, so that the symmetric part of $\mathbbm{k}_{\rm a}$ plays the role of the second fundamental form of the target midsurface, whose first fundamental form is the identity \citep{argento2021shaping, colorado2022morphing}. Given the structure of $\mathbb{C}_{\rm a}$ and the scalings, the six independent compatibility conditions $\mathsf{R}(\mathbb{C}_{\rm a}) = \mathsf{0}$ actually reduce to three, i.e., $R_{1212} = R_{1213} = R_{1223} = 0$, which govern the embeddability of the target surface with these fundamental forms in the Euclidean space. The remaining three conditions appear at higher order in $Z$ and are therefore filtered out by the plate reduction. Thus, in the FvK limit, the problem of compatibility of the target metric corresponds to the \emph{Gauss} and \emph{Codazzi--Mainardi} conditions \citep{ciarlet2005} for the target surface, which reduce respectively to}
\begin{equation} \label{eq:ka_compatibility}
   \rev{ \det\mathbbm{k}_{\rm a} = 0 \quad \textrm{(Gauss)} \, , \qquad \frac{\partial k_{22}}{\partial X} - \frac{\partial k_{12}}{\partial Y} = 0 \, , \;\; \frac{\partial k_{12}}{\partial X} - \frac{\partial k_{11}}{\partial Y} = 0 \, \Rightarrow \, \mathbf{b}_{\rm a} = \mathbf{0} \quad \textrm{(Codazzi--Mainardi)} \, .}
\end{equation}
\rev{ Therefore, if $\mathbbm{a} = \mathbb{O}$, the metric $\mathbb{C}_{\rm a}$ is flat if and only if $\mathbbm{k}_{\rm a}$ does not develop Gaussian curvature and defines an integrable midsurface. The Gauss defect enters the energy through $\mathcal{F}$ via $\det\mathbb{W}_{\rm a}$ (see Eq.~\eqref{eq:Fcal}). Instead, the defect of integrability is represented by $\mathbf{b}_{\rm a}$, whose coupling $\nabla\xi\cdot\mathbf{b}_{\rm a}$ in the bending energy can be interpreted as the energetic cost of accommodating a target curvature that does not satisfy the integrability conditions for surfaces in $\mathbb{R}^3$. 
In the general case $\mathbbm{a}\neq \mathbb{O}$, the same geometric structure persists, but the first fundamental form of the target surface is no longer flat, and the effective target curvature $\mathbbm{h}$ introduced above appears. The leading-order Gauss equation for the target metric becomes} 
\begin{equation} \label{eq:deth}
    \rev{\det\mathbbm{h} = C_{a,\phi}}
\end{equation}
\rev{with $C_{a,\phi}$ defined by}
\begin{equation} \label{eq:Caphi}
    \rev{C_{a,\phi} := \frac{\partial^2}{\partial X\partial Y}\Bigl(a_{12} + a_{21} + a_{31}a_{32}\Bigr) - \frac{\partial^2}{\partial Y^2}\Bigl(a_{11} + \frac{1}{2}a_{31}^2\Bigr) - \frac{\partial^2}{\partial X^2}\Bigl(a_{22} + \frac{1}{2}a_{32}^2\Bigr) \, ,}
\end{equation}
\rev{whereas the Codazzi--Mainardi conditions at leading order read $\operatorname{curl}\mathbbm{h} = \mathbb{O}$. So, in this case, the surface characterized by the target curvature $\mathbbm{h}$ must be integrable, and its Gaussian curvature must match the one prescribed by the metric, as per Eq.~\eqref{eq:deth}. A competition between the effects of $\mathbbm{a}$ and $\mathbbm{k}_{\rm a}$ is then introduced, so that the incompatibility effects induced by one contribution can be compensated by the other, providing a surface that can be realized without residual stresses. 
Finally, the tensor $\mathbb{W}_{\rm a} = \mathbb{H}(\xi) - \mathbbm{h}$ reduces to the linearized Weingarten map of the deformed midsurface when $\mathbbm{h} = \mathbb{O}$, and in general measures the mismatch between the actual curvature $\mathbb{H}(\xi)$ and the target curvature $\mathbbm{h}$ imposed by the active distortions. Hence, the anisotropic term in the bending energy, $\mathbf{m} \cdot \mathbb{W}_{\rm a}\mathbf{m}$, penalizes curvature mismatch along the reinforcement direction $\mathbf{m}$.}\\
Another interpretation of the different terms in the bending energy can be provided by noting that  \[\operatorname{tr}\mathbb{W}_{\rm a} = \Delta \xi - C_m \qquad \textrm{and} \qquad \det\mathbb{W}_{\rm a} = \mathcal{A} + \mathcal{F} + \nabla \xi \cdot \mathbf{b}_{\rm a} = [\xi,\xi] - C_G \, , \]
where $C_G$ is the \emph{target Gaussian curvature} (the complete expression is reported in \ref{sec:Appendix_A}) and $C_m$ represents the target mean curvature of Eq.~\eqref{eq:Cm}, as previously observed. Consequently, 
\begin{equation}
\mathscr{W}_{\rm bend} = \frac{\mu H^3}{6}\int_S \Bigl[ \left(\operatorname{tr}\mathbb{W}_{\rm a}\right)^2 - \det\mathbb{W}_{\rm a}  \Bigr] \, \d A + \frac{k_{44}H^3}{6} \int_S \left(\mathbb{M} : \mathbb{W}_{\rm a}\right)^2 \, \d A  \, ,
\end{equation}
and the isotropic part is coherent with the standard form for the bending energy of a plate \citep{landau2012theory} and with metric-based formulations of non-Euclidean plates \citep{helfrich1973elastic,seifert1997configurations, efrati2009noneuclidean}. However, note that, if $\mathbf{b}_{\rm a} \neq \mathbf{0}$, the contribution of $\det\mathbb{W}_{\rm a}$ is not limited to the boundary, because the Gauss--Bonnet theorem cannot be applied to $C_G$ if it does not represent the curvature of a compatible surface. 
Hence, the tensor $\mathbb{W}_{\rm a}$ encodes information about the target mean $C_m$ and Gaussian $C_G$ curvatures induced by active contractions, which in general may be non-realizable in a physical surface due to incompatibility \citep{dervaux2009morphogenesis, iakunin2018biofilms}. 

\subsection{\rev{Equilibrium equations in the general case}}

Having identified the contribution of each term, we can now deduce the equilibrium equations and boundary conditions by variational arguments, as shown in \ref{sec:Appendix_B}. In particular, the system of FvK equations for an active anisotropic plate reads: 
\begin{align}
 &\operatorname{div} \mathbb{S}_n^{[0]} = \mathbf{0} \, , \label{eq:equilibrium_1} \\
 & D\left(\Delta^2 \xi - \Delta C_m\right) + D_k \operatorname{div}\operatorname{div}\bigl((\mathbb{M}:\mathbb{W}_{\rm a})\mathbb{M}\bigr) + \frac{D}{2} \operatorname{div}\mathbf{b}_{\rm a} - H \operatorname{div}\left(\mathbb{S}_n^{[0]}\nabla\xi\right) = F \, , \label{eq:equilibrium_2}
\end{align}
with $D := \mu H^3 /3$ and $D_k := k_{44} H^3 /3$, while $\operatorname{div}$ stands for the in-plane divergence operator. 

A standard strategy to rewrite the equilibrium equations consists in the introduction of the \emph{Airy potential} $\phi$, defined by 
\begin{equation}
 (\mathbb{S}_n^{[0]})_{11} = \frac{\partial^2 \phi}{\partial Y^2} \, , \qquad (\mathbb{S}_n^{[0]})_{22} = \frac{\partial^2 \phi}{\partial X^2} \, , \qquad (\mathbb{S}_n^{[0]})_{12} = - \frac{\partial^2 \phi}{\partial X\partial Y} \, .   
\end{equation}
Then, Eq.~\eqref{eq:equilibrium_1} is automatically satisfied. A natural governing equation for the Airy potential $\phi$ follows from imposing the compatibility condition associated with the constitutive relation of Eq.~\eqref{eq:reduced_constitutive_relations} between in–plane stress and strain, resulting in the following PDE (see \ref{sec:Appendix_C} for the full derivation):

\begin{equation}
    \frac{1}{E_{Y}}\,\Delta^2 \phi - \frac{1}{E_k}\operatorname{div}\operatorname{div}\bigl[(\widetilde{\mathbb{M}}\otimes\widetilde{\mathbb{M}})\mathbb{H}(\phi) \bigr] = C_{a,\phi} - [\xi,\xi] \, ,
\end{equation}
where $C_{a,\phi}$ has been defined in Eq.~\eqref{eq:Caphi}, 
$\mathbb{H}(\phi)$ is the Hessian matrix of $\phi$, and 
\begin{equation}
\widetilde{\mathbb{M}} := \mathbb{M} - \frac{2}{3}\mathbb{I} \, , \qquad E_Y := 3\mu \, , \qquad E_k := \frac{\mu^2}{k_{44}}\left(1+\frac{4}{3}\frac{k_{44}}{\mu}\right) \, .
\end{equation}
Note that the anisotropic component  vanishes for $k_{44}\to0$, as expected. Introducing the Airy potential in the equation for the transverse deflection yields the final set of equations in terms of $\phi$ and $\xi$:  
\begin{align}
& D(\Delta^2\xi - \Delta C_m) + D_k \operatorname{div}\operatorname{div}\bigl((\mathbb{M}:\mathbb{W}_{\rm a})\mathbb{M}\bigr) + \frac{D}{2}\operatorname{div}\mathbf{b}_{\rm a} - 2H[\phi, \xi] = F \, ,  \label{eq:final_eq_xi}\\[0.5ex]  
& \frac{1}{E_{Y}}\,\Delta^2 \phi - \frac{1}{E_k}\operatorname{div}\operatorname{div}\bigl[(\widetilde{\mathbb{M}}\otimes\widetilde{\mathbb{M}})\mathbb{H}(\phi) \bigr] = C_{a,\phi} - [\xi,\xi] \, .  \label{eq:final_eq_phi}
\end{align}
If $D_k = 0$ (so $1/E_k \to 0$) and $\mathbbm{k}_{\rm a} = \mathbb{O}$, then the set of Eqs.~\eqref{eq:final_eq_xi}--\eqref{eq:final_eq_phi} corresponds to the result obtained in similar mathematical contexts, such as the morphogenesis of plates in \citep{dervaux2009morphogenesis}, petal/leaf growth in \citep{wang2024morphomechanics, liang2009leaf}, or nematic plates in \citep{mihai2020plate}. On the other hand, for $a_{\alpha 3} = a_{3\alpha} = 0$, the system falls into the framework obtained in \citep{lewicka2011foppl} by means of $\Gamma$-convergence, although there are differences between the two methodologies. Our model generalizes these frameworks and the general structure of Eq.~\eqref{eq:Foppl_von_Karman_syst} by accounting for spatially varying anisotropy of the plate and for inhomogeneous active contractions along the thickness, which may be relevant for capturing a wide range of phenomena. 

\rev{To connect with the geometric compatibility discussed in Sect.~\ref{sec:compatibility}, we observe that, if the Gauss condition of Eq.~\eqref{eq:deth} holds and the deflection adapts so that $[\xi,\xi] = \det\mathbbm{h} = C_{a,\phi}$, the right-hand side of Eq.~\eqref{eq:final_eq_phi} vanishes and, in the absence of imposed boundary stresses, $\phi = 0$, thus no residual membrane stresses develop. If instead the Gauss condition is violated, then $[\xi,\xi] = \det\mathbbm{h} \neq C_{a,\phi}$, and membrane stress arises. Conversely, violation of the Codazzi--Mainardi conditions enters directly in Eq.~\eqref{eq:final_eq_xi} through $\operatorname{div}\mathbf{b}_{\rm a}$.} 

\paragraph{\rev{Remark 4 (on the curl of $\mathbbm{h}$)}} \rev{We note that Eq.~\eqref{eq:final_eq_xi} does not depend on $\mathbf{h}_{\rm a} := (-(\operatorname{curl}\mathbbm{h})_{32},\, (\operatorname{curl}\mathbbm{h})_{31})^{\rm T}$, but only on $\mathbf{b}_{\rm a}$. The two are related by $\mathbf{h}_{\rm a} = \mathbf{f}_{\rm a} - \mathbf{b}_{\rm a}$, where $\mathbf{f}_{\rm a} := (-(\operatorname{curl} \mathbbm{f})_{32}, (\operatorname{curl} \mathbbm{f})_{31})^{\rm T}$. However, since $\operatorname{div}\mathbf{f}_{\rm a} = 0$ identically, this difference contributes to the energy only through boundary and does not affect the equilibrium equations, where only $\operatorname{div}\mathbf{b}_{\rm a}$ enters.} 

\subsection{Equilibrium equations for spatially uniform anisotropy}

The system obtained in the previous section describes the governing equations of an active Föppl--von Kármán plate endowed with a generic anisotropic microstructure characterized by the tensor field $\mathbb{M} = \mathbb{M}(X,Y)$, which may vary spatially in general.  There exist, however, situations in which the anisotropic structure can be considered spatially uniform. In this case, the coupling between stress and spatial inhomogeneities vanishes. Indeed, when the tensor field $\mathbb{M}$ is constant in space, we can observe that 
\begin{equation}
\mathbb{M}:\mathbb{W}_{\rm a} = \Delta_{\mathbb{M}}\xi - C_{m,k} \, , 
\end{equation}
where the directional Laplacian $\Delta_{\mathbb{M}}(\cdot)$ and $C_{m,k}$ are defined as 
\begin{equation}
    \begin{split}
         \Delta_{\mathbb{M}}\left(\cdot\right) & := \operatorname{div} \bigl[ \mathbb{M}\nabla(\cdot)\bigr] \, , \\[1ex]
         C_{m,k} & := \operatorname{div}\bigl[ \mathbb{M}\,
        (\mathbf{a}_1+\mathbf{a}_3)\bigr] - \mathbb{M}:\mathbbm{k}_{\rm a} \, \rev{= \mathbb{M} : \mathbbm{h}} \, .
    \end{split}
    \label{eq:operators}
\end{equation}
Thus, the bending energy density writes as:
\begin{equation}
    \mathscr{W}_{\rm bend} = \frac{H^3}{6}\int_S\left\{\mu \left[ \left(\Delta \xi - C_m \right)^2 - \nabla\xi\cdot\mathbf{b}_{\rm a} - \mathcal{A} - \mathcal{F}  \right] + k_{44}\left( \Delta_\mathbb{M}\xi - C_{m,k} \right)^2\right\} \d A \, .
    \label{eq:energy_bend_uniform}
\end{equation}
Moreover, the operators involved in the non-linear system \eqref{eq:final_eq_xi}--\eqref{eq:final_eq_phi} reduce to (see also \ref{sec:Appendix_C}):
\begin{align}
\operatorname{div}\operatorname{div}\bigl((\mathbb{M}:\mathbb{W}_{\rm a})\mathbb{M}\bigr) &= \operatorname{div}\bigl(\mathbb{M}\nabla(\Delta_{\mathbb{M}}\xi - C_{m,k})\bigr) 
= \Bigl(\Delta_\mathbb{M}^2 \xi - \Delta_\mathbb{M} C_{m,k}\Bigr )   \, ,  \\[0.4ex] 
  \operatorname{div}\operatorname{div}\Bigl((\widetilde{\mathbb{M}}\otimes\widetilde{\mathbb{M}})\mathbb{H}(\phi)\Bigr)
   &  =   \left (\widetilde{\mathbb{M}}\otimes\widetilde{\mathbb{M}}\right) 
    :: \nabla\nabla\mathbb{H}(\phi)
     = 
    \left(\Delta_{\mathbb{M}} - \frac{2}{3}\,\Delta\right)^2\phi \, ,  \label{eq:divdiv_MMH}
\end{align}
where $::$ denotes the inner product between fourth-order tensors. 
Hence, the system of equations in the case of homogeneous anisotropy simplifies as: 
\begin{align}
        D\Bigl(\Delta^2\xi - \Delta C_m\Bigr) + D_k\Bigl(\Delta_\mathbb{M}^2 \xi - \Delta_\mathbb{M} C_{m,k}\Bigr) +\frac{D}{2}\operatorname{div}\mathbf{b}_{\rm a} - 2H\,[\xi,\phi]   & = F \, , \label{eq:final_eq_xi_uniform}\\[1ex]
        \frac{1}{E_{Y}}\,\Delta^2 \phi - \frac{1}{E_k}\Bigl(\Delta_{\mathbb{M}} - \frac{2}{3}\Delta\Bigr)^2\phi + [\xi,\xi] - C_{a,\phi} &= 0 \label{eq:final_eq_phi_uniform}\, .
\end{align}

\section{\rev{Buckling of active anisotropic rings}} \label{sec:example}

\rev{To further illustrate the governing equations derived in Sect.~3, we investigate the onset of transverse buckling in an active annular plate characterized by azimuthal  reinforcement and possibly subject to radial compressive stress at its outer boundary. The ring geometry is chosen so that the base state is radially symmetric and admits an analytical in-plane solution, allowing for a clean reduction to a single-mode eigenvalue problem. We specifically examine how the anisotropy ratio $\kappa := k_{44}/\mu = D_k /D$ and active contractions modulate the critical loading threshold, as well as the onset of buckling instabilities driven purely by active contraction in the absence of external loading.}

\rev{Let the reference configuration be defined by the annular domain $(R,\Theta,Z) \in (R_0,\,1)\times[0,\,2\pi)\times(-H,H)$ in cylindrical coordinates, where $0 < R_0 < 1$ denotes the dimensionless inner radius rescaled by the outer radius. The direction of anisotropy is purely azimuthal, i.e., $\mathbf{m} = \mathbf{e}_\Theta$, such that the structural tensor is $\mathbb{M} = \mathbf{e}_\Theta\otimes\mathbf{e}_\Theta$. Note that $\mathbb{M}$ varies in space, which requires us to use the general formulation of Eqs.~\eqref{eq:final_eq_xi}--\eqref{eq:final_eq_phi}. At the inner boundary, $R = R_0$, we prescribe clamped conditions $u^{(0)} = v^{(0)} = \xi = 0$. At the outer boundary, $R = 1$, we apply a radial compressive stress $[\mathbb{S}_n^{(0)}]_{RR} = 2\mu s_0$ and vanishing shear stress $[\mathbb{S}_n^{(0)}]_{R\Theta} = 0$, where $s_0 \leq 0$ serves as the dimensionless loading parameter. An in-plane active deformation $\mathbb{F}_{\rm a} = \operatorname{diag}(1+a_1,\,1+a_2,\,1)$ is prescribed, with $a_1$ and $a_2$ representing the constant radial and hoop active strains, respectively. In this curvilinear setting, such a deformation is generally geometrically incompatible, thereby generating non-trivial residual stresses even in the absence of external loads.}

\rev{For a radially symmetric configuration, the azimuthal displacement vanishes ($u_\Theta = 0$) and $u_R$ only depends on $R$.  The in-plane equilibrium equation \eqref{eq:equilibrium_1}  then reduces in cylindrical coordinates to a second-order ordinary differential equation for the radial displacement $u_R(R)$:} 
  \begin{equation}
      \rev{R^2 u''_R + R\,u'_R - (1+\kappa)\,u_R
      = \tfrac{1}{2}R\!\left[a_1 - a_2(1+2\kappa)\right] \, ,}
  \end{equation}
\rev{whose general solution, depending on $\kappa$, is given by:}
\begin{equation}
    \label{eq:ring_disp_sol}
    \rev{u_R(R) = \begin{cases}
        \displaystyle
        C_1 R^{\,\beta} + C_2 R^{-\beta}
        - \dfrac{a_1 - a_2(1+2\kappa)}{2\kappa}\,R \, ,
        & \kappa \neq 0 \, , \\[10pt]
        \displaystyle
        C_1 R + C_2 R^{-1} - \dfrac{a_1-a_2}{4}\,R\ln R \, ,
        & \kappa = 0 \, ,
    \end{cases}}
\end{equation}
\rev{where $\beta := \sqrt{1+\kappa}$ and the constants $C_1, C_2$ are determined by the in-plane boundary conditions stated above. The resulting base-state stress components $S_{RR}^{(0)}$ and $S_{\Theta\Theta}^{(0)}$ are inserted into the out-of-plane governing equation \eqref{eq:final_eq_xi}.}

\rev{For the active deformation $\mathbb{F}_{\rm a}$, the terms $\Delta C_m$ and $\operatorname{div}\mathbf{b}_{\rm a}$ vanish identically. Consequently, the governing equation \eqref{eq:final_eq_xi} for the transverse displacement $\xi$  simplifies to:}
\begin{equation}
    \label{eq:ring_outofplane}
    \rev{\Delta^2\xi
    + \kappa\,\operatorname{div}\operatorname{div}\bigl((\mathbb{M}:\mathbb{W}_{\rm a})\mathbb{M}\bigr)
    = \frac{3}{\mu H^2}\!\left[
        S_{RR}^{(0)}\frac{\partial^2\xi}{\partial R^2}
        + S_{\Theta\Theta}^{(0)}\!\left(
          \frac{1}{R}\frac{\partial\xi}{\partial R}
          + \frac{1}{R^2}\frac{\partial^2\xi}{\partial\Theta^2}
          \right)
      \right] \, ,}
\end{equation}
\rev{subject to clamped conditions $\xi(R_0,\Theta) = 0$, $ \partial_R\xi(R_0,\Theta) = 0$ at $R=R_0$ and traction-free conditions at $R=1$, as derived in Eqs.~\eqref{eq:BC1}--\eqref{eq:BC2} of \ref{sec:Appendix_B}. Decomposing the solution as $\xi(R, \Theta) = \zeta(R)\cos(m\Theta)$ for integer modes $m \geq 1$, the governing equation for the radial profile $\zeta$ becomes:}
\rev{\begin{align}
    \label{eq:ring_ode_zeta}
    R^4\zeta^{\rm(iv)} + 2R^3\zeta'''
    &- \left(1 + 2m^2 + \kappa + C\, S_{RR}^{(0)} R^2\right)R^2\zeta''
    \nonumber\\
    &+ \left(1 + 2m^2 +\kappa - C\, S_{\Theta\Theta}^{(0)}R^2\right)R\,\zeta'
    \nonumber\\
    &- m^2\bigl[4 - m^2+\kappa(2-m^2) - C\,S_{\Theta\Theta}^{(0)}R^2\bigr]\zeta = 0 \, ,
\end{align}}
\rev{where $C := 6/(\mu H^2)$. The boundary conditions at $R = R_0$ are $\zeta(R_0) = \zeta'(R_0) = 0$, while at $R = 1$ we have:
\begin{equation}
    \label{eq:ring_bc_zeta}
    \zeta'' + \frac{1}{2}\zeta' - \frac{m^2}{2}\zeta = 0 \, , \qquad
    \zeta'''+\zeta'' - \left(1+\frac{3}{2}m^2+\kappa\right)\,\zeta'
    + m^2\left(\frac{5}{2} + \kappa\right)\,\zeta = 0 \, .
\end{equation}}
\rev{To determine the buckling threshold, we treat the boundary value problem \eqref{eq:ring_ode_zeta}--\eqref{eq:ring_bc_zeta} as a linear generalized eigenvalue problem and solve it numerically by Chebyshev spectral collocation. For the case in which the loading parameter is the external compression, with $a_1$ and $a_2$ fixed, the discretization yields a generalized eigenvalue problem whose eigenvalue is the dimensionless critical loading $C^*:=6s_0^*/(\mu H^2)$, where $s_0^*$ is the smallest-magnitude negative real value of $s_0$ for which a non-trivial buckled mode exists. Conversely, when $s_0 = 0$ and the buckling is driven solely by activity, we fix the direction of contraction $\hat{\mathbf{a}} := (\hat a_1, \hat a_2)$ and treat the contractility magnitude $\alpha^{\rm c}$ such that $(a_1, a_2) = \alpha^{\rm c}(\hat a_1, \hat a_2)$ as the control parameter of the eigenvalue problem. The critical contractility is the smallest positive real value of $\alpha^{\rm c}$ for which a non-trivial buckled mode exists.} 

\begin{figure}[t]
\centering
\begin{subfigure}[b]{0.325\textwidth}
\centering
{\includegraphics[width=\textwidth]{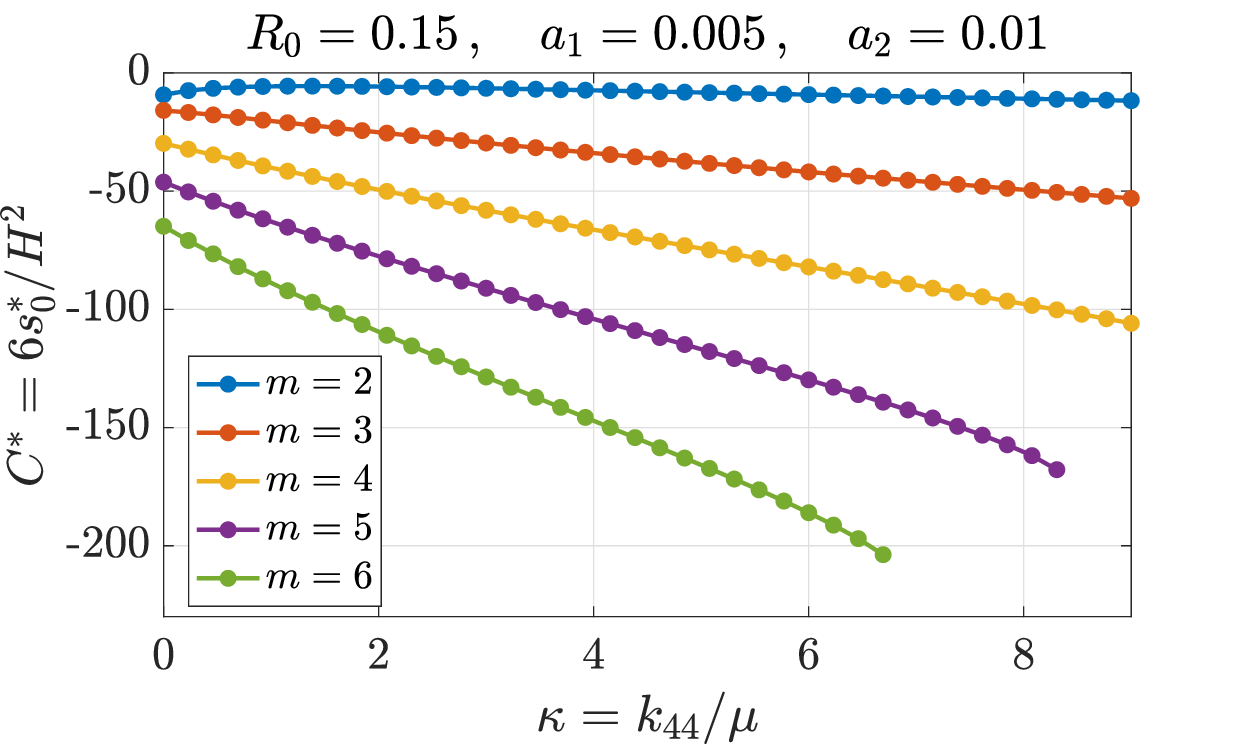}}
\caption{}
\label{fig:Ccrit_vs_anisotropy}
\end{subfigure}
\hfill
\begin{subfigure}[b]{0.325\textwidth}
\centering
{\includegraphics[width=\textwidth]{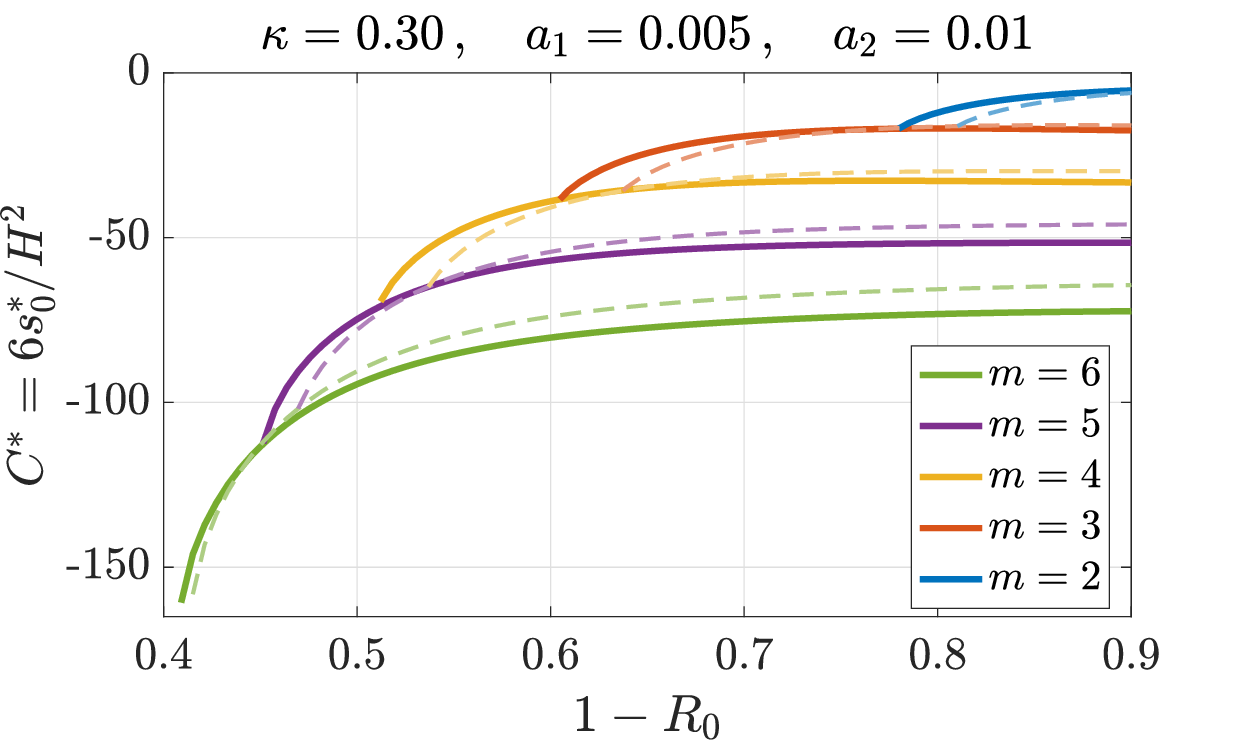}}
\caption{}
\label{fig:Ccrit_vs_ringwidth}
\end{subfigure} 
\hfill
\begin{subfigure}[b]{0.325\textwidth}
\centering
{\includegraphics[width=\textwidth]{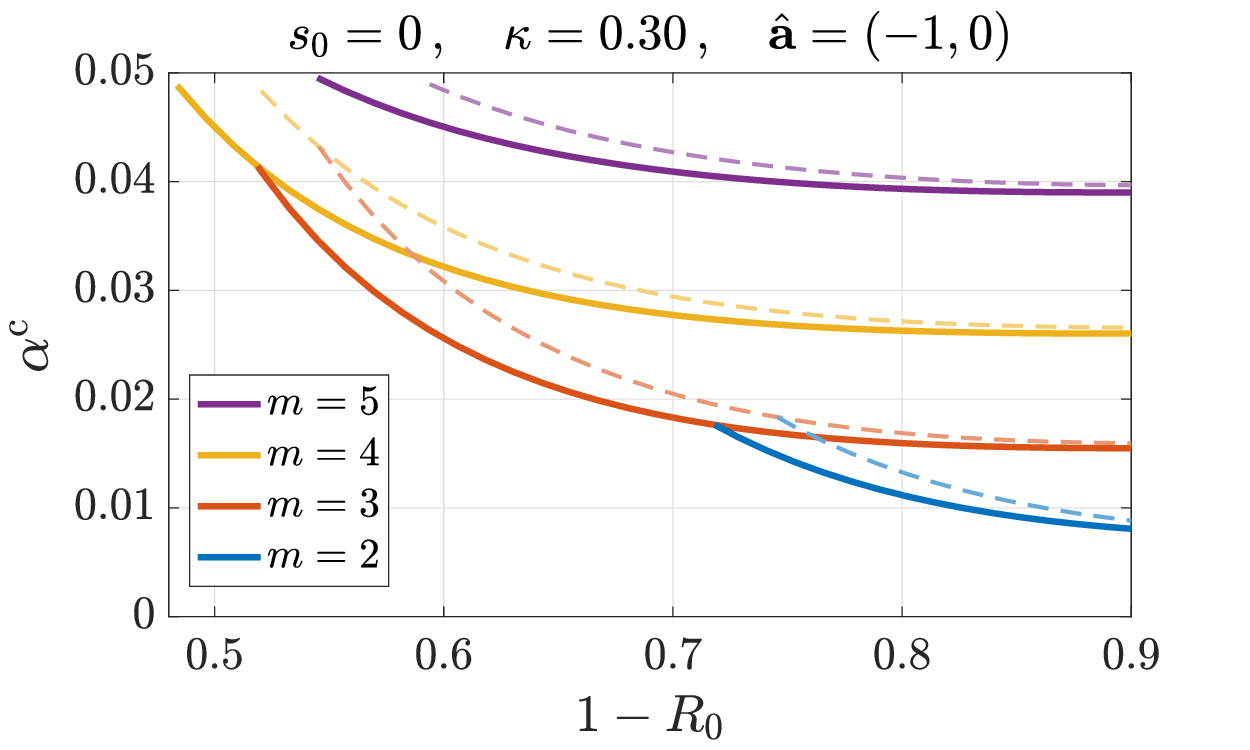}}
\caption{}
\label{fig:critical_activecontraction}
\end{subfigure}
\caption{\rev{(a) Critical loading $C^* = 6s_0^*/H^2$ as a function of the anisotropy ratio $\kappa = k_{44}/\mu$, for modes $m = 2,\ldots,6$ and fixed inner radius $R_0 = 0.15$ and active strains $a_1 = 0.005$, $a_2 = 0.01$. (b) Critical loading $C^*$ as a function of ring width $1-R_0$, for modes $m = 2,\ldots,6$ at fixed $\kappa = 0.3$ (solid lines) and active strains $a_1 = 0.005$, $a_2 = 0.01$. The dashed lines show the isotropic case $\kappa = 0$. (c) Critical active contractility $\alpha^{\rm c}$ for  radial active contraction $\hat{\mathbf{a}} = (-1,0)$ in the absence of external compression ($s_0 = 0$), as a function of ring width $1-R_0$, for modes $m = 2,\ldots,5$ at fixed $\kappa = 0.3$. Solid and dashed lines as in (b).}}
\label{fig:example_figures}
\end{figure}

\rev{Figure~\ref{fig:Ccrit_vs_anisotropy} illustrates the critical loading $C^*$ as a function of $\kappa$ for a fixed $R_0 = 0.15$ and $a_1 = 0.005$, $a_2 = 0.01$, to investigate the influence of anisotropy. For all values of $\kappa$ shown, the mode $m=2$ remains the first to become unstable across the entire range. For each individual mode with $m\geq 3$, the buckling threshold decreases with increasing anisotropic stiffness, indicating that larger compression is required to trigger instability, i.e., azimuthal reinforcement stabilizes these modes. The dominant mode $m = 2$ behaves differently: its sensitivity to $\kappa$ is low, with a slight non-monotonic variation for small values of $\kappa$,  leading to a re-stabilization at small values before the threshold decreases again. Furthermore, higher-order modes (e.g., $m = 5, 6$) cease to appear as isolated instabilities beyond a critical value of $\kappa$ in the plotted range; this behavior results from the coalescence of two distinct real eigenvalues of the ODE, which collapse and become complex, signaling a qualitative change in the solution structure.}

\rev{The dependence of the critical loading $C^*$ on the ring width $1-R_0$ is shown in Fig.~\ref{fig:Ccrit_vs_ringwidth} for $\kappa = 0.3$ (solid lines) and the same active strains. The results highlight modal crossings: as the ring width varies, the dominant instability shifts between consecutive modes. For thick annular bands (large $1-R_0$), the first mode to be triggered is $m=2$, while for thin annular bands (small $1-R_0$) instability first occurs for higher modes. Consequently, each mode is critical only within a specific interval of widths. The comparison between anisotropic (solid lines) and isotropic (dashed lines) cases in Fig.~\ref{fig:Ccrit_vs_ringwidth} shows that azimuthal reinforcement provides a destabilizing effect for narrow rings (small $1-R_0$, thin annular band), requiring lower external compression to trigger buckling, while acting as a stabilizing factor for wide geometries (large $1-R_0$, thick annular band). The isotropic and anisotropic curves cross at mode-specific values of the ring width, separating the destabilizing and stabilizing regime.} 

\rev{Finally, Fig.~\ref{fig:critical_activecontraction} demonstrates that buckling can be induced solely by radial active contraction, i.e., $\hat{\mathbf{a}} = (-1,0)$, in the absence of external mechanical loads ($s_0 = 0$). The critical contractility $\alpha^{\rm c}$ follows a modal structure qualitatively similar to the purely mechanical case, including mode crossings at specific ring widths. We also note that, in this case, anisotropy has a destabilizing effect across all widths, because it requires a smaller contractility value to trigger instability, especially for narrower rings. Conversely, a purely hoop active contraction is not sufficient to destabilize the ring by itself. 
Both the critical loading $C^*$ and contractility $\alpha^{\rm c}$ are found to be within physically relevant ranges that remain consistent with the moderate-deflection assumptions of the present model.}

\section{An application to cell alignment on curved substrates}
\label{sec:applications}

Numerous experiments and a growing body of recent literature have demonstrated that substrate curvature significantly influences cellular behavior. Among the resulting curvature-induced responses, \emph{curvotaxis} (i.e., the curvature-guided reorientation and migration of cells), actin fiber alignment, and directed cell migration are particularly notable, as remarked in the Introduction. In this section, we show how the theoretical model developed above captures the experimentally observed relationship between the orientation angle $\theta$ of the cell and the curvature of the substrate. 

From the biological point of view, the actin-myosin system is the main machinery responsible for the active deformation of the cell. This apparatus depends on the physical properties of the environment and is constantly subject to remodeling \citep{engler2006matrix, hotulainen2006stress}. The activation patterns of the actin stress fibers are concentrated towards the basal surface of the cell, where focal adhesions are located, as shown by numerical simulations in \citep{ronan2012numerical}, and by biological evidence in \citep{feld2020cellular}.
Inspired by these observations, we assume an active deformation $\mathbb{F}_{\rm a}$ in the form 
\begin{equation}
   \rev{ \mathbb{F}_{\rm a} = \mathbb{I} + \left(\frac{H}{2} - Z\right)f_a(\mathbf{X}) \mathbb{M}   \, ,
    \label{eq:active_cells} 
    }
\end{equation}
\rev{where $f_a$ is a non-positive scalar field, representing a measure of the active deformation induced by changes at the microscopic level, and $\mathbb{M} = \mathbf{m}\otimes\mathbf{m}$ is the anisotropic structural tensor, representing the alignment of the cell cytoskeleton identified by the unit vector $\mathbf{m} = (\cos\theta, \sin\theta)^{\rm T}$ in the plane. For simplicity, we assume $\mathbf{m}$ constant in space.} This structure for $\mathbb{F}_{\rm a}$ is built on the assumption that active strains are higher on the basal surface of the cell nearby focal adhesion sites. Indeed, each component of $\mathbb{F}_{\rm a}$ achieves a maximum in magnitude at $Z = -H/2$, hence on the basal surface. The tensor $\mathbb{F}_{\rm a}$ describes directional shortening along the stress fibers, with the field $f_a$ representing deformation along the anisotropy axis $\mathbf{m}$: 
\begin{equation*}
      \mathbb{F}_{\rm a} \mathbf{m} = \left(\mathbb{I} + \left(\frac{H}{2} - Z\right)f_a(\mathbf{X}) \mathbb{M}\right)\mathbf{m} = \left(1 + \left(\frac{H}{2} - Z\right) f_a(\mathbf{X})\right)\mathbf{m} \, ,
\end{equation*}
so that $(\frac{H}{2} - Z) f_a(\mathbf{X})$ is the activity-induced deformation factor. Note that it must be \(-1<f_a(\mathbf{X}) < ~0\) \( \forall \, \mathbf{X}\in\mathcal{S} \) to ensure a valid deformation tensor. Moreover, the scaling assumption on the active strains \eqref{eq:active_scaling} implies that $f_a = \mathcal{O}(\xi/L^2)$ and it has the unit of measure of a curvature. 

We first analyze a cylindrical substrate and compare our model predictions with the relevant literature. The cylinder is a common experimental geometry for studying curvature effects on cells, while also being relevant in physiological contexts, such as blood vessels.  We then examine how more complex geometries (such as ellipsoids and spheres) affect cell alignment within our framework.

\subsection{Cell adherent to a cylindrical substrate}
\label{sec:cylinder}

\begin{figure}[t]
\centering
{\includegraphics[width=0.45\linewidth]{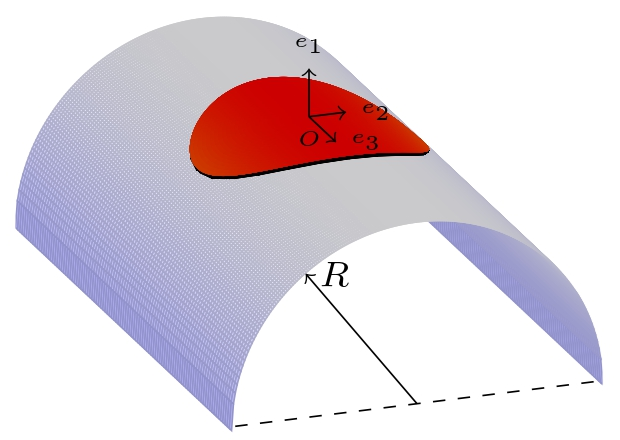}}
\hspace{2cm}
{\includegraphics[width=0.33\linewidth]{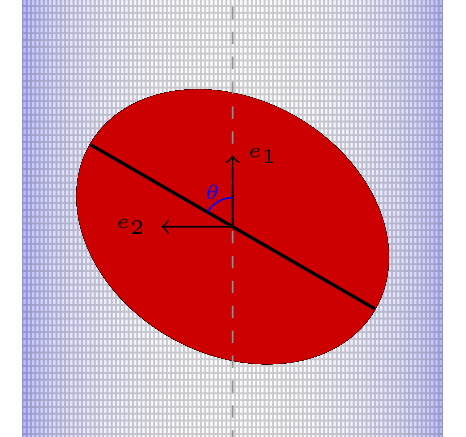}}
\caption{Sketch of the mechanical setup for a cell adhering to a cylindrical surface.}
\label{fig:cylinder_scheme}
\end{figure}

Following the basic assumption proposed in \citep{biton2009cellular}, we model the cell as an active anisotropic plate bent over a cylinder. Exploiting our novel formulation, we directly incorporate the anisotropic cellular structure into the model. This accounts for the internal cytoskeletal structure of the cell, specifically the myosin-activated stress fibers inducing active contraction. Starting from the active strain of Eq.~\eqref{eq:active_cells}, we prescribe the visible deformation and look for alignments that minimize the plate energy. 

In general, the expression of $f_a$ needs to be prescribed constitutively based on experimental observations. Although this field would typically satisfy its own evolution equation \citep{ruiz2014mathematical}, here we do not consider the dynamics of contraction, since we only aim to describe the final alignment of the cell on the curved surface. 
For simplicity, to assess the features of the model and compare the results with the literature, we assume that 
\begin{equation} f_a(\mathbf{X}) = \widehat{f}_a =  \alpha < 0 \quad \forall \mathbf{X}\in \mathcal{S} \, . \label{eq:linear_hyp_fa} 
\end{equation}
The uniform deformation across the cell layer is an oversimplifying assumption, since it implies that the activation level of the stress fibers is uniform below the cell. Instead, activated stress fibers bundles appear to be concentrated in the periphery of the cell, far from the nucleus, and are highly localized \citep{dembo1999stresses, schwarz2002calculation, peschetola2013time}. Still, as we show next, this assumption is sufficient to examine some of the model implications while maintaining an analytical treatment. Note that this choice of $\mathbb{F}_{\rm a}$ corresponds to choosing 
\begin{equation} \label{eq:ak_active}
\mathbbm{a} = \frac{H}{2}\alpha \mathbb{M} \, , \qquad \mathbbm{k}_{\rm a} = - \alpha \mathbb{M} \, .    
\end{equation}
Since both tensors are constant in space, they are trivially compatible. However, recalling Eq.~\eqref{eq:curl_Fa}, the resulting active strain tensor $\mathbb{F}_{\rm a}$ is not compatible, since $\operatorname{curl}\mathbb{F}_{\rm a} \neq \mathbb{O}$. This example shows that choosing both $\mathbbm{a}$ and $\mathbbm{k}_{\rm a}$ compatible does not guarantee a compatible overall active deformation, due to the presence of out-of-plane active contractions. \rev{Nevertheless, the associated target metric $\mathbb{C}_{\rm a}$ is compatible, since $\mathbbm{h} = \alpha\mathbb{M}$ is such that $\det\mathbbm{h} = C_{a,\phi} = 0$ and $\operatorname{curl}\mathbbm{h} = \mathbb{O}$ (see Remark~2 and Eq.~\eqref{eq:deth}).} 

The cell bending is introduced by assuming that it adheres to a cylindrical surface of radius $R$, with $L \ll R$, so that the deflection is prescribed as 
\begin{equation}
\xi(X,Y) = R\cos\left(\frac{Y}{R}\right) - R = \frac{1}{\kappa_1}\Bigl(1 - \cos(\kappa_1 Y)\Bigr) \, ,
\label{eq:cylinder_deflection}
\end{equation}
where $\kappa_1 := -1/R$ denotes the (negative) curvature of the cylinder. With this choice and the active strains of Eq.~\eqref{eq:ak_active}, we have $C_m = C_{m,k} = [\xi,\xi] = C_{a,\phi} = 0$ and $\mathbf{b}_{\rm a} = \mathbf{0}$. Therefore, Eq.~\eqref{eq:final_eq_phi_uniform} gives $\phi = 0$, meaning that no in-plane membrane stresses are developed. Instead, Eq.~\eqref{eq:final_eq_xi_uniform} becomes 
\begin{equation}
-D \, \kappa_1^3 \cos(\kappa_1 Y) - D_k \,  \kappa_1^3 \sin^4\theta \cos(\kappa_1 Y) = F \quad \text{in } \mathcal{S} \, ,
\label{eq:deflection_GovEq_CylindricalBending}
\end{equation}
which may not be satisfied in general. 
It follows that a forcing term $F_{\rm ad}$ must be included in the external load $F$ to sustain this deformation. Alternatively, a Lagrange multiplier $\lambda_{\rm ad}$ enforcing the geometric constraint $\xi = f$ could be introduced. In either case, $F_{\rm ad}$ or $\lambda_{\rm ad}$ represents the adhesive force that maintains the cell in contact with the curved substrate, which is not otherwise accounted for in the model. 

The fact that the Airy potential $\phi = 0$ implies that the in-plane stretching energy $\mathscr{W}_{\rm str} = 0$ and that no residual stresses arise in the plane.  
Thus, the total energy density coincides with the bending energy density $\psi_{\rm bend}$ of the body, which after some algebra becomes 
\begin{equation}
     \psi_{\rm bend} = \frac{\mu H^3}{6} \left[ \left(\kappa_1 \cos(\kappa_1 Y)-\alpha\right)^2 + \alpha \kappa_1 \cos^2\theta\cos(\kappa_1 Y) \right] + \frac{k_{44}H^3}{6} \left[ \kappa_1 \sin^2\theta \cos(\kappa_1 Y) - \alpha \right]^2 \, ,
 \label{eq:tot_energy_DevelopedHere}
\end{equation}
where we substituted the expression of $\xi$ from Eq.~\eqref{eq:cylinder_deflection} and used the definitions of Eq.~\eqref{eq:ak_active} inside the bending energy of Eq.~\eqref{eq:energy_bend_uniform}.

Stationary configurations of the system are found by setting the derivative of the total strain energy density with respect to the fiber alignment angle $\theta$ to zero: 
\begin{equation}
    \frac{\partial \psi_{\rm bend}}{\partial \theta} = \frac{H^3}{6}\kappa_1 \sin2\theta \bigl[ -\alpha \mu - 2\alpha k_{44} + 2 k_{44} \kappa_1 (1-\cos^2\theta) \bigr] = 0 \, ,
    \label{eq:configur_der_ex1}
\end{equation}
where we used the approximation $\kappa_1 \cos(\kappa_1 Y) \approx \kappa_1$ to further simplify the problem. This approximation is justified within the Föppl--von Kármán limit, consistent with the assumptions $\xi = \mathcal{O}(H)$ and with the geometry of the system. Indeed, the coordinate dependence of $\kappa_1 \cos(\kappa_1 Y)$ appears only in higher-order terms. Since $\kappa_1 Y$ is small under the Föppl--von Kármán hypothesis, the additional terms in the expansion are of smaller order and can be neglected. Consequently, we disregard these contributions, which depend on the specific cell geometry and on the coordinate dependence of $\Delta \xi$. Nevertheless, such effects could be incorporated in the model at the cost of an increased computational complexity.

Equation~\eqref{eq:configur_der_ex1} highlights the existence of two trivial stationary configurations, namely $\theta_{\|} = 0$ and $ \theta_\perp = \pi/2$, corresponding to cell alignment along the cylinder axis and in the circumferential direction, respectively. Additionally, there may be a third stationary configuration $\theta^*$ satisfying 
\begin{equation} \label{eq:cylinder_theta_star}
    \cos^2\theta^* = 1 - \frac{\alpha}{\kappa_1} \left(1+\frac{\mu}{2k_{44}}\right) \, ,
\end{equation}
which exists if and only if 
\[
    0 \leq \frac{\alpha}{\kappa_1}\left(1+\frac{\mu}{2k_{44}} \right) \leq 1 \, .
\]
The left inequality is trivially satisfied given the signs of the terms involved, whereas the right inequality restricts the parameter regime for the existence of $\theta^*$, depending on material $(\mu,k_{44})$, active $(\alpha)$, and geometric $(\kappa_1)$ parameters.

Stability of these alignments may be directly assessed through the analysis of the second derivative of the energy, due to the smoothness hypothesis of the fields involved:
\begin{equation}
\frac{\partial^2\psi_{\rm bend}}{\partial \theta^2} = \frac{H^3}{3}\kappa_1 \cos2\theta \bigl[ -\alpha\mu - 2\alpha k_{44} + 2k_{44} \kappa_1 (1-\cos^2\theta) \bigr]  + \frac{H^3}{3}k_{44} \, \kappa_1^2 \sin^2 2\theta \, .
\end{equation}
Therefore, we have for the stationary configurations found above:
\begin{align}
 & \frac{\partial^2 \psi_{\rm bend}}{\partial \theta^2}(\theta_{\|}) = -\frac{H^3}{3}\kappa_1\alpha (\mu + 2k_{44}) < 0 \qquad \textrm{for all values of the parameters;} \\[1.3ex]
 & \frac{\partial^2 \psi_{\rm bend}}{\partial \theta^2}(\theta_{\perp}) = \frac{H^3}{3}\kappa_1 \bigl[ \alpha(\mu + 2k_{44}) - 2k_{44}\kappa_1 \bigr] \, ; \\[1.3ex]
 & \frac{\partial^2 \psi_{\rm bend}}{\partial \theta^2}(\theta^*) = \frac{H^3}{3}k_{44}\kappa_1^2 \sin^2 2\theta^* > 0 \, .
\end{align}
The stability analysis is then immediate. The parallel configuration $\theta_{\|}$, i.e., where the anisotropy of the cell is oriented along the cylinder axis, is never stable. Even though this may seem counterintuitive, it is observed in experiments, as discussed later. On the other hand, recalling that $\kappa_1 < 0$, the stability of the orthogonal configuration $\theta_\perp$ arises if and only if
\begin{equation}
    \alpha(\mu + 2k_{44}) - 2k_{44}\kappa_1 < 0 \quad  \iff \quad \frac{\alpha}{\kappa_1}\left(1+\frac{\mu}{2k_{44}} \right) > 1 \, ,
\end{equation}
which corresponds to the opposite inequality proven for the existence of the oblique alignment $\theta^*$. Therefore, if $\theta^*$ exists, it is always a stable configuration, otherwise the cell aligns with the circumferential direction, orthogonal to the cylinder axis. Thus, the system undergoes a bifurcation depending on the active contractility measure $\alpha$, on the substrate curvature $\kappa_1$ and on the mechanical properties of the cell itself. Each of these parameters plays a crucial role in determining the preferential alignment, and a detailed analysis is able to qualitatively explains some of the experimental observations, as we will show next.


\subsubsection{Discussion}

The bifurcation analysis presented above reveals several noteworthy insights. First, Eq.~\eqref{eq:cylinder_theta_star} clearly separates the influence of material parameters from that of the deformation fields. In particular, material properties emerge through the stiffness ratio $\mu/2k_{44}$, where $\mu$ denotes the isotropic shear modulus and $k_{44}$ characterizes the stiffness along the stress fibers (SF). Since $k_{44}$ describes the reinforcement provided by the SF network, which is typically stiffer than the surrounding structure, it is reasonable to assume that $\mu/k_{44} < 1$.

On the other hand, the ratio $\alpha/\kappa_1$ plays a central role in our model, representing the competition between active contractility and substrate curvature. Its variation captures the range of possible optimal cell orientations on the cylinder. Once the material parameters are fixed, high contractility relative to curvature promotes alignment in the circumferential direction, while high curvature with relatively low contractility favors a more axial  orientation. This admits two complementary interpretations. For a given substrate curvature, more contractile cells tend to align perpendicular to the cylinder axis, while less contractile cell types are expected to align obliquely or more parallel to it. Conversely, for fixed contractility, high curvature drives axial alignment, whereas low curvature yields oblique or perpendicular orientation. 

\begin{figure}[t]
\centering
\begin{subfigure}[b]{0.47\textwidth}
\centering
{\includegraphics[width=\textwidth]{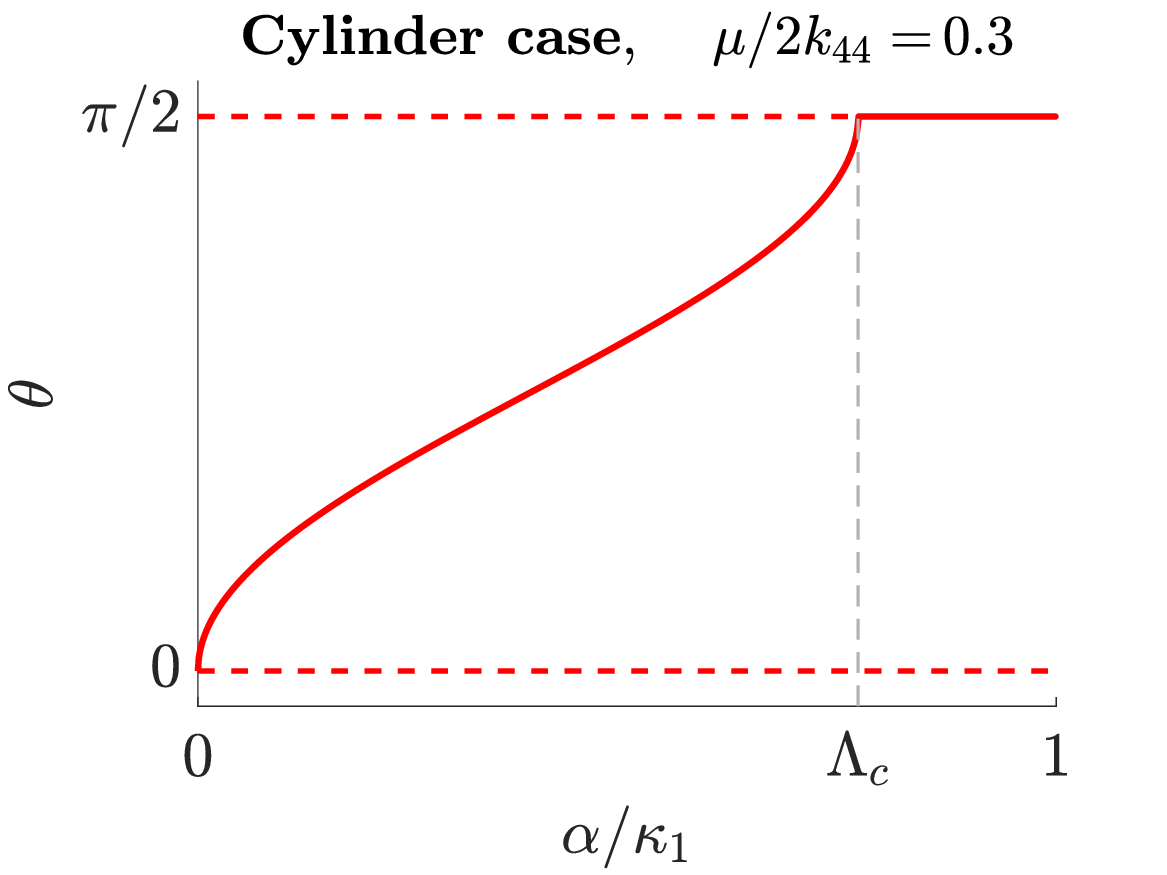}}
\caption{}
\label{fig:cylinder_theta}
\end{subfigure}
\hfill
\begin{subfigure}[b]{0.47\textwidth}
{\includegraphics[width=\textwidth]{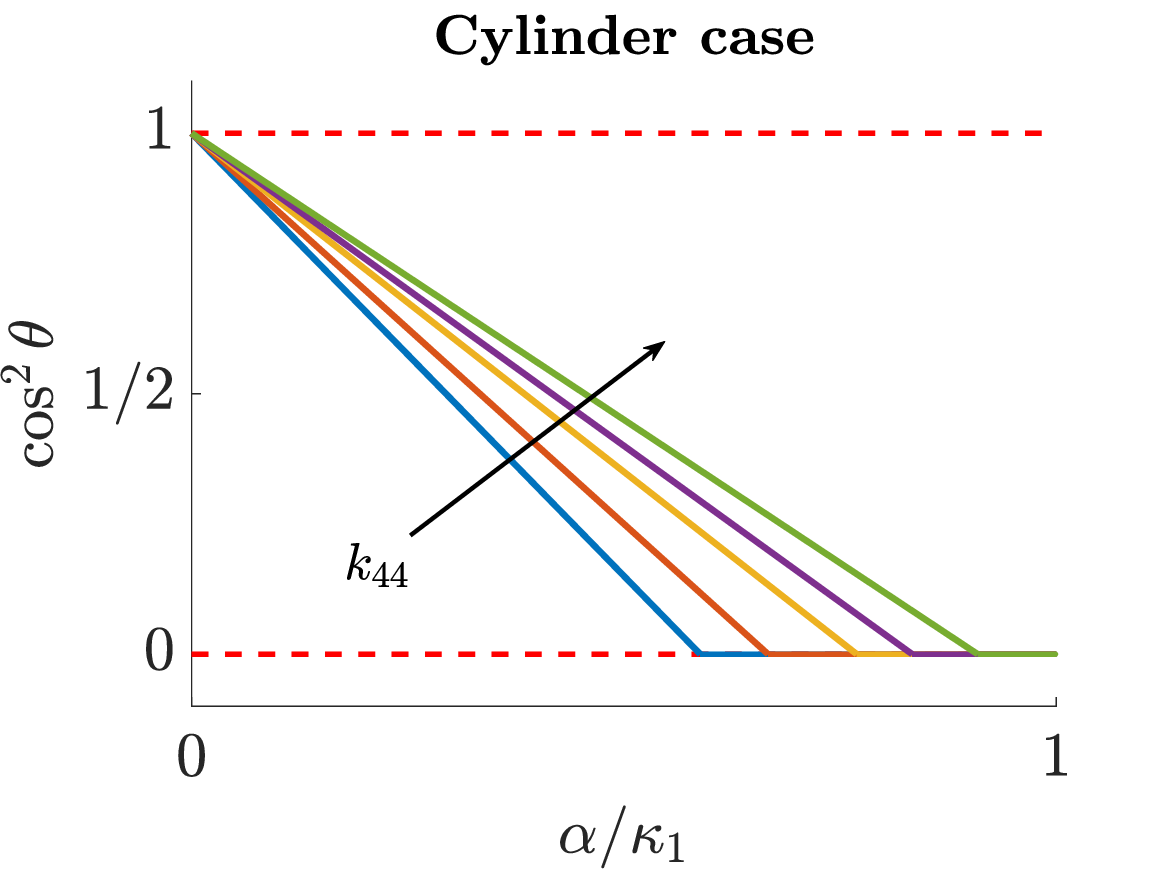}}
\caption{}
\label{fig:cylinder_cos2_theta}
\end{subfigure}
\caption{(a) Bifurcation diagram showing stable (solid lines) and unstable (dashed lines) orientations on a cylinder, as a function of $\alpha/\kappa_1$, for a fixed value of $\mu/2k_{44}= 0.3$. The bifurcation point $\Lambda_c$ is defined in \eqref{eq:bif_point_cylinder}. (b) Bifurcation diagram in the space $(\alpha/\kappa_1, \cos^2\theta)$ for the cylinder case and different values of $k_{44}$.}
\label{fig:bifurcation_diagram_cylinder}
\end{figure}

The bifurcation diagrams in Fig.~\ref{fig:bifurcation_diagram_cylinder} illustrate this behavior. The parallel configuration $\theta_{\|}$ is never strictly optimal but may be approached asymptotically for large  curvature. The bifurcation point at which the oblique configuration $\theta^*$ ceases to exist and the perpendicular configuration $\theta_\perp$ becomes stable is given by:
\begin{equation}
\Lambda_c := \left(1 + \frac{\mu}{2k_{44}}\right)^{-1} \, .
\label{eq:bif_point_cylinder}
\end{equation}
This bifurcation threshold depends only on the material properties of the cell. Cells with a stiff anisotropic architecture exhibit a shift of the bifurcation point toward the upper limit of 1. The fact that this bifurcation depends solely on material characteristics is consistent with general features of cell reorientation and also emerges in other modeling frameworks  \citep{lucci2021nonlinear, livne2014cell}.
Furthermore, the slope of the bifurcation curve in the $(\alpha/\kappa_1, \cos^2\theta)$ plane is also governed by material parameters, as shown in Fig.~\ref{fig:cylinder_cos2_theta}. For a fixed ratio $\alpha/\kappa_1$, increasing the stiffness along the fiber direction shifts the optimal configuration toward parallel alignment.

From a geometric perspective, one should note that the dependence on the cell's own geometric properties  is lost due to the Föppl--von Kármán approximation. This occurs because we assumed deflection of the order of the thickness, which remains small compared to the radius of the cell. However, this approximation breaks down for large displacements $\xi$ or large polarization of the cell, i.e., high aspect ratio between the typical sizes of the cell. In such cases, it may be necessary to introduce a  measure of perceived curvature that accounts for the cell size relative to the cylinder radius. 

These theoretical results qualitatively align with several biological observations. For instance, epithelial cells, typically characterized by a softer cytoskeletal structure and relatively high contractility, tend to align perpendicularly to the axis of the cylinder, as observed in \citep{yevick2015architecture, yu2018substrate}. In contrast, muscle-like cells such as myofibroblasts, which exhibit a stiffer and more elongated architecture, are more sensitive to curvature and tend to align parallel to the cylinder axis to avoid bending, as reported in several studies \citep{bade2017curvature,jin2021emergent}.
However, an important property must be emphasized: while we mathematically treat $\alpha$ (contractility) as an independent variable, it is tightly linked to the cytoskeletal architecture. Since stress fibers contribute both to active contractility and to anisotropic stiffness, any experimental manipulation that alters one may affect the other. For example, increasing contractility by activating Rho proteins \citep{bade2017curvature} or reducing it through pharmacological agents  \citep{frey2025curvature} may produce complex and sometimes counterintuitive behaviors.
Specifically, one might naively expect that reducing $\alpha$ through  chemical treatment should shift the preferred orientation toward the parallel configuration. However, if the drug also disrupts the cytoskeleton, thereby reducing $k_{44}$, the net result may paradoxically reinforce the shift toward perpendicular alignment. This underscores the importance of interpreting biological experiments cautiously, as chemical stimuli often simultaneously affect both active and passive mechanical properties of the cell.

\subsubsection{Comparison with modeling literature}

To the best of our knowledge, only two models have been proposed on this topic. \citet{sanz2009effect} introduced a framework that accounts for a pre-deformed state arising from the underlying geometry and includes an active term operating in a suboptimal bent configuration. However, their formulation did not incorporate any explicit energy dependence on the bending deformation itself. As a result, it could not  reproduce the experimental observations for cylindrical substrates, although it did provide new qualitative insights.

In contrast, \citet{biton2009cellular} successfully captured the experimental trends at a qualitative level. Their model describes the cell as an isotropic elastic plate and emphasizes the interplay between passive bending and stress fibers contractility. To account for the anisotropic effect of stress fibers, they included an additional energy term corresponding to the bending of individual fibers modeled as elastic rods.
In their formulation, the passive bending deformation is prescribed, while the active contractility is represented as a linear compressive strain field that reaches its maximum (in absolute value) at the apical surface of the cell. Following this approach, the total strain energy density is expressed as \citep{biton2009cellular}
\begin{equation}
\psi_{\rm tot} \propto H^3 \left[ \alpha^2 + \kappa_1^2 + \alpha \kappa_1\left(\frac{1}{2} \cos2\theta - \frac{3}{2} \right) - \Gamma \lambda \alpha \kappa_1^2 \sin^4\theta \right] \, ,
\label{eq:total_Denergy_B_S}
\end{equation}
where $\lambda$ is the ratio between the central surface area and the plate thickness, while $\Gamma$ is a dimensionless parameter capturing the effect of stress fiber density, bending stiffness of individual fibers, and the stiffness of the isotropic matrix.

Structural similarities can be observed between Eq.~\eqref{eq:total_Denergy_B_S} and the total energy derived in our formulation (see Eq.~\eqref{eq:tot_energy_DevelopedHere}). Both models include an isotropic bending term, an active term, and a fiber-related bending contribution, all modulated by the orientation angle $\theta$. Notably, both models feature mixed $\alpha \kappa_1$ terms and curvature-squared contributions $\kappa_1^2$ weighted by angular functions and stiffness ratios.
Two key methodological differences distinguish the frameworks. First, in \citep{biton2009cellular}, the bending contribution of stress fibers is introduced as a correction to an isotropic plate model. Our approach instead incorporates anisotropy at the constitutive level by assuming transverse isotropy in the material symmetry group itself, yielding a description of the cell's mechanical response grounded in Continuum Mechanics common approaches. Second,  \citet{biton2009cellular} model active contractility as a macroscopic compressive strain applied along the apical surface of the cell, motivated by experimental observations. Our formulation instead introduces active strains based on the organization and activation pattern of the cytoskeleton, producing a basal-biased activation profile consistent with the biological function of stress fibers, although simplified to maintain analytical tractability. Despite these conceptual differences, both models ultimately admit the same class of macroscopically admissible deformations, differing only by a sign convention in the active term. 

Although our model introduces additional mathematical complexity, it has the advantage of being consistent with theoretical frameworks commonly employed in active matter mechanics, while aligning with both experimental observations and prior theoretical studies. By embedding anisotropy and activity directly at the kinematic and constitutive level before performing dimensional reduction, the present formulation remains very general and offers an alternative systematic description that complements existing approaches. 

\subsection{Cell adherent to an ellipsoidal or spherical substrate} \label{sec:ellipsoid}

The main advantage of the Föppl–von Kármán formulation lies in its intrinsic coupling between the deflection $\xi$ and the Airy stress function $\phi$. In particular, when the substrate is described as a surface with non-zero Gaussian curvature, an elastic deformation is necessarily induced in order to map the reference flat configuration onto the target geometry.

We consider the deformation of a thin plate from a flat configuration to an ellipsoidal surface defined by
\begin{equation}
    \frac{X^2}{A^2} + \frac{Y^2}{B^2} + Z^2 = R^2 \, ,
\end{equation}
where $AR$, $BR$, and $R$ are the semi-axes of the ellipsoid. This equation reduces to several relevant special cases: if $A = B$ we obtain a rotational ellipsoid; if $A = B = 1$ it corresponds to a sphere of radius $R$; while in the limit $A \to +\infty$ it becomes a cylinder (specifically, the same cylinder considered in Sect.~\ref{sec:cylinder} if $B =1$).   The deflection at the apex of the surface is approximated as
\begin{equation}
    \xi(X,Y) = -\frac{1}{R} \left(\frac{X^2}{2A^2} + \frac{Y^2}{2B^2}\right) = \kappa_1 \left(\frac{X^2}{2A^2} + \frac{Y^2}{2B^2}\right) \, .
    \label{eq:xi_ellips}
\end{equation}
For $A = B$, this expression coincides with the deflection of a spheroidal cap.

In contrast to the cylindrical case, the non-zero Gaussian curvature associated with $\xi$ necessarily induces stretching deformations that must be accounted for in the energy balance.
\rev{However, it enters the governing equations through a constant forcing term in the Airy equation. Furthermore, when the mid-surface $\mathcal{S}$ is rotationally symmetric, it can be proven that the stretching energy $\mathscr{W}_{\mathrm{str}}$ defined in Eq. \eqref{eq:Wtot_lin} is independent of the anisotropy orientation angle $\theta$: under the rotation $g_\theta$ that aligns the coordinate axes with the fiber direction, the Airy equation retains the same structure and $\mathcal{S}$ maps onto itself, so that a change of variables in the energy integral yields $\mathscr{W}_{\mathrm{str}}(\phi,\theta) = \mathscr{W}_{\mathrm{str}}(\phi_0,0)$ for all $\theta$ (see \ref{sec:Appendix_D} for additional details).} The stretching energy therefore contributes a constant term that is independent of the director orientation, and its contribution can be neglected in the analysis of preferential alignment that follows.

In this situation, we have $C_{m,k} = C_G = 0$, $C_m = -k_{11}-k_{22}$, $\mathbf{b}_{\rm a} = \mathbf{0}$, and $[\xi, \xi] = \kappa_1^2 / (A^2 B^2)$. Moreover, the choice of the deflection yields $ \Delta^2 \xi = \Delta_{\mathbb{M}}^2\xi = 0$.  Following the same approach as in the cylindrical case, the stationary configurations of the bending energy satisfy
\begin{equation}
    \frac{\partial \psi_{\rm bend}}{\partial \theta} = \frac{H^3}{6} \kappa_1 
    \left( \frac{1}{B^2} - \frac{1}{A^2} \right)\sin2\theta
    \left[-\alpha(\mu + 2k_{44}) + 2k_{44}\kappa_1\left(\frac{1}{B^2} + 
    \biggl(\frac{1}{A^2} - \frac{1}{B^2}\right) \cos^2\theta \biggr) \right] \, .
    \label{eq:stationary_elips}
\end{equation}
It follows that, in the case $A=B$, no preferential orientation exists, because the dependence on $\theta$ in Eq.~\eqref{eq:stationary_elips} vanishes. This is expected and consistent with experimental observations, since at the apex of a rotational ellipsoid the principal curvatures are equal and there are no preferred directions. 
When $A \neq B$, the parallel $\theta_\| = 0$ and orthogonal $\theta_\perp = \pi/2$ orientations are stationary points of the energy. In addition, a third oblique configuration $\theta^*$ may exist, satisfying
\begin{equation} \label{eq:cos2_ellipsoid}
    \cos^2\theta^* = \frac{A^2}{A^2-B^2}\left[1 - 
    \frac{\alpha}{\kappa_1} B^2\left(1+\frac{\mu}{2k_{44}}\right) \right] \, .
\end{equation}
The existence of $\theta^*$ is not guaranteed and depends sensitively on the ellipsoidal geometry and its principal curvatures through $A$, $B$, and $R$. Note that, for $B = 1$ and $A \to +\infty$, Eq.~\eqref{eq:cos2_ellipsoid} reduces to Eq.~\eqref{eq:cylinder_theta_star}, i.e., to the cylindrical case. 

To complete the analysis, we focus on the case $A > B$; the opposite case $B > A$ is symmetric and leads to analogous conclusions.
The oblique solution $\theta^*$ exists if and only if
\begin{equation}
  \frac{\alpha}{\kappa_1} \in \left[\Lambda_0, \Lambda_\perp \right]\, , \qquad \text{where} \quad \Lambda_{\perp} :=\frac{1}{B^2}\biggl(1+\frac{\mu}{2k_{44}} \biggr)^{-1}   \, , 
  \quad 
   \Lambda_0 := \frac{1}{A^2}\biggl(1+\frac{\mu}{2k_{44}} \biggr)^{-1}  \, . 
  \label{eq:existence_ellips}
\end{equation}
The stability analysis through the second derivative yields:
\begin{align}
 & \frac{\partial^2 \psi_{\rm bend}}{\partial \theta^2}(\theta_{\|}) = \frac{H^3}{3}\kappa_1 \left(\frac{1}{B^2}-\frac{1}{A^2}\right)\left[-\alpha(\mu + 2k_{44}) + \frac{2k_{44}\kappa_1}{A^2}\right] \, , \\[1.5ex]
 & \frac{\partial^2 \psi_{\rm bend}}{\partial \theta^2}(\theta_{\perp}) = -\frac{H^3}{3}\kappa_1 \left(\frac{1}{B^2}-\frac{1}{A^2}\right)\left[-\alpha(\mu + 2k_{44}) + \frac{2k_{44}\kappa_1}{B^2}\right] \, ,\\[1.5ex]
 & \frac{\partial^2 \psi_{\rm bend}}{\partial \theta^2}(\theta^*) = \frac{H^3}{3}k_{44}\kappa_1^2 \left( \frac{1}{B^2} - \frac{1}{A^2}\right)^2 \sin^2 2\theta^* \, .
\end{align}
Thus, whenever the oblique configuration $\theta^*$ exists, it is stable. Furthermore, the system exhibits two bifurcations at the critical values $\Lambda_0$ and $\Lambda_\perp$, which coincide with the boundaries of existence of $\theta^*$. Specifically, for $\alpha/\kappa_1<\Lambda_0$, the parallel configuration $\theta_\|$ is stable, while $\theta_\perp$ is unstable and $\theta^*$ does not exist. Conversely, for $\alpha/\kappa_1 > \Lambda_\perp$, the orthogonal configuration $\theta_\perp$ is stable, while $\theta_0$ becomes unstable. In the intermediate regime $\Lambda_0 < \alpha/\kappa_1 < \Lambda_\perp$, the only stable solution is the oblique orientation $\theta^*$.

\subsubsection{Discussion}

\begin{figure}[t]
\centering
\begin{subfigure}[b]{0.47\textwidth}
\centering
{\includegraphics[width=\textwidth]{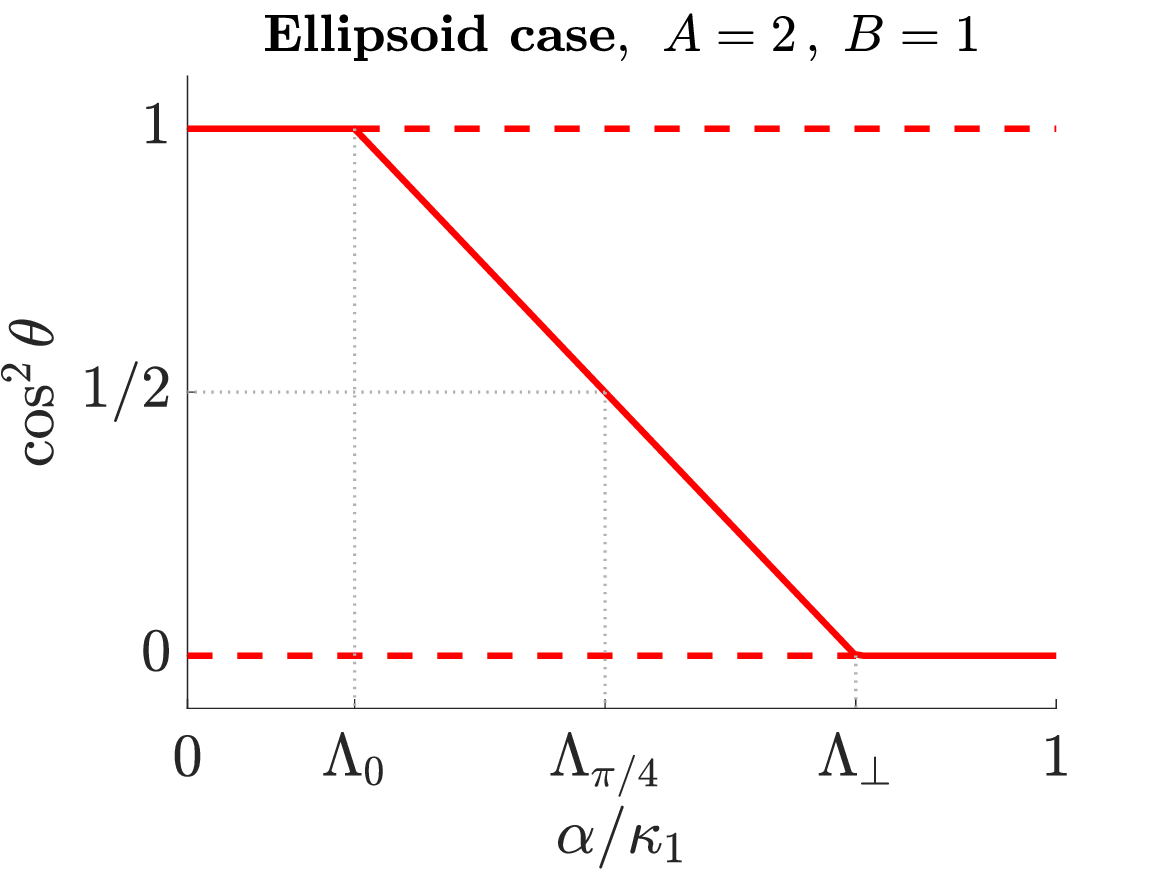}}
\caption{}
\label{fig:bif_ellipsoid_cos2theta}
\end{subfigure}
\begin{subfigure}[b]{0.47\textwidth}
{\includegraphics[width=\textwidth]{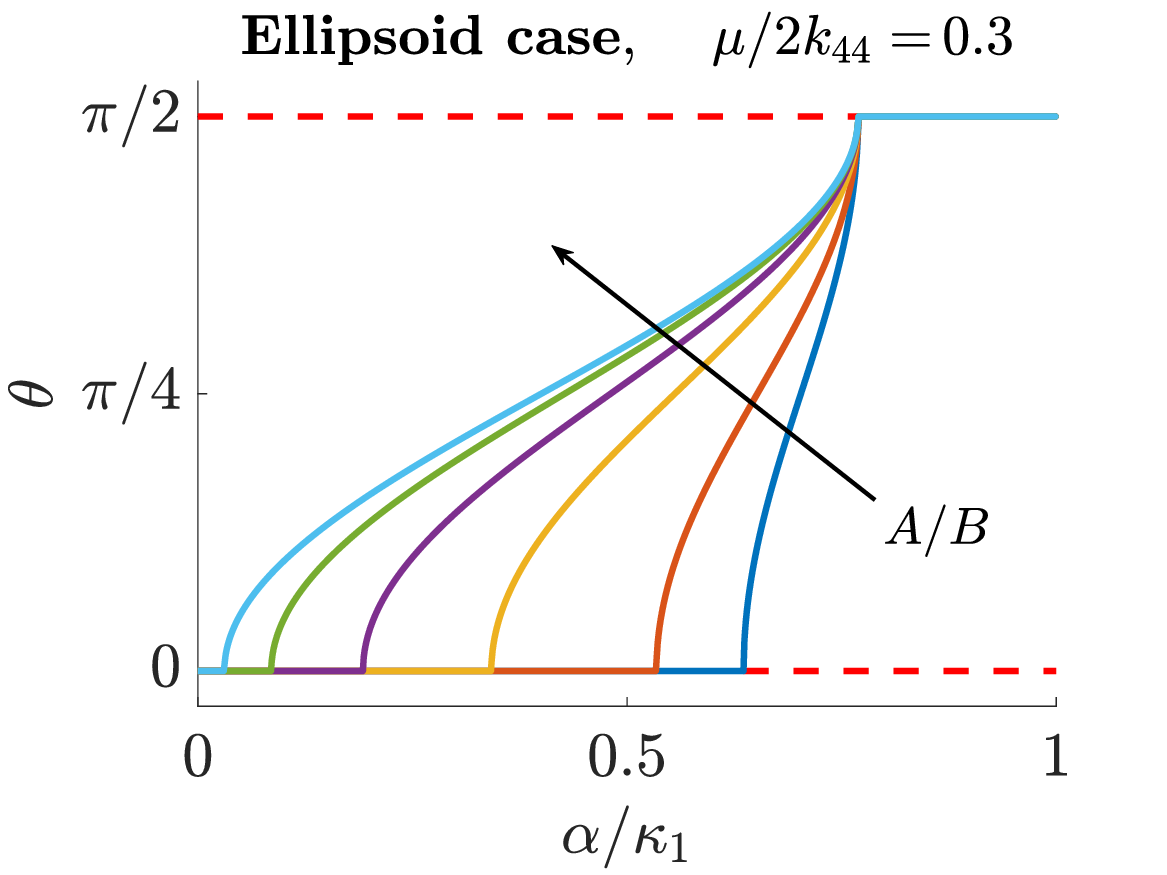}}
\caption{}
\label{fig:bif_ellipsoid_theta}
\end{subfigure}
\caption{ (a) Plot of $\cos^2(\theta)$ as a function of $\Lambda = \alpha/\kappa_1$ for the ellipsoid case, with $A = 2$ and $B = 1$. (b) Plot of the stationary angles $\theta$ as a function of $\Lambda = \alpha/\kappa_1$ for different values of $A/B$. The dashed and solid lines describe unstable and stable configurations, respectively.}
\label{fig:bifurcation_diag_ellips}
\end{figure}

The main qualitative difference from the cylindrical case lies in the fact that, 
for the ellipsoidal geometry, the parallel orientation $\theta_\|$ can be stable over a finite range of $\alpha/\kappa_1$ values, 
whereas on the cylinder it is attained only as a limiting case. 
This behavior is clearly illustrated in Fig.~\ref{fig:bifurcation_diag_ellips}, 
where the parallel orientation becomes optimal for values of $\alpha/\kappa_1$ smaller than the bifurcation point $\Lambda_0$, 
defined in Eq.~\eqref{eq:existence_ellips}. 
This result aligns with the physical intuition of the problem: since the ellipsoidal surface exhibits different curvature in the principal directions, 
active contractility does not favor a single preferred orientation. 
Consequently, the parallel direction is less energetically penalized than in the cylindrical case. 
Furthermore, Fig.~\ref{fig:bif_ellipsoid_theta}  shows that increasing the ratio $A/B$, 
and thus making the ellipsoid approach a cylindrical shape, 
the bifurcation diagram progressively tends to that of the cylinder reported in Fig.~\ref{fig:bifurcation_diagram_cylinder}, as expected. 

\begin{figure}[t]
\centering
\begin{subfigure}[b]{0.45\textwidth}
\centering
{\includegraphics[width=\textwidth]{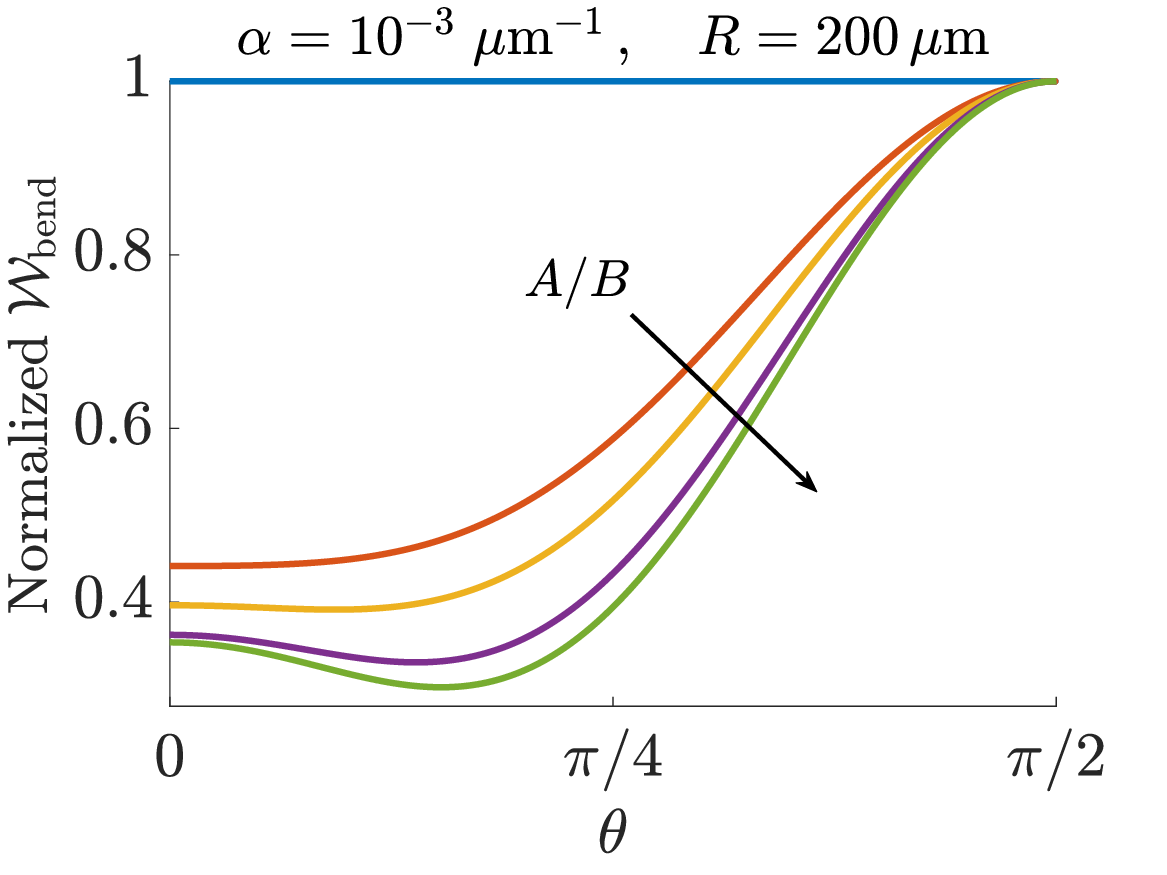}}
\caption{}
\label{fig:W_bend_1e03}
\end{subfigure}
\hfill
\begin{subfigure}[b]{0.45\textwidth}
\centering
{\includegraphics[width=\textwidth]{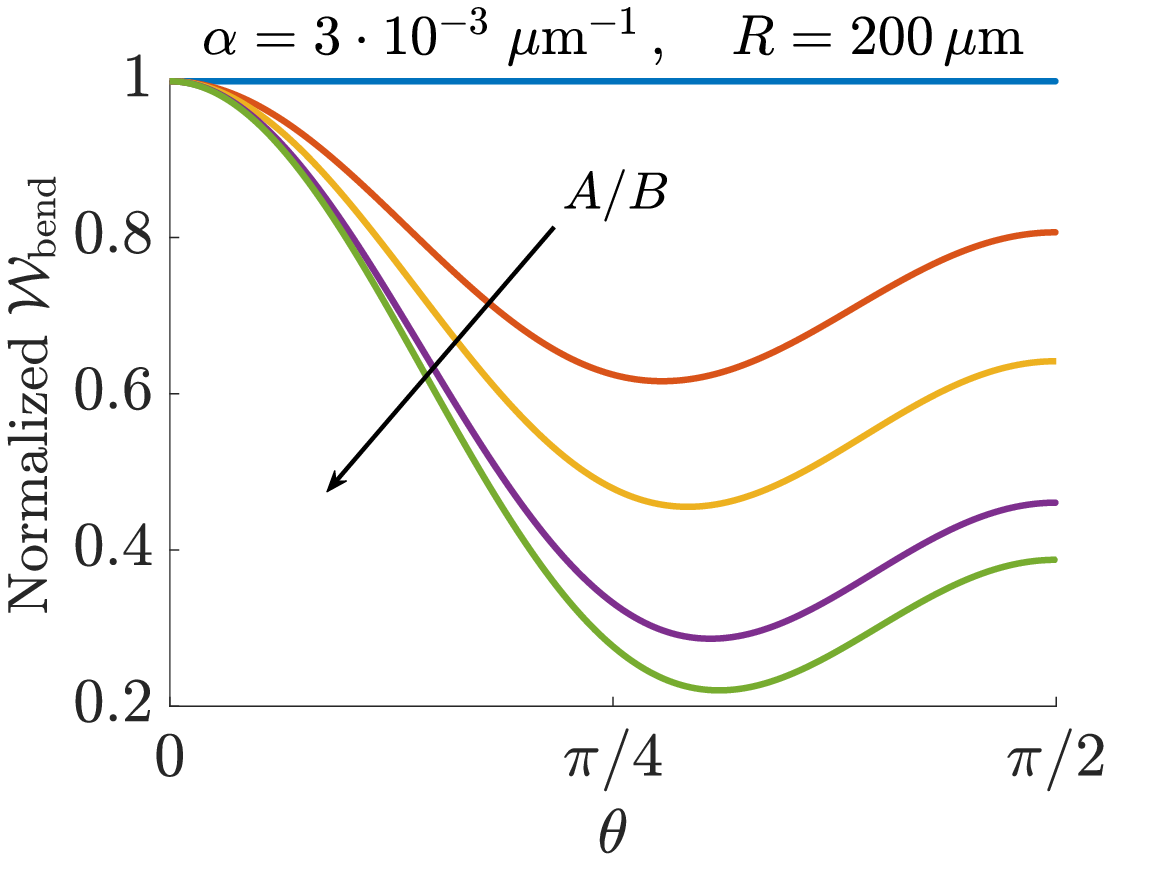}}
\caption{}
\label{fig:W_bend_3e03}
\end{subfigure} 
\begin{subfigure}[b]{0.45\textwidth}
\centering
{\includegraphics[width=\textwidth]{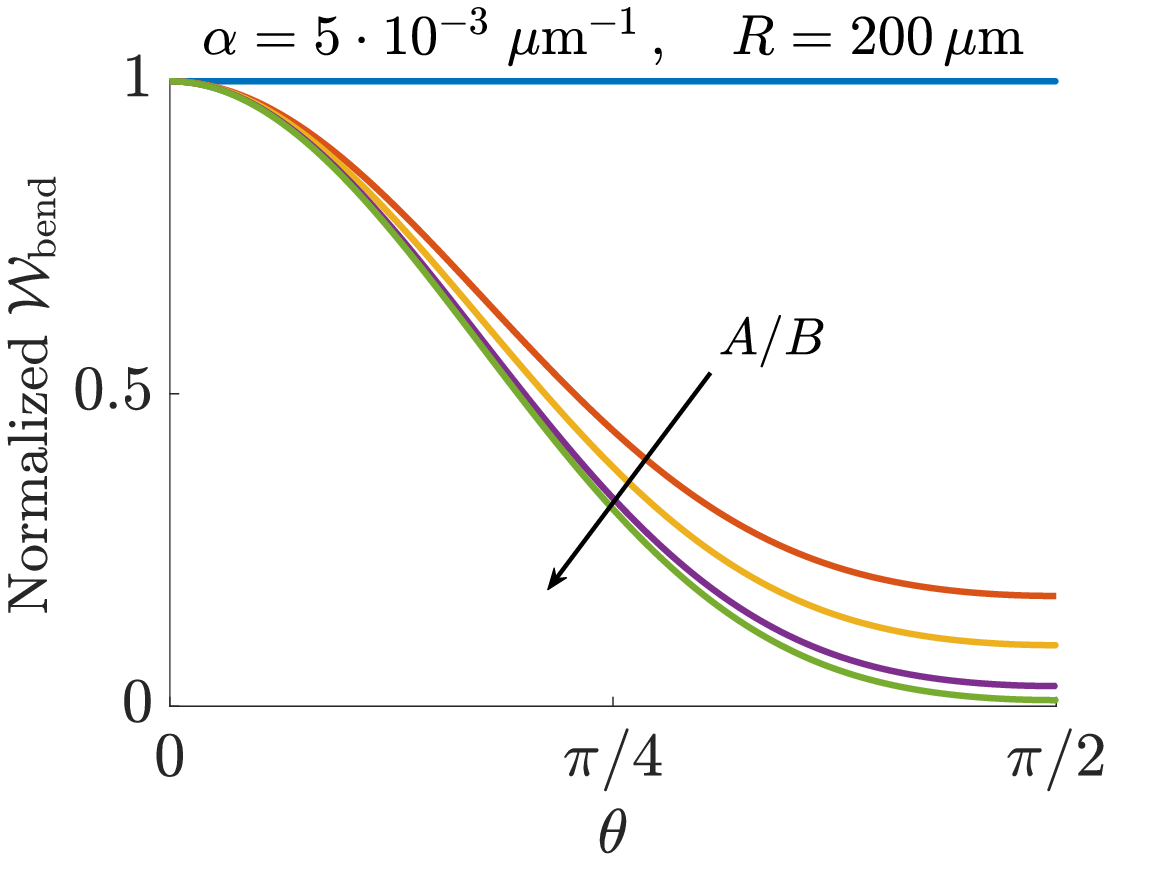}}
\caption{}
\label{fig:W_bend_5e03}
\end{subfigure}
\hfill
\begin{subfigure}[b]{0.45\textwidth}
\centering
{\includegraphics[width=\textwidth]{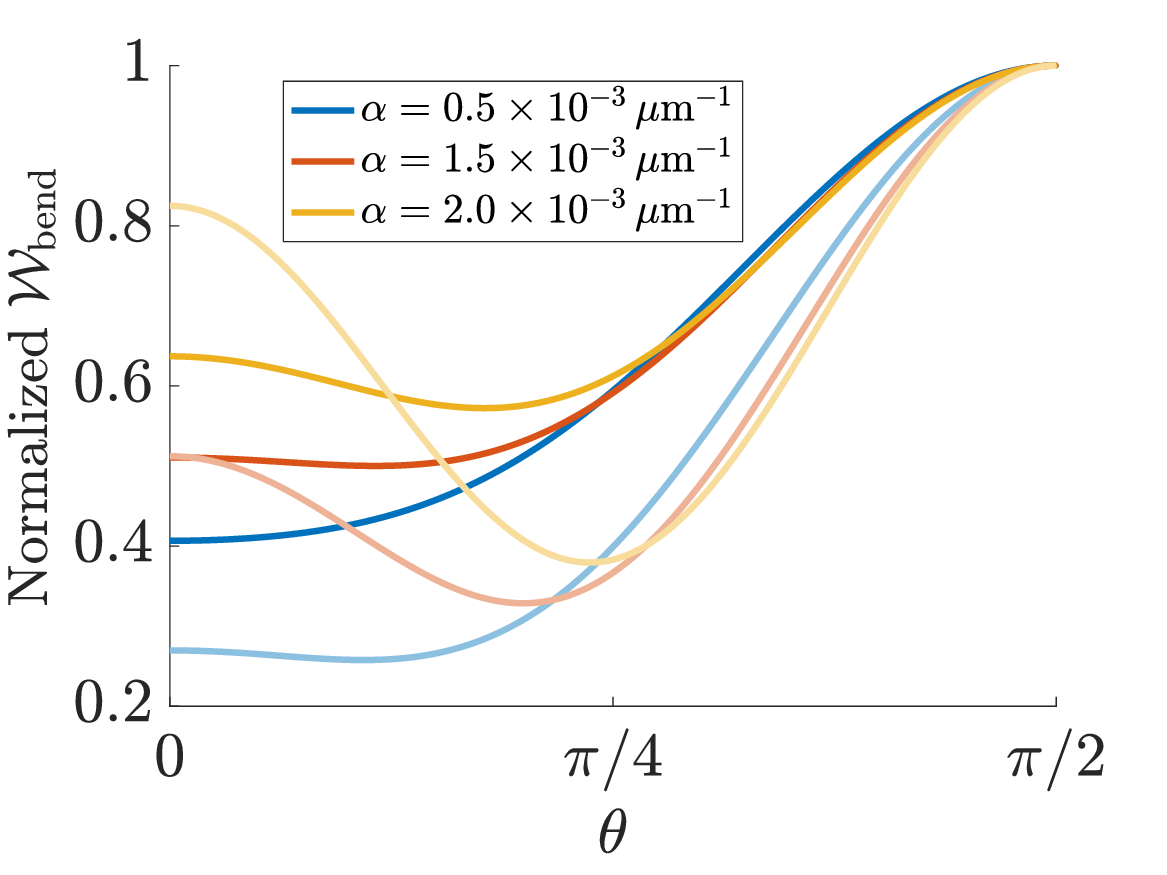}}
\caption{}
\label{fig:W_bend_comparison}
\end{subfigure}
i\caption{Plots of the normalized bending energy $\mathscr{W}_{\rm bend}$ as a function of the angle, for different values of the ratio $A/B$. (a) $\alpha = 10^{-3} \mu$m$^{-1}$. (b) $\alpha = 3\cdot10^{-3} \mu$m$^{-1}$. (c) $\alpha = 5\cdot10^{-3} \mu$m$^{-1}$. (d) Comparison for different values of $\alpha$ and $A/B$. For each value of $\alpha$, darker curves correspond to $A/B = 2$, while lighter curves correspond to $A/B = 20$.}
\label{fig:bend_energy_ellips}
\end{figure}

Fig.~\ref{fig:bend_energy_ellips} reports the normalized bending energy $\mathscr{W}_{\rm bend}$ 
as a function of the orientation angle $\theta \in (0,\pi/2)$ for different values of $\alpha$ and aspect ratios $A/B$ of the ellipsoid. 
The three panels correspond to the regimes where the optimal configuration is parallel (small $\Lambda$), 
oblique (intermediate $\Lambda$), or orthogonal (large $\Lambda$). 
In particular, for the spherical case ($A=B=1$, blue solid line), 
the bending energy is independent of $\theta$, as expected, since the curvature is constant in all directions. 
Conversely, for large $A/B$, the bending energy profile approaches that of the cylindrical case, consistent with the bifurcation results. 
An important feature is the increasing energy gap between stationary orientations as $A/B$ grows, 
indicating that although multiple stationary configurations exist, their energy differences may be small enough 
for other effects, such as thermal fluctuations, to overcome them.

\section{Conclusions} \label{sec:conclusions}

In this work, we developed a theoretical framework for modeling the mechanics of active anisotropic plates, which we applied to the alignment of cells adherent to curved substrates. Building upon the Föppl--von Kármán limit, we introduced a constitutive formulation that incorporates both passive bending imposed by the underlying substrate geometry and active remodeling induced by stress-fiber contractility. Unlike previous isotropic approaches to the problem, our model explicitly accounts for transverse isotropy through the material properties, thereby capturing the directional mechanical response induced by cytoskeletal architecture.

The main contribution of this study is the derivation of a reduced plate model that consistently integrates active anisotropic deformations, possibly varying through the thickness, within a nonlinear continuum mechanics framework. Through asymptotic expansion, we obtained coupled equilibrium equations linking the out-of-plane deflection and in-plane stresses, and demonstrated their ability to qualitatively reproduce key experimental observations on curvature-induced cell alignment. In particular, the model predicts the existence of bifurcation regimes for the preferred fiber orientation depending on the interplay between active contractility, substrate curvature, and material properties. Our theoretical results align with experimental observations and demonstrate the model's flexibility in analyzing curvature effects across different prototypical surfaces.  

Although our main application concerns cell mechanics, the proposed framework is very general and can describe the mechanics of active fiber-reinforced plates with potentially non-homogeneous fiber distribution, \rev{as observed in the example of Sect.~\ref{sec:example}}. Applications are envisaged in the fields of soft robotics, tissue engineering, and active composite materials, among others. 

Despite its flexibility, the present framework has some limitations. First, the formulation is restricted to transversely isotropic materials. Capturing the full range of cellular responses and material behavior would require extension to an orthotropic model. As shown in \citep{lucci2021nonlinear}, modeling cell orientation under cyclic stretching requires adopting  orthotropic constitutive relations, which significantly increases the number of parameters. Second, the analysis is limited to moderately large deflections consistent with the Föppl--von Kármán regime, thus neglecting possible effects of extreme bending or large in-plane strains. \rev{Although the membrane assumption employed here accounts for elastic thickness changes via the incompressibility constraint, it lacks an independent descriptor for significant through-thickness distension. More comprehensive thin-body theories that explicitly track thickness variations, such as those in \citep{deseri2008derivation, deseri2013stretching, dicarlo2001shells}, could capture additional physical effects in regimes of highly inhomogeneous active deformation.} Most notably, the cell alignment model derived herein is restricted to substrates with negative curvature. For positively curved geometries, experimental evidence often indicates cell detachment and loss of adhesion, phenomena that lie beyond the scope of the current framework.

Future work will focus on overcoming these limitations by incorporating adhesion dynamics, extending the formulation to handle positive curvature regimes, and refining the constitutive description to include orthotropic effects. \rev{In particular, treating the cell deflection as a variational unknown subject to a finite adhesion energy, rather than prescribing it from the substrate geometry, would allow the model to capture shape transitions arising from a stretching-versus-bending competition, complementing the orientation analysis presented here.} Additionally, spatially dependent fiber distributions and non-uniform active deformation fields, potentially stronger near the cell periphery, could be incorporated. Nevertheless, the proposed model provides a rigorous and versatile tool for studying the coupled influence of substrate geometry and active remodeling on cellular alignment, and can be readily adapted to more refined descriptions. 

\section*{Acknowledgements}
All authors acknowledge the support of the National Group of Mathematical Physics (GNFM) of the Italian National Institute of Advanced Mathematics (INdAM). GL is currently supported by the post-doctoral fellowship \emph{BE-FOR-ERC 2024} from Sapienza University of Rome. GL also acknowledges the support of the GNFM--INdAM project \emph{Progetto Giovani 2025} (CUP: E5324001950001). This work has been conducted under the National Plan for Complementary Investments to the PNRR, project “D34H—Digital Driven Diagnostics, prognostics and therapeutics for sustainable Health care” (project code: PNC0000001), Spoke 4 funded by the Italian Ministry of University and Research. The authors wish to thank Prof. Paola Nardinocchi for insightful comments and discussions on the work. 

\appendix
\section{Leading order of the bending energy}
\label{sec:Appendix_A}

In the following, we compute the leading order of the bending energy introduced in Sect.~\ref{sec:plate_model}, following the procedure outlined in \citep{dervaux2009morphogenesis}. Given the assumptions made in Sect.~\ref{sec:setup}, the leading order of the bending energy \(\mathscr{W}_{\rm bend}\) provided in Eq.~\eqref{eq:Wtot_lin} can be expanded using the effective constitutive equations \eqref{eq:reduced_constitutive_relations}:
\begin{align}
    \mathscr{W}_{\rm bend} & = \frac{\mu H^3}{6} \int_S \left[ (\mathbb{E}_{\rm e}^{[1]})_{11}^2 + (\mathbb{E}_{\rm e}^{[1]})_{22}^2 + (\mathbb{E}_{\rm e}^{[1]})_{12}^2 + (\mathbb{E}_{\rm e}^{[1]})_{11}(\mathbb{E}_{\rm e}^{[1]})_{22}\right]  \, \d A + \frac{k_{44}H^3}{6} \int_S (J_4^{[1]})^2 \, \d A \\[1ex] 
    &:= \mathscr{W}_{\mu} + \mathscr{W}_k \, , \notag
\end{align}
 where we have denoted by $\mathscr{W}_\mu$ the isotropic part and by $\mathscr{W}_k$ the anisotropic one, and $J_4^{[1]} = \mathbb{E}_{\rm e}^{[1]}:\mathbb{M}$. By using the expressions of $\mathbb{E}_{\rm e}^{[1]}$ provided in Eq.~\eqref{eq:E1}, after some algebra we arrive at the following expression for the isotropic bending energy: 
\begin{align}
 \mathscr{W}_\mu &= \frac{\mu H^3}{6}  \int_S \Biggr\{  \left(\frac{\partial^2 \xi}{\partial X^2} + \frac{\partial^2 \xi}{\partial Y^2} - \frac{\partial}{\partial X}(a_{13}+a_{31}) - \frac{\partial}{\partial Y}(a_{23} + a_{32}) + k_{11} + k_{22}\right)^2 \\[0.9ex]
& +\frac{\partial}{\partial X}\left[\frac{\partial \xi}{\partial X}\left(\frac{\partial}{\partial Y}(a_{23}+a_{32}) - k_{22}\right) - \frac{\partial \xi}{\partial Y}\left(\frac{\partial}{\partial Y}(a_{13}+a_{31}) - k_{12}\right)\right] \notag \\[0.9ex] 
& +\frac{\partial}{\partial Y}\left[\frac{\partial \xi}{\partial Y}\left(\frac{\partial}{\partial X}(a_{13}+a_{31}) - k_{11}\right) - \frac{\partial \xi}{\partial X}\left(\frac{\partial}{\partial X}(a_{23}+a_{32}) - k_{12}\right)\right] \notag \\[0.9ex] 
& +\frac{1}{4}\Biggl[-\frac{\partial}{\partial Y}(a_{13} + a_{31}) - \frac{\partial}{\partial X}(a_{23}+a_{32}) + 2k_{12} \Biggr]^2 - k_{11}k_{22} 
+ k_{11}\frac{\partial}{\partial Y}(a_{23}+a_{32}) + k_{22}\frac{\partial}{\partial X}(a_{13}+a_{31}) \notag \\[0.9ex] 
& - \frac{\partial}{\partial X}(a_{13}+a_{31})\frac{\partial}{\partial Y}(a_{23}+a_{32}) - \frac{\partial^2\xi}{\partial X^2}\frac{\partial^2\xi}{\partial Y^2} + \left(\frac{\partial^2\xi}{\partial X \partial Y}\right)^2 + \frac{\partial \xi}{\partial X} \left(\frac{\partial k_{22}}{\partial X} - \frac{\partial k_{12}}{\partial Y}\right) + \frac{\partial \xi}{\partial Y}\left(\frac{\partial k_{11}}{\partial Y} - \frac{\partial k_{12}}{\partial X}\right)\Biggr\} \, \d A 
\notag 
\end{align}
which can be rewritten compactly as 
\begin{equation} \label{eq:W_mu}
\mathscr{W}_\mu = \frac{\mu H^3}{6} \int_S \Bigl[ (\Delta \xi - C_m)^2 - \nabla \xi \cdot \mathbf{b}_{\rm a} - \mathcal{A} - \mathcal{F}\Bigr] \, \d A \, ,   
\end{equation}
where $C_m$ and $\mathbf{b}_{\rm a}$ have been defined in Eqs.~\eqref{eq:Cm}--\eqref{eq:Wa_ba}, while $\mathcal{A}$ and $\mathcal{F}$ are given respectively by 
\begin{align}
\mathcal{A} &:= [\xi, \xi] + \frac{\partial}{\partial X} \Biggl[ \frac{\partial \xi}{\partial Y} \left( \frac{\partial}{\partial Y}(a_{13} + a_{31}) - k_{12} \right) - \frac{\partial \xi}{\partial X} \left(\frac{\partial}{\partial Y}(a_{23} + a_{32}) - k_{22}\right)\Biggr] \label{eq:Acal}\\[0.7ex]
&\phantom{= [\xi, \xi]} + \frac{\partial}{\partial Y} \Biggl[ \frac{\partial \xi}{\partial X} \left( \frac{\partial}{\partial X}(a_{23} + a_{32}) - k_{12} \right) - \frac{\partial \xi}{\partial Y} \left(\frac{\partial}{\partial X}(a_{13} + a_{31}) - k_{11}\right)\Biggr] \notag \\[2ex]
&= \rev{[\xi,\xi] - \operatorname{div}(\mathbbm{h}^*\nabla\xi) - \nabla \xi \cdot \mathbf{f}_{\rm a} \, ,} \notag
\end{align}
\rev{with $\mathbbm{h}^* := (\det\mathbbm{h})\mathbbm{h}^{-1}$ the cofactor of the target curvature tensor $\mathbbm{h}$ and $\mathbf{f}_{\rm a} = (-(\operatorname{curl} \mathbbm{f})_{32}, (\operatorname{curl} \mathbbm{f})_{31})^{\rm T}$,} 
and 
\begin{align}
\mathcal{F} := -&\frac{1}{4}\Biggl[-\frac{\partial}{\partial Y}(a_{13} + a_{31}) - \frac{\partial}{\partial X}(a_{23}+a_{32}) + 2k_{12}\Biggr]^2 + k_{11}k_{22} \label{eq:Fcal} \\[1ex] 
&- k_{11}\frac{\partial}{\partial Y}(a_{23}+a_{32}) - k_{22}\frac{\partial}{\partial X}(a_{13}+a_{31}) + \frac{\partial}{\partial X}(a_{13}+a_{31})\frac{\partial}{\partial Y}(a_{23}+a_{32}) \, \, \rev{= \det \mathbbm{h}} \notag \, .
\end{align}
The operator $[\cdot,\cdot]$ computes the Gaussian curvature at leading order and is defined by
\begin{equation}
[\xi, \xi] = \frac{\partial^2 \xi}{\partial X^2}\frac{\partial^2 \xi}{\partial Y^2} - \left(\frac{\partial^2 \xi}{\partial X \partial Y}\right)^2 = \frac{1}{2}\operatorname{div} \begin{pmatrix}
    \dfrac{\partial \xi}{\partial X}\dfrac{\partial^2 \xi}{\partial Y^2} - \dfrac{\partial \xi}{\partial Y}\dfrac{\partial^2 \xi}{\partial X \partial Y} \\[2.5ex]
    \dfrac{\partial \xi}{\partial Y}\dfrac{\partial^2 \xi}{\partial X^2} - \dfrac{\partial \xi}{\partial X}\dfrac{\partial^2 \xi}{\partial X \partial Y}
\end{pmatrix} \, .
\end{equation}
\rev{Note that $\mathcal{A}$ is a purely boundary contribution: $[\xi,\xi]$ and $\operatorname{div}(\mathbbm{h}^*\nabla\xi)$ are divergence terms, while $\nabla\xi\cdot\mathbf{f}_{\rm a}$ reduces to a boundary integral, since $\operatorname{div}\mathbf{f}_{\rm a} = 0$ identically.} 

Moreover, we observe that the isotropic part of the bending energy can be rephrased as 
\begin{equation}
    \mathscr{W}_\mu = \frac{\mu H^3}{6} \int_S \Bigl[ (\operatorname{tr}\mathbb{W}_{\rm a})^2 - \det\mathbb{W}_{\rm a} \Bigr] \, \d A = \frac{\mu H^3}{6}\int_S \Bigl[(\Delta \xi - C_m)^2 - ([\xi, \xi] - C_G)\Bigr] \, \d A \, ,
\end{equation}
where $\mathbb{W}_{\rm a}$ has been defined in Eq.~\eqref{eq:Wa_ba} and 
\begin{align}
C_G :=   &-\frac{\partial}{\partial X} \Biggl[ \frac{\partial \xi}{\partial Y} \left(   \frac{\partial}{\partial Y}(a_{13} + a_{31}) - k_{12} \right) -   \frac{\partial \xi}{\partial X} \left(\frac{\partial}{\partial Y}(a_{23} + a_{32}) - k_{22}\right)\Biggr] \\[2ex] 
& -\frac{\partial}{\partial Y} \Biggl[ \frac{\partial \xi}{\partial X} \left(   \frac{\partial}{\partial X}(a_{23} + a_{32}) - k_{12} \right) -   \frac{\partial \xi}{\partial Y} \left(\frac{\partial}{\partial X}(a_{13} + a_{31}) - k_{11}\right)\Biggr] - \mathcal{F} - \nabla \xi \cdot \mathbf{b}_{\rm a} \notag \\[2ex] 
&= \rev{\operatorname{div}(\mathbbm{h}^*\nabla\xi) - \det\mathbbm{h} + \nabla \xi \cdot (\mathbf{f}_{\rm a} - \mathbf{b}_{\rm a})} \notag \, .
\end{align}

A similar reasoning can be applied to compute the anisotropic component $\mathscr{W}_k$. Note that 
\begin{equation}
J_4^{[1]} = \mathbb{E}_{\rm e}^{[1]}:\mathbb{M} = m_1^2 (\mathbb{E}_{\rm e}^{[1]})_{11} + 2 m_1 m_2 (\mathbb{E}_{\rm e}^{[1]})_{12} + m_2^2 (\mathbb{E}_{\rm e}^{[1]})_{22} \, .
\end{equation}
Thus, the anisotropic part of the bending energy reads: 
\begin{align}
\mathscr{W}_k = \frac{k_{44}H^3}{6}\int_S \Biggl\{ & m_1^2  \left[ \frac{\partial^2 \xi}{\partial X^2} - \frac{\partial}{\partial X}(a_{13}+a_{31}) + k_{11} \right] + m_2^2 \left[ \frac{\partial^2 \xi}{\partial Y^2} - \frac{\partial}{\partial Y}(a_{23}+a_{32}) + k_{22}\right]  \\[0.9ex]
& + 2 m_1 m_2 \left[ \frac{\partial^2 \xi}{\partial X \partial Y} - \frac{1}{2}\frac{\partial}{\partial X}(a_{23}+a_{32}) - \frac{1}{2}\frac{\partial}{\partial Y}(a_{13}+a_{31}) + k_{12} \right]\Biggr\}^2 \, \d A \, , \notag
\end{align}
which can be rewritten compactly as 
\begin{equation} \label{eq:W_k}
\mathscr{W}_k = \frac{k_{44}H^3}{6} \int_S \left(\mathbb{M} : \mathbb{W}_{\rm a}\right)^2 \, \d A \, . 
\end{equation}
Finally, the total bending energy is given by $\mathscr{W}_{\rm bend} = \mathscr{W}_\mu + \mathscr{W}_k$, as shown in Eq.~\eqref{eq:final_bending_energy}.

\section{Variation of the energy}
\label{sec:Appendix_B}

In this section, we derive the equilibrium equations and boundary conditions of the plate model through variational arguments. The variation of the stretching energy can be computed directly by using the definition of the central surface strain \(\mathbb{E}_{\rm e}^{[0]}\), given in Eq.~\eqref{eq:E0}, and the symmetry of the second Piola–Kirchhoff stress tensor \(\mathbb{S}_n^{[0]}\):
\begin{equation} 
\delta (\mathbb{S}_n^{[0]}:\mathbb{E}_{\rm e}^{[0]}) = 2\mathbb{S}^{[0]}_n : \,\delta\mathbb{E}_{\rm e}^{[0]} = 2(\mathbb{S}^{[0]}_n)_{\alpha\beta} \left( \frac{\partial \delta u_\alpha^{[0]}}{\partial X_\beta} + \frac{\partial \delta \xi}{\partial X_\beta}\frac{\partial \xi}{\partial X_\alpha}\right) \, .
\end{equation}
Thus, upon integration by parts, the variation of the stretching energy is written as
\begin{align}
\delta &\mathscr{W}_{\rm str}  = \frac H2 \int_S \delta (\mathbb{S}_n^{[0]}:\mathbb{E}_{\rm e}^{[0]}) \, \d A \label{eq:stretching_variation} \\[2ex] 
&= -H \int_S \left[ \frac{\partial (\mathbb{S}^{[0]}_n)_{\alpha\beta}}{\partial X_\beta}\delta u_\alpha^{[0]} + \frac{\partial}{\partial X_\beta}\Bigl((\mathbb{S}^{[0]}_n)_{\alpha\beta}\frac{\partial\xi}{\partial X_\alpha}\Bigr) \delta \xi \right] \, \d A + H\oint_{\partial S} \left[ (\mathbb{S}^{[0]}_n)_{\alpha\beta} \, n_\beta \, \delta u_\alpha^{[0]} + (\mathbb{S}_n^{[0]})_{\alpha\beta} \frac{\partial \xi}{\partial X_\alpha} n_\beta \, \delta \xi \right] \d \ell \notag \\[2ex]
    &= -H \int_S \left[ \operatorname{div}(\mathbb{S}^{[0]}_n) \cdot \delta \mathbf{u}^{[0]}  + \operatorname{div}(\mathbb{S}^{[0]}_n \nabla \xi) \, \delta \xi \right] \d A + H \oint_{\partial S} \left[ \mathbb{S}^{[0]}_n \mathbf{n} \cdot \delta \mathbf{u}^{[0]} + \mathbb{S}^{[0]}_n \nabla \xi \cdot \mathbf{n} \, \delta \xi \right] \d\ell \, , \notag
\end{align}
being $\mathbf{n}$ the outward unit normal to the boundary of the central surface $\mathcal{S}$. 


Regarding the bending energy, we begin with the isotropic part of Eq.~\eqref{eq:W_mu}. We find that 
\begin{equation}
\begin{split}
\int_S &\frac{1}{2}\,\delta \Bigl((\Delta \xi - C_m)^2\Bigr) \, \d A  = \int_S (\Delta\xi - C_m)\,\Delta\delta\xi \, \d A = \int_S (\Delta\xi - C_m)\,\operatorname{div}\nabla\delta\xi \, \d A \\[1ex]
& = \int_S \operatorname{div}\Bigl((\Delta\xi - C_m)\nabla\delta\xi\Bigr) \, \d A - \int_S \nabla\delta\xi\cdot\nabla(\Delta\xi - C_m) \, \d A \\[1ex]
& = \int_S \operatorname{div}\Bigl((\Delta\xi - C_m)\nabla\delta\xi\Bigr) \, \d A - \int_S \operatorname{div}\Bigl(\delta\xi\,\nabla(\Delta\xi - C_m)\Bigr) \, \d A + \int_S \Bigl(\Delta^2\xi - \Delta C_m\Bigr)\delta\xi \, \d A \\[1ex] 
& = \underbrace{\oint_{\partial S} (\Delta \xi - C_m)\nabla \delta \xi \cdot \mathbf{n} \, \d\ell}_{:= \mathcal{C}_1} - \underbrace{\oint_{\partial S} \delta \xi \, \nabla(\Delta \xi - C_m) \cdot \mathbf{n} \, \d\ell}_{:= \mathcal{C}_2} + \int_S \Bigl(\Delta^2\xi - \Delta C_m\Bigr)\delta\xi \, \d A \, ,
\end{split}
\label{eq:variation_b_iso}
\end{equation}
where we have introduced the 2D bilaplacian $\Delta^2$. Note that $\mathcal{C}_1$ and $\mathcal{C}_2$ represent boundary terms. 

Then, we compute 
\begin{equation} \label{eq:variation_ba}
\int_S \delta (\nabla \xi \cdot \mathbf{b}_{\rm a}) \, \d A = \underbrace{\oint_{\partial S} \delta \xi \,  \mathbf{b}_{\rm a} \cdot \mathbf{n} \, \d\ell}_{:= \mathcal{C}_3} - \int_S \delta \xi \, \operatorname{div} \mathbf{b}_{\rm a} \, \d A \, . 
\end{equation}

For the last term, recalling Eq.~\eqref{eq:Acal} and following the procedure outlined in \citep{landau2012theory, dervaux2009morphogenesis, mihai2020plate}, we have 
\begin{align}
\mathcal{C}_4 & := \int_S \delta \mathcal{A} \, \d A \\[0.5ex]
& = \oint_{\partial S} \left(\mathbf{n}\cdot \nabla \delta \xi\right)  \Biggl[n_1^2 \left(\frac{\partial^2 \xi}{\partial Y^2} - \frac{\partial}{\partial Y}(a_{23}+a_{32}) + k_{22}\right) + n_2^2 \left( \frac{\partial^2 \xi}{\partial X^2} - \frac{\partial}{\partial X}(a_{13} + a_{31}) + k_{11}\right) \notag \\[0.4ex]
& \qquad \qquad \qquad \qquad + n_1 n_2 \left( -2\frac{\partial^2 \xi}{\partial X \partial Y} + \frac{\partial}{\partial Y}(a_{13}+a_{31}) + \frac{\partial}{\partial X}(a_{23}+a_{32}) - 2k_{12} \right)
\Biggr] \d\ell \notag \\[1ex]
& + \oint_{\partial S} \delta \xi \, \mathbf{t} \cdot \nabla \Biggl[ \left(n_1^2-n_2^2\right)\frac{\partial^2 \xi}{\partial X \partial Y} - n_1^2 \left(\frac{\partial}{\partial Y}(a_{13} + a_{31}) - k_{12}\right) + n_2^2 \left(\frac{\partial}{\partial X}(a_{23}+a_{32}) - k_{12}\right)  \notag \\[0.4ex]
& \qquad \qquad \qquad \qquad  + n_1 n_2 \left( \frac{\partial^2 \xi}{\partial Y^2} - \frac{\partial^2 \xi}{\partial X^2} - \frac{\partial}{\partial Y}(a_{23}+a_{32}) + \frac{\partial}{\partial X}(a_{13}+a_{31}) + k_{22} - k_{11} \right)\Biggr] \d\ell \, , \label{eq:C4} \\[1ex]
& \rev{= \oint_S \left(\mathbf{n}\cdot \nabla \delta \xi\right) \left(\mathbf{n}\cdot\mathbb{W}_{\rm a}^* \, \mathbf{n}\right) \, \d\ell - \oint_S \delta\xi \, \mathbf{t} \cdot \nabla \left[ \mathbf{n}\cdot\mathbb{W}_{\rm a}^*\,\mathbf{t} \right] \, \d\ell - \oint_S \delta\xi \, \mathbf{n}\cdot \mathbf{f}_{\rm a} \, \d\ell \, ,}
\end{align}
where $\mathbf{n} = \left(n_1, n_2\right)^{\rm T}$ and $\mathbf{t} := \left(-n_2, n_1\right)^{\rm T}$ is the tangent vector to the boundary. \rev{The contribution of $\mathbf{f}_{\rm a}$ is therefore confined to the boundary, since $\operatorname{div}\mathbf{f}_{\rm a} = 0$, so that the bulk contribution only includes $\operatorname{div}\mathbf{b}_{\rm a}$.} 

Similarly, the variation of the anisotropic term $\mathscr{W}_k$ of Eq.~\eqref{eq:W_k} is computed after some algebra as 
\begin{align} \label{eq:delta_aniso}
 \int_S \frac{1}{2} \delta \Bigl( (\mathbb{M}:\mathbb{W}_{\rm a})^2\Bigr) \, \d A & = \int_S (\mathbb{M}:\mathbb{W}_{\rm a}) (\mathbb{M}:\delta \mathbb{W}_{\rm a}) \, \d A \\[1ex]
 &= \underbrace{\oint_{\partial S} (\mathbb{M}:\mathbb{W}_{\rm a}) \mathbb{M}\nabla \delta \xi \cdot \mathbf{n} \, \d\ell}_{:= \mathcal{C}_5} - \underbrace{\oint_{\partial S} \delta \xi \,\operatorname{div}\bigl((\mathbb{M}:\mathbb{W}_{\rm a})\mathbb{M}\bigr) \, \d\ell}_{:= \mathcal{C}_6} + \int_S \delta \xi \operatorname{div}\operatorname{div} \Bigl((\mathbb{M}:\mathbb{W}_{\rm a})\mathbb{M}\Bigr)\d A \, , \notag
\end{align}
where the operator $\operatorname{div}\operatorname{div}$ is defined by $\operatorname{div}\operatorname{div}\mathbb{Q} = \frac{\partial Q_{ij}}{\partial X_i \partial X_j}$ for a second-order tensor $\mathbb{Q}$. 

Putting together all the terms: 
\begin{align} \label{eq:deltaW_bend}
\delta \mathscr{W}_{\rm bend} & = \frac{\mu H^3}{6}\left( 2\int_S \delta \xi \, (\Delta^2\xi - \Delta C_m) \, \d A + \int_S \delta \xi  \operatorname{div}\mathbf{b}_{\rm a} \, \d A + 2\mathcal{C}_1 - 2\mathcal{C}_2 - \mathcal{C}_3 - \mathcal{C}_4 \right) \\[1ex]
& + \frac{k_{44}H^3}{6}\left( 2\int_S \delta \xi \, \operatorname{div}\operatorname{div}\Bigl((\mathbb{M}:\mathbb{W}_{\rm a}\Bigr) \, \d A + 2\mathcal{C}_5 - 2\mathcal{C}_6 \right)  \notag \, .
\end{align}
Therefore, assuming that $F$ denotes the density of external forces acting along the normal to the surface, which scales as $F = \mathcal{O}(\xi^4/L^4)$, and by imposing $\delta \mathscr{W}_{\mathrm{tot}} = \delta \mathscr{W}_{\mathrm{str}} + \delta\mathscr{W}_{\mathrm{bend}} - \delta F = 0$ in the bulk for all admissible variations $\delta \mathbf{u}^{[0]}$ and $\delta \xi$, one obtains the two equilibrium equations:
\begin{align}
 &\operatorname{div} \mathbb{S}_n^{[0]} = \mathbf{0} \, , \\
 & D\left(\Delta^2 \xi - \Delta C_m\right) + D_k \operatorname{div}\operatorname{div}\bigl((\mathbb{M}:\mathbb{W}_{\rm a})\mathbb{M}\bigr) + \frac{D}{2} \operatorname{div}\mathbf{b}_{\rm a} - H \operatorname{div}\left(\mathbb{S}_n^{[0]}\nabla\xi\right) = F \, ,
\end{align}
where we have used Eqs.~\eqref{eq:stretching_variation}--\eqref{eq:deltaW_bend} and introduced the bending stiffness $D := \mu H^3 / 3$ and the anisotropic bending stiffness $D_k := k_{44} H^3 / 3$.

Regarding the boundary conditions, from $\delta\mathscr{W}_{\rm str} = 0$ we get $\mathbb{S}_n^{[0]}\mathbf{n} = \mathbf{0}$, in the absence of external boundary forces. Then, the remaining part reads: 
\begin{equation}
\frac{\mu H^3}{6} (2\mathcal{C}_1 - 2\mathcal{C}_2 - \mathcal{C}_3 - \mathcal{C}_4) + \frac{k_{44}H^3}{6}(2\mathcal{C}_5 - 2\mathcal{C}_6) = 0 \, .    
\end{equation}
Upon expanding 
\begin{equation} \label{eq:C5}
\mathcal{C}_5 = \oint_{\partial S}(\mathbb{M}:\mathbb{W}_{\rm a})(\mathbf{m}\cdot\mathbf{n})^2 \, \mathbf{n}\cdot\nabla \delta \xi \, \d\ell - \oint_{\partial S} \delta \xi \, \mathbf{t}\cdot\nabla\Bigl[(\mathbb{M}:\mathbb{W}_{\rm a})(\mathbf{m}\cdot\mathbf{n})(\mathbf{m}\cdot\mathbf{t}) \Bigr] \, \d\ell \, ,
\end{equation}
by combining the results of Eqs.~\eqref{eq:variation_b_iso}--\eqref{eq:variation_ba}--\eqref{eq:C4}--\eqref{eq:delta_aniso}--\eqref{eq:C5} and considering that the variations $\delta \xi$, $\mathbf{n}\cdot\nabla\delta\xi$ are independent, we arrive at the final system of boundary conditions: 
\begin{align}
    & 2\mu (\Delta \xi - C_m) - \mu\Biggl[n_1^2 \left(\frac{\partial^2 \xi}{\partial Y^2} - \frac{\partial}{\partial Y}(a_{23}+a_{32}) + k_{22}\right) + n_2^2 \left( \frac{\partial^2 \xi}{\partial X^2} - \frac{\partial}{\partial X}(a_{13} + a_{31}) + k_{11}\right)  \label{eq:BC1} \\[0.4ex] 
    & \qquad + n_1 n_2 \left( -2\frac{\partial^2 \xi}{\partial X \partial Y} + \frac{\partial}{\partial Y}(a_{13}+a_{31}) + \frac{\partial}{\partial X}(a_{23}+a_{32}) - 2k_{12} \right)
\Biggr] + 2k_{44}(\mathbb{M}:\mathbb{W}_{\rm a})(\mathbf{m}\cdot\mathbf{n})^2 = 0 \, , \notag \\[1ex]
& 2\mu \, \mathbf{n}\cdot\nabla(\Delta \xi - C_m) + \mu \, \mathbf{t} \cdot \nabla \Biggl[ \left(n_1^2-n_2^2\right)\frac{\partial^2 \xi}{\partial X \partial Y} - n_1^2 \left(\frac{\partial}{\partial Y}(a_{13} + a_{31}) - k_{12}\right) + n_2^2 \left(\frac{\partial}{\partial X}(a_{23}+a_{32}) - k_{12}\right)  \notag \\[0.4ex]
& \quad  + n_1 n_2 \left( \frac{\partial^2 \xi}{\partial Y^2} - \frac{\partial^2 \xi}{\partial X^2} - \frac{\partial}{\partial Y}(a_{23}+a_{32}) + \frac{\partial}{\partial X}(a_{13}+a_{31}) + k_{22} - k_{11} \right)\Biggr] + \mu \mathbf{n}\cdot\mathbf{b}_{\rm a} + 2k_{44} \, \mathbf{n}\cdot\operatorname{div}\bigl((\mathbb{M}:\mathbb{W}_{\rm a})\mathbb{M}\bigr) \notag\\[0.4ex]
& \quad + 2k_{44} \, \mathbf{t}\cdot \nabla \bigl[(\mathbb{M}:\mathbb{W}_{\rm a})(\mathbf{m}\cdot\mathbf{n})(\mathbf{m}\cdot\mathbf{t})\bigr] = 0 \, . \label{eq:BC2}
\end{align}

\section{Formulation in terms of Airy potential} \label{sec:Appendix_C}

The equation for the Airy potential $\phi$ can be computed by eliminating the in-plane displacements \( u^{[0]}_\alpha \) from the in-plane strain equations and by exploiting the effective constitutive equation defined in Eq.~\eqref{eq:reduced_constitutive_relations}. In particular, recalling Eq.~\eqref{eq:E0}, one can observe that
\begin{equation} \label{eq:phi_LHS}
\frac{\partial^2 (\mathbb{E}_{\rm e}^{[0]})_{11}}{\partial Y^2} + \frac{\partial^2 (\mathbb{E}_{\rm e}^{[0]})_{22}}{\partial X^2} - 2\frac{\partial^2 (\mathbb{E}_{\rm e}^{[0]})_{12}}{\partial X \partial Y} = -\frac{\partial^2}{\partial Y^2}\Bigl(a_{11} + \frac{1}{2}a_{31}^2\Bigr) - \frac{\partial^2}{\partial X^2}\Bigl(a_{22} + \frac{1}{2}a_{32}^2\Bigr) + \frac{\partial^2}{\partial X \partial Y}\Bigl(a_{12} + a_{21} + a_{31}a_{32}\Bigr) - [\xi,\xi].
\end{equation}
This expression does not depend on the in-plane displacements \( u^{[0]}_\alpha \). The next step is to express the in-plane strain \( \mathbb{E}_{\rm e}^{[0]} \) as a function of the in-plane stresses \( \mathbb{S}_n^{[0]} \). Hence, it is necessary to invert the effective constitutive relation \eqref{eq:reduced_constitutive_relations}. One way to view this problem is to write it using Voigt notation, namely,
\begin{equation}
 \begin{bmatrix}
(\mathbb{S}^{[0]}_n)_{11} \\[0.5ex]
(\mathbb{S}^{[0]}_n)_{22} \\[0.5ex]
(\mathbb{S}^{[0]}_n)_{12} 
\end{bmatrix} = \left( \underbrace{\mu
\begin{bmatrix}
4 & 2 & 0 \\
2 & 4 & 0\\
0 & 0 & 1
\end{bmatrix}}_{:= \mathbb{V}} + \underbrace{4k_{44} 
\begin{bmatrix}
\mathbb{M}_{11}^2 & \mathbb{M}_{11}\mathbb{M}_{22} & \mathbb{M}_{11}\mathbb{M}_{12} \\[0.3ex]
\mathbb{M}_{11}\mathbb{M}_{22} & \mathbb{M}_{22}^2 & \mathbb{M}_{12}\mathbb{M}_{22} \\[0.3ex]
\mathbb{M}_{11}\mathbb{M}_{12} & \mathbb{M}_{12}\mathbb{M}_{22} & \mathbb{M}_{12}^2
\end{bmatrix}}_{:= \mathbf{v}\otimes\mathbf{w}}\right)
\begin{bmatrix}
(\mathbb{E}_{\rm e}^{[0]})_{11} \\[0.5ex]
(\mathbb{E}_{\rm e}^{[0]})_{22} \\[0.5ex]
2(\mathbb{E}_{\rm e}^{[0]})_{12}
\end{bmatrix}   \, .
\end{equation}
The Sherman–Morrison formula then allows us to compute the inverse of the matrix block \(\mathbb{K}\) above as
\begin{equation}
\mathbb{K}^{-1} = \mathbb{V}^{-1} - \frac{\mathbb{V}^{-1}\mathbf{v}\otimes\mathbb{V}^{-1}\mathbf{w}}{1 + \mathbf{v}\cdot\mathbb{V}^{-1}\mathbf{w}} \, ,
\end{equation}
to express the in-plane strain \( \mathbb{E}_{\rm e}^{[0]} \) in terms of the stress $\mathbb{S}_n^{[0]}$. Then, the denominator on the right-hand side is given by:
\begin{equation}
1 + \mathbf{v}\cdot\mathbb{V}^{-1}\mathbf{w}  =  1 + \frac{4}{3}\frac{k_{44}}{\mu} := \widetilde{E}_k,
\end{equation}
which does not depend on the fiber angle \(\theta\). Consequently, the resulting relation is expressed as
\begin{equation}
\begin{split}
(\mathbb{E}_{\rm e}^{[0]})_{11} &= \frac{1}{3\mu}\Bigl( (\mathbb{S}_n^{[0]})_{11} -\frac{1}{2}(\mathbb{S}_n^{[0]})_{22} \Bigr) - \frac{k_{44}}{\mu^2 \widetilde{E}_k} \biggl[ (\mathbb{S}_n^{[0]})_{11}\left(\cos^2\theta - \frac{1}{3}\right)^2 + (\mathbb{S}_n^{[0]})_{22} \left(\cos^2\theta\sin^2\theta - \frac{2}{9}\right)\\
&\hspace{16em} + 2(\mathbb{S}_n^{[0]})_{12} \sin\theta\cos\theta \left(\cos^2\theta - \frac{1}{3}\right) \biggr], \\[1ex]
(\mathbb{E}_{\rm e}^{[0]})_{22} &= \frac{1}{3\mu}\Bigl( (\mathbb{S}_n^{[0]})_{22} -\frac{1}{2}(\mathbb{S}_n^{[0]})_{11}\Bigr) - \frac{k_{44}}{\mu^2 \widetilde{E}_k} \biggl[ (\mathbb{S}_n^{[0]})_{11}\left(\cos^2\theta\sin^2\theta - \frac{2}{9}\right) + (\mathbb{S}_n^{[0]})_{22} \left(\sin^2\theta - \frac{1}{3}\right)^2 \\
&\hspace{16em} +2(\mathbb{S}_n^{[0]})_{12} \sin\theta\cos\theta \left(\sin^2\theta - \frac{1}{3}\right) \biggr], \\[1ex]
(\mathbb{E}_{\rm e}^{[0]})_{12} &= \frac{1}{2\mu}(\mathbb{S}_n^{[0]})_{12} - \frac{k_{44}}{\mu^2 \widetilde{E}_k} \biggl[ (\mathbb{S}_n^{[0]})_{11} \sin\theta\cos\theta\left(\cos^2\theta - \frac{1}{3}\right) + (\mathbb{S}_n^{[0]})_{22} \sin\theta\cos\theta \left(\sin^2\theta - \frac{1}{3}\right) \\
&\hspace{10em} + 2(\mathbb{S}_n^{[0]})_{12}\sin^2\theta\cos^2\theta\biggr] \, .
\end{split}
\end{equation}
After some algebraic manipulation, one can rearrange these terms to obtain
\begin{equation} \label{eq:phi_RHS}
\frac{\partial^2 (\mathbb{E}_{\rm e}^{[0]})_{11}}{\partial Y^2} + \frac{\partial^2 (\mathbb{E}_{\rm e}^{[0]})_{22}}{\partial X^2} - 2\frac{\partial^2 (\mathbb{E}_{\rm e}^{[0]})_{12}}{\partial X \partial Y} = \frac{1}{3\mu}\Delta^2 \phi - \frac{1}{E_k} \operatorname{div}\operatorname{div}\Biggl(\left[ \left(\mathbb{M}-\frac{2}{3}\mathbb{I}\right)\otimes\left(\mathbb{M}-\frac{2}{3}\mathbb{I}\right)\right]\mathbb{H}(\phi)\Biggr) \, ,
\end{equation}
where $\mathbb{H}$ denotes the Hessian matrix. Combining Eq.~\eqref{eq:phi_LHS} and Eq.~\eqref{eq:phi_RHS} yields the result presented in Eq.~\eqref{eq:final_eq_phi}. 

Note that the $\operatorname{div}\operatorname{div}$ term can be expanded using the Leibniz rule as: 
\begin{equation}\label{eq:Leibniz_MM}
\operatorname{div}\operatorname{div}\Bigl((\widetilde{\mathbb{M}}\otimes\widetilde{\mathbb{M}})\mathbb{H}(\phi)\Bigr) = \operatorname{div}\operatorname{div}\left(\widetilde{\mathbb{M}}\otimes\widetilde{\mathbb{M}}\right) : \mathbb{H}(\phi) + 2 \operatorname{div}\left(\widetilde{\mathbb{M}}\otimes\widetilde{\mathbb{M}}\right) \tripledot \nabla\mathbb{H}(\phi) + \left(\widetilde{\mathbb{M}}\otimes\widetilde{\mathbb{M}}\right) :: \nabla\nabla\mathbb{H}(\phi) \, ,  
\end{equation}
where we have denoted by $\tripledot$ the scalar product between third-order tensors and by $::$ the one between fourth-order tensors. In particular, the gradient of the Hessian and the divergence of $\widetilde{\mathbb{M}}\otimes\widetilde{\mathbb{M}}$ are the third-order tensors defined respectively by
\begin{equation}
 [\nabla\mathbb{H}(\phi)]_{ijk} = \frac{\partial^3 \phi}{\partial X_i \partial X_j \partial X_k} \, , \qquad \operatorname{div}[(\widetilde{\mathbb{M}}\otimes\widetilde{\mathbb{M}})]_{ijk} = \frac{\partial}{\partial X_h} \left(\widetilde{M}_{ij} \,\widetilde{M}_{kh}\right)  \, . 
\end{equation}
Clearly, if $\mathbb{M}$ does not depend on space, the first two terms on the right-hand side of Eq.~\eqref{eq:Leibniz_MM} vanish, yielding the result presented in Eq.~\eqref{eq:divdiv_MMH}.

\section{\rev{Ellipsoidal case: Invariance of the stretching energy}}\label{sec:Appendix_D}
\noindent \rev{In this Appendix, with reference to Sect.~\ref{sec:ellipsoid}, we show that, for a rotationally symmetric mid-surface $\mathcal{S}$, the stretching energy is independent of the angle $\theta$. We assume for instance that $\mathcal{S}$ is a thin disk, which is a reasonable idealization for cells that are not strongly elongated. The latter case lies beyond the scopes of this work, as discussed in Sect.~\ref{sec:cylinder}.} 

\rev{From Eqs.~\eqref{eq:active_cells} and \eqref{eq:xi_ellips}, on an ellipsoidal surface with principal curvatures determined by $A$, $B$, and $R$, one has $[\xi,\,\xi] = \frac{\kappa_1^2}{A^2 B^2},$ and $C_{a,\phi} = 0$, so that Eq.~\eqref{eq:final_eq_phi_uniform} governing the Airy potential $\phi$ reduces to
\begin{equation}
    \label{eq:airy_ellipsoid}
    \frac{1}{E_Y}\,\Delta^2\phi
    \;-\; \frac{1}{E_k}\!\left(\Delta_{\mathbb{M}}
      - \frac{2}{3}\Delta\right)^{\!2}\phi
    \;=\; - \frac{\kappa_1^2}{A^2 B^2} \, ,
\end{equation}}
\rev{supplemented by clamped boundary conditions on $\partial\mathcal{S}$.}

\rev{Let $\phi_0$ denote the solution of Eq.~\eqref{eq:airy_ellipsoid} for the
reference orientation $\theta = 0$, in which the anisotropy axis is aligned with the $X$-direction so that $\mathbb{M} = \mathbf{e}_1\otimes \mathbf{e}_1$. For a general angle $\theta \in [0, \pi)$, introduce the in-plane rotation}
\begin{equation}
    \rev{g_\theta \colon (X,Y) \;\longmapsto\; (\widetilde{X},\widetilde{Y}) \, ,
    \qquad
    \begin{pmatrix}\widetilde{X}\\\widetilde{Y}\end{pmatrix}
    =
    \begin{pmatrix}\cos\theta & \sin\theta \\ -\sin\theta & \cos\theta\end{pmatrix}
    \begin{pmatrix}X\\Y\end{pmatrix} \, ,}
\end{equation}
\rev{which maps physical coordinates $(X,Y)$ to fiber-aligned coordinates $(\widetilde{X}, \widetilde{Y})$ so that the fiber direction is mapped onto $\widetilde{\mathbf{e}}_1$. Under $g_\theta$, the differential operators transform as}
\begin{equation}
    \rev{\Delta \;\longmapsto\; \widetilde{\Delta} \, ,
    \qquad
    \Delta_{\mathbb{M}} \;\longmapsto\;
    \frac{\partial^2}{\partial\widetilde{X}^2} \, ,}
\end{equation}
\rev{since the Laplacian is rotationally invariant and the directional Laplacian $\Delta_{\mathbb{M}}$ reduces to the second derivative along the fiber direction in the rotated frame. Equation~\eqref{eq:airy_ellipsoid} therefore takes the same form in the rotated frame:} 
\begin{equation}
    \label{eq:airy_rotated}
  \rev{  \frac{1}{E_Y}\,\widetilde{\Delta}^2\phi
    \;-\; \frac{1}{E_k}\!\left(\frac{\partial^2}{\partial\widetilde{X}^2}
      - \frac{2}{3}\widetilde{\Delta}\right)^{\!2}\phi
    \;=\; - \frac{\kappa_1^2}{A^2 B^2} \, ,}
\end{equation}
\rev{which is structurally identical to Eq.~\eqref{eq:airy_ellipsoid} at
$\theta = 0$. Since the cell $\mathcal{S}$ is mapped onto itself by any rotation and the clamped boundary conditions are
preserved, by uniqueness the solution of Eq.~~\eqref{eq:airy_rotated} at angle $\theta$, expressed in physical coordinates, is the pull-back of $\phi_0$ under $g_\theta$:}
\begin{equation}
\rev{ \phi\bigl(X,Y;\theta\bigr)
    \;=\; \phi_0\left(g_\theta(X,Y)\right) \, .}
\end{equation}
\rev{Under the assumption of homogeneous anisotropic angle $\theta$, the stretching energy associated with $\phi$ is}
\begin{equation}
    \label{eq:W_stretch}
    \rev{\mathscr{W}_{\mathrm{str}}(\phi;\theta)
    \;=\;
    \int_{\mathcal{S}} \left\{
      \frac{1}{3\mu}\Bigl[(\Delta\phi)^2 -3\,[\phi,\phi]\Bigr]
      \;-\;
      \frac{1}{E_k}\!\left(\Delta_{\mathbb{M}}\phi
        - \frac{2}{3}\Delta\phi\right)^{\!2}
    \right\} \d A \, .}
\end{equation}
\rev{Applying the change of variables $(\widetilde{X},\widetilde{Y}) = g_\theta(X, Y)$ and using the chain rule, together with the rotational symmetry of $\mathcal{S}$ (so that the Jacobian
is unity and the integration domain is unchanged), one obtains}
\begin{equation}
    \label{eq:W_invariant}
   \rev{ \begin{aligned}
        \mathscr{W}_{\mathrm{str}}(\phi,\theta)
        &\;=\;
        \int_{\mathcal{S}} \left\{
          \frac{1}{3\mu}\Bigl[(\widetilde\Delta\phi)^2
            -3\,[\phi,\phi]\Bigr]
          \;-\;
          \frac{1}{E_k}\!\left(\frac{\partial^2\phi}{\partial\widetilde{X}^2}
            - \frac{2}{3}\widetilde\Delta\phi\right)^{\!2}
        \right\} \d\widetilde{A} \\[6pt]
        &\;=\;
        \int_{\mathcal{S}} \left\{
          \frac{1}{3\mu}\Bigl[(\Delta\phi_0)^2 -3\,[\phi_0,\phi_0]\Bigr]
          \;-\;
          \frac{1}{E_k}\!\left(\frac{\partial^2\phi_0}{\partial X^2}
            - \frac{2}{3}\Delta\phi_0\right)^{\!2}
        \right\} \d A \\[4pt]
        &\;=\; \mathscr{W}_{\mathrm{str}}(\phi_0,0),
    \end{aligned}}
\end{equation}
\rev{for all $\theta\in[0,\pi)$. Hence, under the assumption that the reference cell domain is rotationally symmetric, the stretching energy is independent of the anisotropy orientation angle.}

\bibliographystyle{elsarticle-harv} 
\bibliography{refs}

\end{document}